\documentclass[]{config/antgroup}
\usepackage{config/antgroup}
\usepackage{xcolor}
\definecolor{mygray}{RGB}{120,120,120}
\newcommand*\justify{%
  \fontdimen2\font=0.4em% interword space
  \fontdimen3\font=0.2em% interword stretch
  \fontdimen4\font=0.1em% interword shrink
  \fontdimen7\font=0.1em% extra space
  \hyphenchar\font=`\-% allowing hyphenation
}

\renewcommand{\texttt}[1]{%
\begingroup
\ttfamily
\begingroup\lccode`~=`/\lowercase{\endgroup\def~}{/\discretionary{}{}{}}%
\begingroup\lccode`~=`[\lowercase{\endgroup\def~}{[\discretionary{}{}{}}%
\begingroup\lccode`~=`.\lowercase{\endgroup\def~}{.\discretionary{}{}{}}%
\catcode`/=\active\catcode`[=\active\catcode`.=\active
\justify\scantokens{#1\noexpand}%
\endgroup
}

\PassOptionsToPackage{numbers, compress}{natbib}

\usepackage[utf8]{inputenc}         %
\usepackage[T1]{fontenc}            %
\usepackage{graphicx}
\usepackage{todonotes}
\usepackage{multirow}
\usepackage{hyperref}
\usepackage[capitalize,noabbrev,nameinlink,sort]{cleveref}
\usepackage{subcaption}
\usepackage{multicol}
\usepackage{array}
\usepackage{enumitem}
\usepackage{algorithm}
\usepackage{algpseudocode}
\usepackage{tabularx}
\usepackage{listings}
\usepackage{makecell}
\usepackage{xspace}
\usepackage{colortbl}
\usepackage{fontawesome5}
\usepackage{pifont}
\usepackage[normalem]{ulem}
\useunder{\uline}{\ul}{}
\usepackage{tablefootnote}
\usepackage{tcolorbox}
\usepackage{arydshln}
\usepackage[toc, page, header]{appendix}
\usepackage{tabularx}
\usepackage{listings}

\lstdefinelanguage{yaml}{
  keywords={true,false,null,yes,no},
  keywordstyle=\color{blue}\bfseries,
  basicstyle=\ttfamily\small,
  sensitive=false,
  comment=[l]{\#},
  commentstyle=\color{gray}\ttfamily,
  stringstyle=\color{orange},
  morestring=[b]',
  morestring=[b]"
}

\usepackage{amsmath}
\usepackage{amssymb}
\usepackage{amsfonts}                               % blackboard math symbols
\usepackage{amsthm}
\usepackage[mathcal]{eucal}
\usepackage{mathrsfs}
\usepackage{bm}                                     % bm command
\usepackage{blkarray}                               % to support matrix
\usepackage{nicefrac}                               % compact symbols for 1/2, etc.
\usepackage{bbm}

\usepackage{wrapfig}
\usepackage{graphicx}                               % include pdf figures
\usepackage{caption}
\usepackage{tikz}
\usepackage{circuitikz}
\usetikzlibrary{patterns,snakes}
\usetikzlibrary{positioning,calc,fit,decorations.pathmorphing,shapes.geometric, shapes.gates.logic.US, calc}
\usetikzlibrary{arrows,arrows.meta,decorations.markings,shapes,shapes.arrows}
\usetikzlibrary{decorations,decorations.pathreplacing}
\usetikzlibrary{backgrounds}
\usepackage{filecontents}                           % support to pgfplots
\usepackage{pgfplots}
\usepackage{pgfplotstable}
\usepgfplotslibrary{groupplots}
\usepackage{scalefnt}
\pgfplotsset{compat=newest}
\usepackage{xcolor}

\definecolor{firstcolor}{HTML}{C3423F}
\definecolor{secondcolor}{HTML}{2A4B8C}
\definecolor{aworld_blue}{HTML}{4e81ff}
\definecolor{aworld_cyan}{HTML}{41d7fa}
\definecolor{aworld_teal}{HTML}{5fede4}
\definecolor{coral}{RGB}{255,127,80}
\definecolor{darkgreen}{RGB}{0,100,0}
\definecolor{darkyellow}{RGB}{204,153,0}
\definecolor{salmon}{RGB}{250,128,114}
\definecolor{darkred}{RGB}{150,0,0}

\hypersetup{
  colorlinks=true,
  linkbordercolor=aworld_blue,
  citebordercolor=aworld_cyan,
  urlbordercolor=aworld_teal
}

\definecolor{improvementblue}{RGB}{55,126,184}    % Blue (#377eb8)
\definecolor{degradationorange}{RGB}{230,85,13}   % Orange (#e6550d)

\def\eqref#1{equation~\ref{#1}}
\def\1{\bm{1}}

\DeclareMathAlphabet{\mathsfit}{\encodingdefault}{\sfdefault}{m}{sl}
\SetMathAlphabet{\mathsfit}{bold}{\encodingdefault}{\sfdefault}{bx}{n}

\begin{document}

\title{
 Privacy-Preserving EHR Transformation with Mathematical Guarantees: A Human–AI Co-Designed Solution}

\author{
  Maolin Wang\textsuperscript{1,\dag},
  Beining Bao\textsuperscript{1,\dag},
  % SciencePal\textsuperscript{1},
  Gan Yuan\textsuperscript{2},
  Hongyu Chen\textsuperscript{1},
  Bingkun Zhao\textsuperscript{1},
  Baoshuo Kan\textsuperscript{1},
  Jiming Xu \textsuperscript{4},
    Qi Shi\textsuperscript{1},
  Yinggong Zhao\textsuperscript{1},
  Yao Wang\textsuperscript{1},
  Wei-Ying Ma\textsuperscript{1},
    Jun Yan\textsuperscript{1,2,3}
}

{\renewcommand\thefootnote{}
\footnotetext{\textsuperscript{\dag}Equal Contribution.}
}

\affiliation[1]{Hong Kong Institute of AI for Science (HKAI-Sci), City University of Hong Kong\quad\quad\quad\quad\quad\quad\quad\quad\quad\quad}
\affiliation[2]{Department of Biostatistics, City University of Hong Kong\quad\quad\quad\quad\quad\quad\quad\quad\quad\quad\quad\quad\quad\quad\quad\quad\quad\quad\quad\quad\quad}
\affiliation[3]{Institute of Digital Medicine, City University of Hong Kong\quad\quad\quad\quad\quad\quad\quad\quad\quad\quad\quad\quad\quad\quad\quad\quad\quad\quad\quad\quad\quad}
\affiliation[4]{YIDU TECH co. Ltd.}

\begin{center}
{\Large\color{gray}Technical Report: Toward Making Clinical Data Available and Visible}
\end{center} 

\maketitle

\vspace{-2em}
\begin{center}
\href{https://github.com/HKAI-Sci/EHR-Privacy-Geometric-Operators}{\faGithub\hspace{0.3em}~ Code \& Projects}
\end{center}
\vspace{0.5em}

%%%%%%%%%%%%%%%%%%%%%%%%%%%%%%%%%%%%%%%%%%%%%%%%%%%%%%%%%%%%%%%%%%
\begin{abstract}
Electronic health records (EHRs) and other real-world clinical data are essential for clinical research, medical artificial intelligence, and life science, but their sharing is severely limited by privacy, governance, and interoperability constraints. These barriers create persistent data silos that hinder multi-center studies, large-scale model development, and broader biomedical discovery.
Existing privacy-preserving approaches, including multi-party computation and related cryptographic techniques, provide strong protection but often introduce substantial computational overhead, reducing the efficiency of large-scale machine learning and foundation-model training. In addition, many such methods make data usable for restricted computation while leaving them effectively invisible to clinicians and researchers, limiting their value in workflows that still require direct inspection, exploratory analysis, and human interpretation.
We propose a real-world-data transformation framework for privacy-preserving sharing of structured clinical records. Instead of converting data into opaque representations, our approach constructs transformed numeric views that preserve medical semantics and major statistical properties while, under a clearly specified threat model, provably breaking direct linkage between those views and protected patient-level attributes. Through collaboration between computer scientists and the AI agent \textbf{SciencePal}, acting as a constrained tool inventor under human guidance, we design three transformation operators that are non-reversible within this threat model, together with an additional mixing strategy for high-risk scenarios, supported by theoretical analysis and empirical evaluation under reconstruction, record linkage, membership inference, and attribute inference attacks.

Although we implement and evaluate the framework on intensive care unit (ICU) data as one representative clinical use case, it is intended for structured EHR and real-world clinical datasets in general, offering a practical and computationally feasible path toward privacy-preserving clinical data access, sharing, and exchange, and ultimately supporting multi-center studies, cross-institutional training of large-scale models, and translational life-science research.
\end{abstract}
%%%%%%%%%%%%%%%%%%%%%%%%%%%%%%%%%%%%%%%%%%%%%%%%%%%%%%%%%%%%%%%%%%

\section{Introduction}
\label{sec:intro}

\subsection{From Model Bottlenecks to Data Bottlenecks}

Electronic health records (EHRs) and other forms of real-world clinical data have become central resources in modern clinical AI. 
Large neural networks and foundation models are now widely applied to diagnosis prediction, risk stratification, clinical decision support, and multimodal patient summarization across diverse healthcare settings~\citep{Johnson23MIMICIV,johnson2016mimic,harutyunyan2019multitask}.
For many hospitals and research groups, having a sufficiently powerful model is no longer the primary obstacle: open-source and commercial models are increasingly abundant, and cloud compute is relatively accessible~\citep{rajpurkar2017chexnet}.

In practice, a more pressing concern repeatedly surfaces in our collaborations with hospital IT and data governance teams: the real bottleneck lies not in the model, but in the data. 
Under the joint constraints of regulation, ethics, and institutional governance, high-quality clinical data containing longitudinal records, time series, laboratory results, medications, and treatment histories are often required to remain strictly within hospital intranets or controlled environments~\citep{van2025privacy,conduah2025data,goldberger2000physiobank}. They cannot be freely copied to research clouds, let alone uploaded to third-party platforms~\citep{kaissis2020secure,rieke2020future,aggarwal2008general}. 
Even for internal research teams, many hospitals expose only restricted query interfaces or sandboxed environments, while external collaborators often have access only to heavily aggregated or de-identified summaries.

These arrangements are essential for patient privacy, but they create a shared practical difficulty for clinical research and deployment: models can move, data cannot.
Algorithm engineers struggle to perform systematic exploratory data analysis (EDA), quality checks, and pipeline debugging on the true data distribution.
Clinicians lack sufficiently rich cohort-level views to judge whether a study population is well defined or whether a statistical pattern is clinically plausible.
At the current stage of clinical AI, data availability and governability, rather than model capacity alone, constitute the primary bottleneck~\citep{deng2022explainable,rojas2018predicting}.
Although this challenge applies broadly to structured EHR and real-world clinical datasets, in the following sections we use ICU data as a representative and practically important showcase.

\subsection{Mainstream Approaches: Usable but Invisible}

To enable joint analysis and modeling without exposing plaintext data, the privacy-preserving computation community has developed a rich set of methods, including multi-party computation (MPC), homomorphic encryption (HE), trusted execution environments (TEE), federated learning, and differentially private (DP) query or training mechanisms. These approaches have substantially advanced the state of secure data use and are especially valuable in settings where formal protection guarantees are paramount and direct human access to raw records is neither necessary nor desirable. Over the past decade, this toolbox has expanded from classical secure query answering to deep and federated models, with practical implementations of neural networks over encrypted data and large-scale federated optimization now available in both research prototypes and open-source frameworks~\citep{gilad2016cryptonets,yuan2025approxhe,mcmahan2017learning,kairouz2021fl,munjal2023systematic}.

From a systems perspective, however, many of these methods occupy what may be called the \emph{usable but invisible} design quadrant: models and algorithms can train or infer on encrypted values, protected hardware, distributed parameters, or privacy-filtered outputs, while humans see little to none of the underlying individual records or per-point values. This property is a strength in many high-value applications, such as privacy-sensitive multi-institutional model training, secure cross-site analytics, and the release of protected aggregate statistics~\citep{mcmahan2017learning,dwork2014algorithmic,munjal2023systematic}. In such scenarios, the primary requirement is trustworthy computation rather than visible data access, and cryptographic or DP-based mechanisms are an appropriate---and often necessary---engineering choice.

Several lines of work nonetheless point out that this design point can become more restrictive in everyday clinical research and real-world model development. First, many cryptographic and privacy-preserving pipelines introduce substantial overhead in computation, communication, and system complexity. Empirical evaluations of HE- and MPC-based deep learning schemes, as well as surveys of approximate-HE PPML systems and federated learning, report significant slowdowns and communication costs compared with plaintext training, even for moderate-size neural networks~\citep{gilad2016cryptonets,yuan2025approxhe,kairouz2021fl}. While such costs may be acceptable for narrowly scoped secure analytics, they can become a practical burden for large-scale machine learning workflows, repeated ablation studies, and especially foundation-model training on high-dimensional clinical data, where practitioners already face tight GPU budgets and long wall-clock times. Differentially private training introduces an additional layer of trade-offs: DP-SGD protects training examples but typically requires more iterations and can reduce accuracy, with the degradation being particularly pronounced for complex models and under-represented subgroups~\citep{bagdasaryan2019dpimpact}. These efficiency and utility costs do not negate the value of these methods, but they do make it harder to adopt them as the default for exploratory, iteration-heavy large-model development.

Second, many real-world research tasks still depend on data being not only computable, but also inspectable. Algorithm engineers often need to run detailed exploratory data analysis, quality checks, failure diagnosis, and pipeline debugging on actual numeric tables in order to identify missingness patterns, temporal misalignment, distribution shift, or data processing errors~\citep{pineau2021improving}. Qualitative studies of high-stakes AI similarly find that practitioners spend substantial effort on ``data work''---cleaning, reformatting, and visually inspecting raw and intermediate datasets---and that breakdowns in this process can trigger cascading downstream failures~\citep{sambasivan2021datacascades}. In clinical domains, secondary analyses of electronic health records (EHRs) and other real-world data repeatedly emphasize that these sources are heterogeneous, noisy, and biased, and thus require intensive preprocessing and medical validation before robust modeling is possible~\citep{liu2022rwd}. Clinicians rely on visible cohort-level data views to judge whether a study population is medically coherent and whether an observed pattern is clinically plausible, while regulators and hospital IT teams may require data representations that are interpretable enough to support practical auditing and governance.

As a result, even when hospitals possess sizable EHR warehouses and substantial computing infrastructure, there remains a gap between methods that are secure in principle and data views that are truly usable in day-to-day scientific practice. This gap is particularly important for structured clinical records and real-world data, where privacy protection must coexist with interpretability, workflow efficiency, and the operational realities of collaborative research and large-model development~\citep{artsi2025large}.

\subsection{Human--AI Co-Design with SciencePal}

Privacy-preserving yet clinically interpretable data views require transformation operators that satisfy multiple, sometimes competing, requirements: they should preserve medical meaning and cohort-level statistics, make it infeasible to reverse sensitive individual information under our threat model, and remain simple and cheap enough to deploy on routine hospital infrastructure. To search this design space, we adopt a human--AI co-design paradigm.

Rather than letting an AI agent use high-compute tools, we ask the AI to \emph{design} low-compute tools~\citep{shin2025r,niu2025flowmodularizedagenticworkflow,shahin2025agents}. The AI role in this process is played by \textbf{SciencePal}\footnote{\url{https://sciencepal.ai}}, an AI research assistant that provided preliminary ideas and candidate proposals at each stage. Human researchers define the problem, specify constraints, and validate every claim; SciencePal explores the space of possible operators and attack sketches under these constraints. The overall protocol proceeds in three iterative phases.

In the first phase, human researchers specify qualitative and quantitative constraints that any candidate operator must satisfy. At a high level, cohort-level means and variances should be preserved, perturbations should be uniformly bounded by a small, human-controlled privacy knob, most data points should actually be perturbed rather than left unchanged, and per-column computation should scale linearly with the number of records and run on commodity CPUs without long-lived shared keys. Formal definitions and proofs of these properties are given later in the paper, but here they serve as a checklist guiding the design search.

In the second phase, SciencePal acts as a constrained tool inventor. Given the human specification, it searches for operator families in the literature that approximately satisfy these requirements and, upon identifying a gap, proposes new geometric operators and mixing strategies that are intended to meet them. These proposals come with informal geometric intuition, standardized forms, Python pseudo-code, and explicit claims about which constraints are satisfied. Crucially, these are \emph{preliminary ideas}: SciencePal contributes conceptual scaffolding, not verified tools.

In the third phase, the human team acts as both reviewer and attacker. We formalize the candidate operators, prove or refute the stated properties, and discard those that violate key constraints. On a MIMIC-IV ICU subset~\citep{Johnson23MIMICIV}, we instantiate a Privacy Evaluation Protocol analogous to cryptographic security games, with three leakage levels (L0/L1/L2: no, few, or many paired samples) and four attack families (A: reconstruction, B: record linkage, C: membership inference, D: attribute inference)~\citep{shokri2017membership,carlini2021extracting,dick2023confidence}. We use these attacks to identify operators and parameter ranges that, under our threat model, provably prevent exact recovery of protected information while still preserving medical semantics and statistical structure, and to separate them from candidates that, although geometrically well-structured, remain highly invertible under strong attacks~\citep{yeom2018privacy,hu2022membership}.

\subsection{Summary of the Framework and Contributions}

Our framework can be viewed as a solution for transforming structured clinical and real-world data into views that remain clinically interpretable and statistically faithful, while making it theoretically impossible, under our threat model, to exactly reverse protected individual-level information. Privacy strength and utility loss are controlled by a single, human-interpretable privacy knob, enabling practitioners to tune the trade-off instead of relying on opaque de-identification heuristics. This makes ``usable and visible'' privacy-preserving numeric views a realistic option for many clinical and research workflows.

Formally, we cast column-level transformations as constrained motions on a \emph{mean-variance manifold} with an \(\alpha\)-hierarchy. In standardized space, each column lies on the intersection of a zero-mean hyperplane and a fixed-norm sphere, and all operators are required to preserve cohort-level means and variances while keeping perturbations within a unified \(\ell_\infty\) bound \(\alpha\)~\citep{freksen2021introduction}. This geometric view allows us to reason about which parts of the data distribution are preserved, which are randomized, and how irreversibility arises.

Three base operators emerge from this process.

\textbf{T1: Local triplet rotations.} Along the time axis, each standardized sequence is partitioned into short windows, and small random orthogonal rotations are applied in the local mean-zero subspace, preserving short-range autocorrelation while injecting controlled noise~\citep{yoon2019time}.

\textbf{T2: Noise plus manifold projection.} Controlled noise is added in standardized space and then reprojected by re-centering and re-scaling onto the mean-variance manifold, which is particularly effective at preserving marginal distributions and multi-column correlation~\citep{goodfellow2014generative,xu2019modeling}.

\textbf{T3: Global Householder reflection (negative case).} We apply a single Householder reflection in the mean-zero subspace. This transformation nearly preserves all first and second order statistics as well as autocorrelation, but remains highly invertible under strong reconstruction attacks, and therefore serves as a negative control~\citep{narayanan2008robust}.

Early experiments with fixed bounds in physical units revealed severe imbalance: low-variance variables were over-perturbed while high-variance variables were barely affected~\citep{roman2023evaluating}. Switching to a unified z-score bound \(\alpha\) automatically adapts perturbation magnitude to each variable's scale, making the privacy-utility trade-off curve more interpretable. We further observe that merely increasing \(\alpha\) yields gradual rather than step-like privacy gains: even at large \(\alpha\), some variables still admit high reconstruction \(R^2\)~\citep{li2022local}. To address this, we introduce a per-stay orthogonal Q-mixing extension for a small set of high-risk variables. For each stay, a \(48 \times 48\) orthogonal mixing is applied to local time windows before applying T1 or T2. Under this construction, and with \(\alpha\) fixed at \(1.0\), linear reconstruction \(R^2\) for sensitive variables such as heart rate and glucose is driven close to zero, while preserving distributional shape and downstream prediction performance within acceptable ranges~\citep{goncalves2020generation,yuan2023ehrdiff}. Together with our formal analysis, this provides both theoretical and empirical evidence that, at appropriate settings, the transformed views do not allow attackers in our model to reverse key privacy-sensitive attributes.

The main contributions of this paper are:

\begin{enumerate}[leftmargin=*]
  \item \textbf{A real-world-data transformation solution that keeps data usable and visible under privacy constraints.} We introduce a unified geometric framework for column-level EHR and real-world clinical data transformations on a mean-variance manifold, providing a single privacy knob \(\alpha\) that trades small, quantifiable utility loss for provable non-reversibility of protected information under our threat model, while preserving medical semantics and key cohort-level statistics.

  \item \textbf{A family of non-reversible operators under our threat model and a tunable privacy-utility mechanism.} Building on this framework, we propose operators T1, T2, and a negative-control operator T3, together with a per-stay orthogonal Q-mixing extension that sharply reduces linear reconstruction \(R^2\) for high-risk variables without changing \(\alpha\). We accompany these with a Privacy Evaluation Protocol with L0/L1/L2 leakage levels and A/B/C/D attack families to systematically evaluate privacy and utility.

  \item \textbf{An EHR-Privacy-Agent system for multi-scenario deployment.} We build a three-layer system (operators, modules, skills) running nightly CPU-only de-identification over in-hospital caches, with configurable privacy profiles for in-hospital research, teaching, and strong-privacy export scenarios, making privacy-preserving yet interpretable numeric views operationally feasible.

  \item \textbf{A human--AI co-design methodology with SciencePal.} We demonstrate a methodology in which SciencePal provides preliminary geometric ideas and experimental sketches, while human researchers specify constraints, prove properties, design attacks, and select deployable tools. Failures such as T3 are preserved as systematic negative cases, and the full division of labor is documented in the appendix, illustrating how an AI ``scientist'' can expand the search space while humans retain responsibility for the final scientific claims.
\end{enumerate}
\section{Background: Privacy-Preserving Data Release and Analysis}
\label{sec:background}

\subsection{Privacy-Preserving Data Publishing and Mining}

Classical work on privacy-preserving data publishing (PPDP) assumes that a data owner wishes to release a static dataset for offline analysis while reducing the risk of re-identification or sensitive attribute inference. Early work focuses on tabular microdata through two main lines of techniques. Equivalence-class based anonymization, including $k$-anonymity, $\ell$-diversity, and $t$-closeness, generalizes or suppresses quasi-identifiers so that each released record falls into an equivalence class of size at least $k$~\citep{sweeney2002k}. Perturbation and permutation of numerical attributes, where values are blurred with noise, permuted, or aggregated into cluster centers, provides an alternative route to obscuring individual values~\citep{Marino19DataSifter,hammer2024semi,agrawal2000privacy}.

Privacy-preserving data mining (PPDM) concentrates on performing mining tasks without directly releasing the dataset. Randomized response allows each data provider to locally randomize her own record before sending it to the analyst~\citep{warner1965randomized}. Federated and distributed learning frameworks split training across multiple data holders, aggregating intermediate results through secure protocols without exposing raw records~\citep{mcmahan2017learning,kerrigan2020differentially}.

Together, PPDP and PPDM provide a rich toolbox for analyzing without fully exposing raw data~\citep{fung2010privacy,aggarwal2004condensation}. However, most of this literature implicitly assumes that the analyst cares primarily about a derived statistical or mining result rather than a tabular view that human experts can directly use for EDA and visualization. Direct human access to per-column values, temporal structure, and cohort-level shapes is rarely treated as a first-class requirement.
Two mismatches with the in-hospital structured EHR scenario (instantiated here on ICU data) are worth noting. Classical PPDP targets one-time release of a static dataset to external researchers, whereas we need repeated long-term in-hospital reuse: clinicians and engineers perform EDA, visualization, and pipeline debugging on cohorts repeatedly inside a secure hospital environment. Additionally, traditional PPDM views humans as consumers of final results rather than as participants who inspect data quality and structure midway through the workflow. In our setting, clinicians and IT staff both need to see the approximate shape and correlation structure of each column directly on the privacy-enhanced view.

\subsection{Cryptographic Protocols in the Usable-but-Invisible Quadrant}

Over the past decade, cryptography and systems security have produced powerful protocols that enable joint computation without revealing plaintext data. Secure multi-party computation (MPC) splits raw data into shares distributed across multiple parties, which jointly perform training or inference without reconstructing plaintext~\citep{borges2025using}. Homomorphic encryption (HE) allows models to operate directly on encrypted data via supported algebraic operations~\citep{goldreich2004foundations}. Trusted execution environments (TEE) run sensitive computation inside hardware enclaves with memory isolation and remote attestation, so that external parties cannot inspect plaintext data or intermediate states~\citep{ohrimenko2016oblivious}.

From a systems viewpoint, these methods lie in a usable-but-invisible design quadrant: models operate on protected representations while humans have essentially no access to per-record or per-timepoint values. This is ideal for multi-institutional training of prediction models, statistical agencies releasing differential-privacy-constrained official statistics, and joint analysis in finance or advertising where raw data must not be shared. In these settings, humans consume models or aggregate statistics rather than raw views, and not seeing plaintext is part of the design goal.

For the in-hospital structured EHR use case (instantiated in our experiments on ICU data), this quadrant is too narrow. Algorithm engineers need column-level numeric tables to perform EDA, quality checks, and pipeline debugging. Clinicians need to see cohort-level vital sign trajectories to apply medical intuition for sanity checks. Data governance and compliance teams must audit what a pipeline has done to specific variables such as heart rate or lactate, rather than simply trusting a black-box secure protocol. These needs intrinsically call for a visible data view, even if numerically transformed, as long as it remains clinically interpretable. The usable-but-invisible cryptographic toolbox is necessary but not sufficient: it solves cross-institutional secure modeling but does not cover the many daily clinical research workflows that depend on human inspection and intuition.

\subsection{Differential Privacy and Synthetic Data}

Differential privacy (DP) provides a formal privacy definition centered on the indistinguishability of neighboring databases and has been widely deployed for releasing statistics and training models~\citep{dwork2006calibrating}. For EHR and tabular data, there is a growing body of work on DP-GAN, DP-VAE, and PATE-style generators that aim to produce datasets which statistically resemble real data while satisfying $\varepsilon$-DP constraints~\citep{liu2024survey,tian2024reliable,jordon2018pate}. In one-shot external release scenarios, these methods offer formal privacy bounds and high-level utility, preserving certain marginals and correlation structure for statistical analysis and model development.

However, transplanting DP-synthetic schemes directly into the in-hospital structured EHR setting (such as ICU time-series data) raises several practical issues. To satisfy global DP constraints, generative models often sacrifice fidelity on minority patterns and rare events, yet for clinical researchers the mean, variance, and range of individual lab tests are first-class objects in EDA. DP is also defined in terms of global randomness, so individual time-point deviations in synthetic data can be large and hard to bound; in contrast, one of our explicit goals is a unified z-score $\ell_\infty$ bound $\alpha$ so that clinicians can reason about perturbations in terms of standard deviations. DP-GAN and DP-VAE systems further require training relatively large models, consuming substantial privacy budgets during hyperparameter tuning, which is operationally expensive for a nightly pipeline running on hospital CPU clusters. Finally, many synthetic pipelines focus on statistical metrics and downstream task performance~\citep{jordon2022synthetic,loni2025review}, whereas actual EHR QA often requires inspecting questions such as what HR trajectories look like for a cohort or whether the tail of lactate behaves strangely; this demands geometrically controlled perturbations at the column and temporal levels, not just an overall resemblance.

\subsection{Positioning: An In-Hospital Visible View}

Putting the above together, existing privacy technologies fall into two main paradigms. The first is usable but invisible: cryptographic protocols allow models to operate safely on encrypted or protected data, but humans cannot see plaintext. The second is synthetic but DP-guaranteed: generative models and DP-sanitized releases provide formal privacy bounds for one-shot external publication, but make it difficult to precisely control column-level perturbations and temporal structure.

This paper addresses a third, complementary quadrant: constructing, within a controlled in-hospital environment, an EHR view that is usable and visible, yet non-reversible at the individual-record level under our no-key, structure-aware threat model. We do not attempt to replace MPC, HE, or TEE for cross-institutional training, nor do we compete with DP-synthetic GANs at the level of formal $\varepsilon$-DP definitions. We focus instead on a gap in current practice: the lack of a geometrically controllable, EDA-friendly, nightly-generatable privacy-enhanced tabular view for in-hospital clinical research and QA pipelines~\citep{riou2024ensuring,jahan2024synthetic}. Against this backdrop, we now turn to related work on geometric transformations and synthetic EHR, asking which ideas can be adapted into candidate column-level operator families for our Human-AI co-design protocol.

\section{Problem Formulation and Geometric View}
\label{sec:problem-formulation}

\subsection{Data and Pipeline Setup: From Structured EHR to Column-wise Operators}

We consider a typical structured EHR warehouse containing longitudinal clinical measurements. Without loss of generality, we use intensive care unit (ICU) time-series data as a running example. Let \(\mathcal{S}\) denote the set of stays (ICU admissions in our running example) and \(\mathcal{V}\) the set of variables of interest, such as heart rate (HR), blood pressure, lactate, and glucose. For each ICU stay \(s \in \mathcal{S}\) and each variable \(v \in \mathcal{V}\), after preprocessing we obtain a time series
\[
  x_{s,v} = (x_{s,v,1}, \dots, x_{s,v,n_{s,v}}) \in \mathbb{R}^{n_{s,v}},
\]
where \(n_{s,v}\) is the number of sampled time points for stay \(s\) on variable \(v\).

In real systems, these time series are typically irregularly sampled and contain missing values. To focus on the numerical transforms themselves, without entangling interpolation strategies and missingness mechanisms, we adopt the following simplifying convention.

\paragraph{Assumption: Gridding and alignment.}
For each variable \(v\), we resample all stays on a pre-specified time grid, for example one point per hour. After interpolation and imputation, we concatenate all observations of that variable across stays into a single column vector
\[
  x_v \in \mathbb{R}^{n_v},
\]
where \(n_v\) is the total number of time points for variable \(v\) over the cohort. Analyses on a single stay can be viewed as selecting the corresponding subsequences inside this large vector.

Under this simplified view, the raw dataset can be written as
\[
  D_{\text{raw}} = \{ x_v \in \mathbb{R}^{n_v} : v \in \mathcal{V} \},
\]
where each column \(x_v\) represents the time unfolding of one clinical physiologic or laboratory variable (ICU in our running example) over the entire cohort.

We assume a standard hospital-side processing pipeline. Each night, an \emph{EHR-Privacy-Agent} running inside the hospital intranet extracts the latest batch of stays from the production system and forms an incremental view of \(D_{\text{raw}}\). On an intranet server operating in a CPU-only environment, a family of \emph{column-wise operators} is applied to each variable \(v\), mapping the raw column \(x_v\) to a perturbed column \(y_v\). Collecting all perturbed columns yields
\[
  D_{\text{priv}} = \{ y_v : v \in \mathcal{V} \},
\]
which is then written into an in-hospital research and teaching cache that can be accessed by clinical researchers, quality analytics teams, and the EHR-Privacy-Agent itself. Throughout this process, the original \(D_{\text{raw}}\) remains in a tightly controlled production system, is never exported to the research environment, and is never directly exposed to LLMs or external tools.

The core object studied in this paper is therefore a family of column-wise random operators
\[
  T_\alpha : \mathbb{R}^{n_v} \to \mathbb{R}^{n_v},
  \qquad
  x_v \mapsto y_v = T_\alpha(x_v; \omega),
\]
where \(\alpha \ge 0\) is a unified privacy control parameter and \(\omega\) denotes the internal randomness used by the operator. In what follows, all geometric discussions are carried out in the space of a single column vector \(x_v\). The multi-variable, multi-stay setting can be understood as repeatedly applying the same operator family to different columns and subsequences.

\subsection{Column-wise Mean--Variance Manifold: Geometrizing the EHR Transformation Problem}

To strike a balance between easy interpretability and resistance to reconstruction, we want each column-wise operator to satisfy two basic intuitions. First, when clinicians or engineers perform EDA and modeling on the perturbed column \(y_v\), they should still observe scales and fluctuation magnitudes comparable to those of the original data. Second, the operator may induce nontrivial geometric motion in high dimensions, but it should preserve low-order statistics, especially the first and second moments, as much as possible.

To formalize these ideas, we introduce a unified geometric viewpoint, which we refer to as the \emph{mean--variance manifold}. We begin by recalling the basic statistical quantities associated with a column vector.

For any column vector \(x \in \mathbb{R}^n\), its empirical mean is defined as
\[
  \mu(x) = \frac{1}{n} \sum_{i=1}^n x_i,
\]
which captures the overall level of the signal across time.

Correspondingly, the empirical variance is given by
\[
  \sigma^2(x) = \frac{1}{n} \sum_{i=1}^n (x_i - \mu(x))^2,
\]
measuring the magnitude of fluctuations around the mean. The standard deviation is then defined as
\[
  \sigma(x) = \sqrt{\sigma^2(x)}.
\]
In practice, to ensure numerical stability, we clip extremely small values of \(\sigma(x)\) from below to avoid division by zero or unstable scaling.
Based on these quantities, we define the z-score standardized version of \(x\) as
\[
  z(x) = \frac{x - \mu(x)\mathbf{1}}{\sigma(x)} \in \mathbb{R}^n,
\]
where \(\mathbf{1}\) denotes the all-ones vector. This transformation recenters the vector to have zero mean and rescales it to unit variance, effectively removing the original physical units and placing all variables on a common, dimensionless scale.

For any vector \(x\) with \(\sigma(x) > 0\), the standardized vector \(z = z(x)\) lies in
\[
  \mathcal{M}(0,1) = \{ z \in \mathbb{R}^n \mid \mu(z) = 0,\ \sigma^2(z) = 1 \}.
\]
Geometrically, this set can be viewed as the intersection of the zero-mean hyperplane
\[
  H = \{ z \in \mathbb{R}^n \mid \mu(z) = 0 \}
\]
with the sphere of radius \(\sqrt{n}\),
\[
  S = \{ z \in \mathbb{R}^n \mid \|z\|_2 = \sqrt{n} \},
\]
so that \(\mathcal{M}(0,1) = H \cap S\).
Thus, after standardization, each clinical time-series column lies on the same high-dimensional manifold with zero mean and unit variance. This viewpoint is useful for two reasons. First, any transform that starts and ends on \(\mathcal{M}(0,1)\) will, after de-standardization, exactly preserve the original column mean and variance. Specifically, if \(z' \in \mathcal{M}(0,1)\), then
\[
  y = \mu(x)\mathbf{1} + \sigma(x) z'
\]
satisfies
\[
  \mu(y) = \mu(x), \qquad \sigma^2(y) = \sigma^2(x).
\]
This provides a natural geometric route to enforce our later C1 constraint.

Second, differences in physical units across variables, such as mmHg, mg/dL, and bpm, are absorbed into the standardization step. In z-score space, every variable is measured on the same dimensionless scale of number of standard deviations, which makes it possible to introduce a single privacy knob \(\alpha\) across all variables.

Accordingly, each column-wise operator \(T_\alpha\) can be decomposed into three stages. We first standardize the raw column via
\[
  x \mapsto z(x) \in \mathcal{M}(0,1).
\]
We then apply a random transform on the manifold,
\[
  z \mapsto z' = \widetilde{T}_\alpha(z; \omega),
  \qquad
  \widetilde{T}_\alpha : \mathcal{M}(0,1) \to \mathcal{M}(0,1).
\]
Finally, we de-standardize through
\[
  z' \mapsto y = \mu(x)\mathbf{1} + \sigma(x) z'.
\]
The true design freedom lies in the choice of \(\widetilde{T}_\alpha\). The operators T1, T2, T3, as well as the later Q-mix extensions, can all be understood as different realizations of \(\widetilde{T}_\alpha\).

\subsection{A Unified Privacy Knob \texorpdfstring{\(\alpha\)}{alpha}: Controlling \texorpdfstring{\(\ell_\infty\)}{l-infinity} Perturbations in z-Score Space}

Once the mean--variance geometry is fixed, we still need a unified privacy scale across variables. The practical question is simple: how many standard deviations of per-point deviation do we allow, on average, in the transformed data.
To answer this, we introduce a scalar parameter \(\alpha \ge 0\) and constrain perturbations in z-score space through an \(\ell_\infty\) bound.

% \paragraph{Constraint C2 (Unified \(\ell_\infty\) bound).}
For almost all realizations of the internal randomness \(\omega\),
\[
  \bigl\| \widetilde{T}_\alpha(z; \omega) - z \bigr\|_\infty \le \alpha.
\]

In words, each time point in standardized space may move by at most \(\alpha\) standard deviations. Returning to physical units, any coordinate \(x_i\) in the original column must satisfy
\[
  |y_i - x_i| \le \alpha \cdot \sigma(x),
\]
where \(\sigma(x)\) is the cohort-level standard deviation of that column. Hence, \(\alpha\) can be interpreted as a universal ceiling on per-point deviations, measured in units of standard deviations.

In our early experiments, we adopted a more naive strategy and imposed a fixed maximum deviation \texttt{maxdiff} in the original physical space, for example \(\pm 0.5\) or \(\pm 1.0\) in the corresponding units. This quickly revealed two serious problems. For low-variance variables, the same \texttt{maxdiff} could correspond to several standard deviations, severely distorting near-constant trajectories and harming interpretability as well as downstream model stability. For high-variance variables, by contrast, the same \texttt{maxdiff} amounted to only a tiny z-score shift and provided little additional protection against reconstruction or membership attacks.

These observations show that a uniform physical-space bound leads to highly heterogeneous privacy strength across variables. We therefore abandon that design and instead enforce a unified \(\ell_\infty\) bound \(\alpha\) in z-score space, using number of standard deviations as the sole privacy scale. This choice automatically adapts perturbation strength to each variable's intrinsic variability, allows privacy--utility trade-offs across variables to be compared along the same \(\alpha\)-axis, and enables a systematic study of what we later call the \(\alpha\)-hierarchy.

Empirically, we observe that as \(\alpha\) increases, reconstruction difficulty and membership proxies for T1 and T2 improve monotonically. At the same time, we also find that simply increasing \(\alpha\) is not always sufficient to produce a qualitative jump in privacy for some especially sensitive variables, which motivates the introduction of Q-mix later in the paper.

\subsection{Full Design Constraints C1--C4 for Column-wise Operators}

Building on the mean--variance geometry and the unified parameter \(\alpha\), we summarize the engineering and statistical design requirements of our column-wise operators as follows.

\subsubsection*{Design goals: semantics, non-reversibility, and deployability}

Before formalizing the constraints C1--C4, we summarize the three high-level design goals that motivate them.

\textbf{G1: Semantic and statistical faithfulness.} Transformed data should preserve medically meaningful semantics and key cohort-level statistics, so that clinicians and engineers can perform EDA, QA, and downstream modeling on the transformed views almost as they would on the raw data.

\textbf{G2: Theoretical non-reversibility under a clear threat model.} Under a specified, realistic threat model, the transformation should prevent exact reversal of protected individual-level information and allow us to reason about privacy-utility trade-offs through a single, interpretable privacy knob.

\textbf{G3: Practical deployability on routine hospital infrastructure.} The transformation pipeline must be simple and lightweight enough to run nightly on CPU-only hospital servers, without large generative models or specialized hardware, and produce numeric views that remain directly inspectable by humans.

In the rest of this section, constraints C1--C4 and the threat model C5 formalize these three goals in geometric and algorithmic terms.

\paragraph{C1: Column-wise mean--variance preservation.}
For any variable \(v\), let \(x_v\) be the original column and \(y_v\) the perturbed column. We require
\[
  \mu(y_v) = \mu(x_v), \qquad \sigma^2(y_v) = \sigma^2(x_v)
\]
to hold numerically up to machine precision. Geometrically, this means moving on the mean--variance manifold and then de-standardizing.

\paragraph{C2: Unified z-score \(\ell_\infty\) bound \(\alpha\).}
For the standardized vector \(z_v = z(x_v)\), the operator \(\widetilde{T}_\alpha\) must satisfy
\[
  \bigl\| \widetilde{T}_\alpha(z_v; \omega) - z_v \bigr\|_\infty \le \alpha,
\]
with the same \(\alpha\) shared across all variables as the single privacy knob.

\paragraph{C3: Full variability.}
To avoid pseudo-anonymization patterns in which only a small subset of points is perturbed while most points remain unchanged, we require that for the overwhelming majority of time points, the transformed series differs from the original in at least one coordinate. Intuitively, most points in a time series should actually move.

\paragraph{C4: CPU-friendly and streaming complexity.}
From an engineering standpoint, each column-wise operator should have computational complexity \(O(n_v)\) or \(O(n_v \log n_v)\) with a small constant factor, so that nightly jobs over hundreds of thousands of stays remain feasible in a CPU-only environment. Ideally, the operator should also admit a streaming implementation, enabling possible near-real-time extensions.

Together, conditions C1 through C4 define the feasible design space for the family of operators studied in this paper. The fundamental operators, T1, T2, and T3, along with the later Q-mix extensions, are all required to meet these conditions. However, for Q-mix, a slightly relaxed interpretation of C4 is adopted, as it is applied only to a limited subset of high-risk variables, allowing for some flexibility in its implementation while still maintaining robust privacy guarantees.

\subsection{Threat Model C5: No-key, Structure-aware Adversary and L0/L1/L2}\label{sec:threat-model}

Having specified the operator space, we now clarify the class of attackers that we consider to be within scope. Our goal is not to construct a formally complete cryptosystem, but rather to design one-way numerical transformations that, under a realistic yet relatively strong threat model, prevent exact recovery of protected individual-level information and substantially degrade the success of reconstruction- and membership-type attacks compared with using the raw data.

\subsubsection{Kerckhoffs's Principle and the No-key Assumption}

For the numerical transformation pipeline implemented by the EHR-Privacy-Agent, we adopt a transparency assumption analogous to Kerckhoffs's principle in cryptography. In our setting, the adversary is assumed to know the form of the operator family, including the matrix formulations and pseudo-code of T1, T2, T3, and Q-mix, as well as the end-to-end transformation process. Public hyperparameters are also assumed to be known, including \(\alpha\), window lengths, Q-mix dimensions, and which variables have Q-mix enabled.

At the same time, the system does not rely on a long-term shared key. Internally, the operators may still use pseudo-random number generators and seeds, for example when constructing a per-stay, per-variable orthogonal matrix \(Q_{s,v}\) from a tuple such as \((\text{hospital\_internal\_seed}, s, v)\). However, these seeds are treated as one-time secrets confined to the hospital computing environment. They exist only inside the in-hospital pipeline, are never exposed through external interfaces, exported files, or logs, and cannot be reliably inferred or reused by an adversary even if some paired \((D_{\text{raw}}, D_{\text{priv}})\) samples are leaked.

We therefore refer to the adversary as a \emph{no-key, structure-aware adversary}. Such an adversary fully understands the structure and implementation of the operator family, but does not have access to the internal seeds or one-time randomness used in the hospital pipeline. This should be distinguished from traditional encryption with a fixed long-term key. Our goal is not to build a reversible encryption mechanism, but rather a one-way, single-use numerical transformation pipeline. In this sense, no-key means that we do not assume or provide a reusable long-term symmetric key abstraction, even though internal randomness is still present.

\subsubsection{Leakage Levels: L0 / L1 / L2}

Under the no-key assumption, the adversary's power depends crucially on how many input--output pairs can be observed. Let \(x\) denote an original column vector or subsequence and \(y = T_\alpha(x; \omega)\) the corresponding perturbed result. We consider three leakage levels.

At level L0, the adversary observes only many perturbed columns or time series \(\{y_v\}\), without any direct access to the corresponding originals \(\{x_v\}\). This models external researchers or third parties who only see the released privacy-enhanced data. At level L1, the adversary can access a small number of paired samples \((x,y)\), on the order of \(0.01\%\) of stays or time points, corresponding to situations such as sporadic internal leaks, manual spot checks, or accidentally exposed audit samples. At level L2, the adversary can observe a substantial fraction of paired samples, for example about \(20\%\) of stays or time points, which reflects a stronger insider scenario or a multi-center setting in which one party has access to both raw and privacy-enhanced views.
Across these leakage levels, we focus on four classes of attack tasks. The A-family covers pointwise and sequence reconstruction, where the goal is to predict \(x\) or key statistics derived from \(x\) given \(y\). The B-family concerns record linkage and re-identification, where an attacker uses an external raw record \(x^{\text{ext}}\) to identify the most likely matching record in \(\{y\}\). The C-family addresses membership inference, namely deciding whether a particular individual's data appear in \(\{x\}\) or contributed to the privacy-enhanced release~\citep{shokri2017membership,zhang2022membership}. The D-family concerns attribute inference and indistinguishability-style games, for example when the adversary is given \((x_0,x_1)\) and \(y = T_\alpha(x_b)\) and must infer the hidden bit \(b\), in analogy with IND-style formulations in cryptography~\citep{annamalai2024linear}.

In this paper, we mainly instantiate A-family reconstruction attacks in single-column experiments and use membership proxies as approximate indicators for C-family risks~\citep{carlini2021extracting,shokri2017membership}. A complete treatment of the B/C/D families, as well as the fully multivariate setting, is left for future work.

From a cryptographic viewpoint, our reconstruction and membership attacks can be regarded as simplified chosen-plaintext-style evaluations. Depending on the leakage level, the adversary may access some number of \((x,y)\) pairs and attempts to infer \(x\) or some sensitive function of \(x\) from \(y\). We do not claim security in the strict IND-CPA or IND-CCA sense. Rather, these attacks serve as empirical lower bounds on privacy risk within the chosen adversarial model.

\subsection{Formal Problem Statement}

We can now summarize the \textbf{central problem} of this paper.

\paragraph{Problem: Column-wise geometric operator design.}
Given a structured EHR dataset \(D_{\text{raw}}\) (we use ICU time-series EHR data as a running example), the column-wise operator design constraints C1--C4, and the no-key, structure-aware threat model C5, design a family of random operators \(\{T_\alpha\}_{\alpha \ge 0}\) such that, under leakage levels L0/L1/L2 and for selected attack families, in particular A-family reconstruction and membership proxies, the adversary's success rate in recovering protected individual-level information is substantially worse than when using the raw data, while simultaneously preserving downstream task performance within an acceptable utility range and maintaining deviations in column-wise distributional shape, multivariate correlation structure, and temporal structure at levels that remain interpretable and suitable for clinicians and engineers performing EDA and QA.

We next describe a SciencePal co-design protocol that explains how these operators are proposed, filtered, and iteratively refined through a loop in which humans specify constraints, the AI assistant SciencePal proposes candidate tools, and humans then verify and attack the resulting designs.

\section{SciencePal Co-design / AI-making Protocol}
\label{sec:human-ai-codesign}

\begin{figure*}[t]
  \centering
  % e.g. figures/codesign_protocol_math.pdf
  \includegraphics[width=\textwidth]{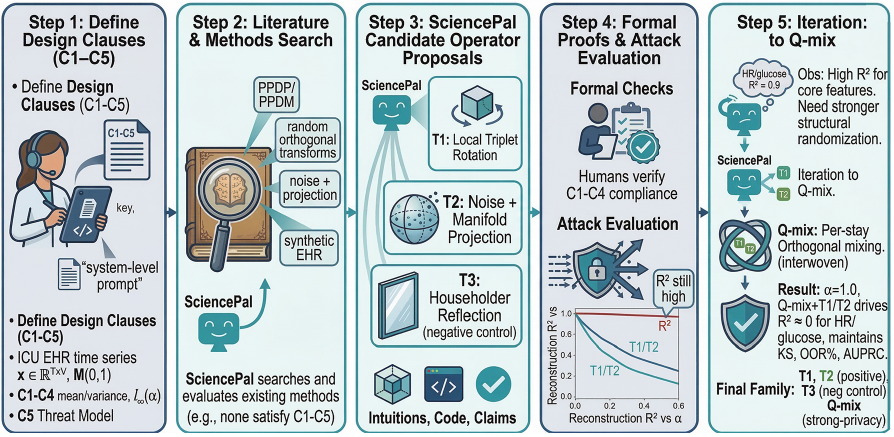}
  \caption{SciencePal co-design protocol for geometric operator families.
\textbf{Step 1}: Humans specify operator clauses C1--C4 and threat model C5.
\textbf{Step 2}: SciencePal searches PPDP/PPDM and related methods, finding no operator that satisfies all clauses.
\textbf{Step 3}: SciencePal proposes candidate operators T1, T2, and T3.
\textbf{Step 4}: Humans prove properties and run attack-based evaluation, keeping T1/T2 and rejecting T3.
\textbf{Step 5}: Observing residual high reconstruction risk
  for some variables, humans and SciencePal co-design per-stay Q-mix extensions, yielding the final operator family.}
  \label{fig:codesign-math}
\end{figure*}

As discussed in Section~\ref{sec:related-work}, we did not find any existing column-wise operator family in the literature that simultaneously satisfies C1--C4 and C5. This motivates an explicit SciencePal co-design workflow, in which the AI assistant SciencePal is treated as a constrained tool-builder rather than an autonomous decision-maker. In this section, we formalize that SciencePal co-design / AI-making protocol. The overall five-step protocol is summarized in Figure~\ref{fig:codesign-math}.

\subsection{Step 1: Human Specification of Geometric and Engineering Requirements}

The process begins with human researchers formulating a precise operator-design specification based on structured EHR use cases (instantiated on ICU data) and hospital engineering constraints. This specification includes data and compute requirements, namely that data remain strictly inside the hospital intranet, nightly jobs must run in a CPU-only environment, and operators must not depend on large generative models or GPU inference. It also includes the geometric and statistical constraints C1--C3, namely exact preservation of column-wise mean and variance, unified \(\ell_\infty\) control in z-score space through \(\alpha\), and full variability so that most points are genuinely perturbed. In addition, the specification includes the complexity requirement C4, which asks for \(O(n_v)\) or near-\(O(n_v)\) algorithms with small constants and, where possible, streaming implementations. Finally, it includes the threat model C5, under which the adversary is no-key and structure-aware, and the operators are expected to offer meaningful protection under L0/L1/L2 leakage against at least a subset of the A/B/C/D attack families.

This specification is written as a mixture of natural language and mathematical notation in the form of operator-design clauses and is then used as a system-level prompt for subsequent interactions with SciencePal.

\subsection{Step 2: SciencePal Literature Search and Candidate Operator Proposals}

Given this specification, SciencePal is asked to perform two tasks. The first is literature search. It surveys the PPDP/PPDM, random orthogonal transform, noise-plus-projection, time-series privacy, and synthetic EHR literature for operator families that might already satisfy C1--C4 while remaining compatible with structured EHR (including ICU time-series data) and the mean--variance manifold perspective. This search concludes that, although many related ingredients exist, including random orthogonality, Householder reflections, and noise-plus-normalization schemes, no off-the-shelf method fully fits the deployment constraints once CPU-only execution and a unified \(\alpha\) knob are imposed.
The second task is candidate operator design. After establishing the lack of a fully compatible existing scheme, SciencePal is prompted to propose new operator families under the given specification. These proposals include geometric intuitions on the mean--variance manifold, explicit mathematical forms and sufficient conditions such as orthogonality and projection steps, pseudo-code level implementation sketches, and self-declared claims regarding C1--C4. At this stage, several candidate structures are proposed, of which three representative families eventually evolve into the basic operators T1, T2, and T3.

\subsection{Step 3: Human Geometric Proofs and First-stage Filtering}

In the third stage, human researchers act as reviewers and carry out formal geometric and statistical analysis of the candidate operators proposed by SciencePal. This includes checking whether each manifold-based definition is self-consistent, deriving its effects on column-wise means and variances to verify C1, analyzing perturbation norms in z-score space and their dependence on \(\alpha\) to verify C2, and evaluating whether the operator can be implemented in \(O(n_v)\) or \(O(n_v \log n_v)\) time with possible streaming support to verify C4.

Candidates that clearly violate C1--C4, for example by introducing uncontrolled nonlinearities or breaking mean/variance preservation in corner cases, are discarded. Candidates with plausible geometric and computational properties are retained for attack-based evaluation. In this paper, the three retained basic operators are T1, a local triplet rotation; T2, a noise-plus-projection transform onto the mean--variance manifold; and T3, a global Householder reflection that is ultimately kept as a negative case.

\subsection{Step 4: Attack-driven Evaluation and Positive/Negative Labeling}

In the fourth stage, the surviving candidates are implemented on real ICU time-series EHR data. The evaluation spans three dimensions. First, we measure geometric and statistical fidelity through numerical errors in column-wise means and variances, as well as metrics such as \(1-\mathrm{KS}\), Frobenius distance between correlation matrices, autocorrelation functions (ACFs), and Ljung--Box tests. Second, we examine downstream utility by tracking changes in AUROC and AUPRC for several ICU prediction tasks. Third, we assess resistance to attacks through reconstruction experiments and membership proxies under L0/L1/L2 leakage.

These experiments allow us to label operators as either positive or negative cases from a privacy perspective. T1 and T2 emerge as positive cases: for moderate \(\alpha\), they preserve means, variances, and correlation structure almost perfectly while substantially increasing the difficulty of reconstruction and membership attacks. T3, by contrast, becomes a negative case. Although it appears nearly perfect on conventional statistical metrics, including means, variances, correlations, and ACFs, it remains highly invertible under L2 combined with A-family linear reconstruction attacks, with \(R^2\) close to 1. We therefore regard T3 as suitable only for very light perturbation or didactic in-hospital use, but not for strong-privacy external release.

Such negative cases are themselves important outputs of the AI-making process. They show that geometric elegance and statistical similarity alone are insufficient. Privacy claims must be tested explicitly through attack experiments.

\subsection{Step 5: Iteration and Extension: From the \texorpdfstring{\(\alpha\)}{alpha}-Hierarchy to Q-mix}

After the initial operator family T1/T2/T3 and its \(\alpha\)-hierarchy had been established, our single-column attack experiments revealed an important phenomenon. As \(\alpha\) increases, the \(R^2\) of T1 and T2 under A-family reconstruction decreases monotonically, but for some especially sensitive variables, particularly HR and glucose, the \(R^2\) values can still remain in the range \(0.6\) to \(0.9\) even when the marginal distributions already show visibly substantial shifts. In other words, privacy improves gradually but does not exhibit a qualitative jump.
This observation triggered a new round of SciencePal co-design. Human researchers sharpened the requirement by asking whether, while keeping \(\alpha\), C1--C4, and the basic T1/T2 structures unchanged, it would be possible to add a stronger structural randomization mechanism for a small subset of high-risk variables so that reconstruction \(R^2\) under L2 and A-family attacks would drop close to zero. In response, SciencePal proposed several forms of per-stay orthogonal mixing. Building on these ideas, we designed and implemented a per-stay orthogonal Q-mixing extension: for each stay \(s\) and variable \(v\), a \(48 \times 48\) orthogonal matrix \(Q_{s,v}\) is constructed, applied to local time windows to mix coordinates, and then combined with T1 or T2.
The human team subsequently proved that Q-mix satisfies C1--C4 in the slightly relaxed deployment sense noted earlier. With \(\alpha = 1.0\) unchanged, Q-mix combined with T1 or T2 can reduce the \(R^2\) of HR and glucose reconstruction under L2 plus linear A-family attacks to values close to zero, while keeping distributional shape and downstream predictive performance within acceptable ranges.

It is important to emphasize that Q-mix is not a new basic operator family. Rather, it is a deployment-level per-stay orthogonal extension module that is enabled only for a small subset of sensitive variables in strong-privacy scenarios such as external release or multi-center sharing. For most in-hospital research and teaching settings, the default remains T1 or T2, with optional light T3 only for very mild perturbation.

In this sense, the SciencePal co-design / AI-making protocol not only yields the three basic operators T1, T2, and T3, but also reveals the limitations of the pure \(\alpha\)-hierarchy and motivates Q-mix as a stronger privacy extension. The protocol itself is therefore one of the methodological contributions of the paper.

\section{Operator Families: Column-wise T1/T2/T3 and Per-stay Q-mix Extensions}
\label{sec:operator-families}

\subsection{Design Guide and Notation}

\begin{figure*}[t]
\centering
% e.g. figures/operators_overview.pdf or .png
\includegraphics[width=\textwidth]{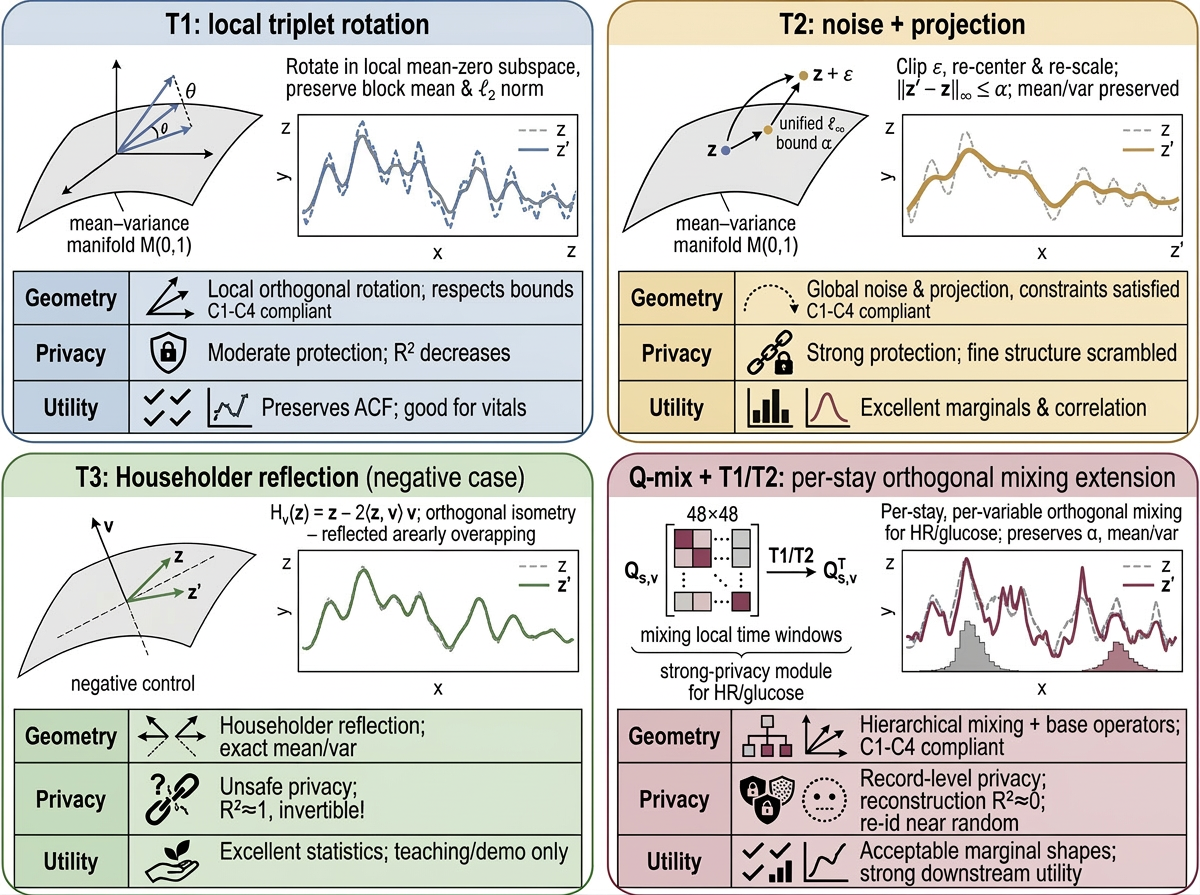}
\caption{
Overview of the column-wise geometric operators studied in this work.
\textbf{T1} applies local triplet rotations on $\mathcal{M}(0,1)$,
preserving mean/variance and short-range autocorrelation with moderate privacy.
\textbf{T2} adds bounded z-score noise with global $\ell_\infty$ limit $\alpha$,
then recenters to $\mathcal{M}(0,1)$, giving stronger privacy while largely
preserving marginals and correlations.
\textbf{T3} is a global Householder reflection that preserves low-order statistics
but is highly invertible ($R^{2} \approx 1$), and serves as a negative control.
\textbf{Q-mix} wraps T1/T2 with per-stay orthogonal mixing on selected high-risk
variables, keeping fixed while sharply reducing reconstruction risk and retaining
acceptable utility.
}
\label{fig:operators-overview}
\end{figure*}

We summarize the geometry, privacy, and utility trade-offs of the four operators in Figure~\ref{fig:operators-overview}.

For convenience, throughout this section we fix a single variable \(v\) and omit its subscript. We therefore work with a single column vector \(x \in \mathbb{R}^n\), its standardized form
\[
  z = z(x) = \frac{x - \mu(x)\mathbf{1}}{\sigma(x)} \in \mathcal{M}(0,1),
\]
and a randomized operator in standardized space of the form
\[
  z' = \widetilde{T}_\alpha(z; \omega),
  \qquad
  z, z' \in \mathcal{M}(0,1),
\]
where \(\omega\) denotes the internal randomness. The corresponding operator in the original physical space is
\[
  T_\alpha(x; \omega)
  = \mu(x)\mathbf{1} + \sigma(x)\,\widetilde{T}_\alpha\bigl(z(x); \omega\bigr).
\]

Under this formulation, C1 is automatically satisfied as long as \(\widetilde{T}_\alpha\) maps \(\mathcal{M}(0,1)\) back to itself. The remaining constraints, namely C2 through C4, depend on the specific construction of \(\widetilde{T}_\alpha\).

We next introduce three basic operator families. T1 performs local triplet rotations, T2 applies noise followed by projection, and T3 is a global Householder reflection that serves as a negative case. We then introduce Q-mix as a per-stay orthogonal extension that wraps around T1 and T2.

\subsection{T1: Local Triplet Rotation}
\label{sec:T1}

\subsubsection{Intuition}

T1 performs local, low-dimensional, small-angle random rotations along the time axis. The aim is to scramble fine-scale temporal patterns while preserving local temporal structure, such as short-range autocorrelation, together with global mean and variance. Concretely, the standardized vector \(z \in \mathbb{R}^n\) is partitioned into disjoint windows of length three, and a random orthogonal transformation is applied within each three-dimensional subspace.

\subsubsection{Algorithm}

Let the column length be \(n\), and partition the index set into triplets
\[
  B_1 = \{1,2,3\},\;
  B_2 = \{4,5,6\},\;
  \dots
\]
For a final segment with fewer than three points, one may either replicate neighboring points or use a reduced-size rotation; this has negligible effect on C1 and C2 and is omitted here for brevity.

For each triplet \(B\), define the local vector \(z_B = (z_i)_{i \in B} \in \mathbb{R}^3\). We first compute its local mean \(\mu_B = \frac{1}{3}\sum_{i \in B} z_i\) and decompose
\[
  z_B = \mu_B \mathbf{1}_3 + r_B,
  \qquad
  \mathbf{1}_3 = (1,1,1)^\top,
  \qquad
  \mathbf{1}_3^\top r_B = 0.
\]
Thus \(r_B\) lies in the two-dimensional subspace orthogonal to \(\mathbf{1}_3\). Let \(\{u_1,u_2\}\) be an orthonormal basis of this subspace and write \(r_B = a_1 u_1 + a_2 u_2\). We then sample
\[
  \theta_B \sim \mathrm{Unif}[-\theta_{\max}, \theta_{\max}]
\]
and rotate the coefficient pair \((a_1,a_2)\) in the plane:
\[
  \begin{pmatrix}
    a_1' \\[2pt]
    a_2'
  \end{pmatrix}
  =
  \begin{pmatrix}
    \cos\theta_B & -\sin\theta_B \\
    \sin\theta_B & \cos\theta_B
  \end{pmatrix}
  \begin{pmatrix}
    a_1 \\[2pt]
    a_2
  \end{pmatrix}.
\]
The transformed block is then reconstructed as
\[
  z_B' = \mu_B \mathbf{1}_3 + a_1' u_1 + a_2' u_2.
\]

Applying this construction to all triplets yields a perturbed vector \(z' \in \mathbb{R}^n\). Since each local transform preserves both the block mean and the block norm, the global mean of \(z'\) remains zero and the global \(\ell_2\) norm remains \(\sqrt{n}\). Hence \(z' \in \mathcal{M}(0,1)\). At the same time, the relative magnitudes and phases inside each triplet are rotated into new configurations, which perturbs higher-order local temporal patterns.

\subsubsection{Satisfying C1--C4}

T1 satisfies C1 because it maps \(\mathcal{M}(0,1)\) to itself. To enforce C2, we calibrate \(\theta_{\max}(\alpha)\) offline so that for any triplet input \(z_B\) and any admissible angle \(\theta \in [-\theta_{\max},\theta_{\max}]\), the local perturbation satisfies \(\|z'_B - z_B\|_\infty \le \alpha\). In practice, this is done through analytic bounds combined with numerical search over representative triplet geometries. Once \(\theta_{\max}(\alpha)\) is fixed, the full-column perturbation also satisfies \(\|z' - z\|_\infty \le \alpha\).

T1 also satisfies C3. Whenever \(\theta_B \neq 0\), at least one component of the residual vector changes, so \(z'_B \neq z_B\). Since \(\theta_B\) is sampled from a continuous distribution, the probability of drawing exactly zero is negligible, and almost all non-degenerate triplets are perturbed. Finally, T1 satisfies C4 because each triplet requires only constant-size vector operations and one \(2 \times 2\) rotation, yielding overall \(O(n)\) complexity with a small constant and a natural streaming implementation.

\subsection{T2: Noise plus Projection}
\label{sec:T2}

\subsubsection{Intuition}

Where T1 emphasizes local geometric perturbations, T2 acts more like a global but controlled noise injection followed by a projection back onto the mean--variance manifold. The idea is to add bounded perturbations in z-score space and then restore the output to zero mean and unit variance.

\subsubsection{Algorithm}

Given \(z \in \mathcal{M}(0,1)\), we sample Gaussian noise
\[
  \varepsilon \sim \mathcal{N}(0, \tau^2 I_n)
\]
and center it by subtracting its empirical mean, yielding
\[
  \varepsilon^{(0)} = \varepsilon - \mu(\varepsilon)\mathbf{1}.
\]
We then clip its coordinates using a threshold \(c(\alpha)\),
\[
  \varepsilon_i^{(1)}
  =
  \operatorname{clip}\bigl(\varepsilon_i^{(0)}, -c(\alpha), c(\alpha)\bigr),
\]
so that \(\|\varepsilon^{(1)}\|_\infty \le \alpha/2\). Defining \(\tilde{z} = z + \varepsilon^{(1)}\), we project back to \(\mathcal{M}(0,1)\) by re-centering and renormalizing:
\[
  \tilde{z}^{(0)} = \tilde{z} - \mu(\tilde{z})\mathbf{1},
  \qquad
  z' = \sqrt{n}\,\frac{\tilde{z}^{(0)}}{\|\tilde{z}^{(0)}\|_2}.
\]

By construction, the output \(z'\) lies on \(\mathcal{M}(0,1)\). In practice, for a given \(\alpha\), the parameters \((\tau, c(\alpha))\) are calibrated offline and then validated numerically in both single-column and multi-column experiments.

\subsubsection{Satisfying C1--C4}

C1 follows immediately from the final projection step, which enforces zero mean and norm \(\sqrt{n}\). For C2, the clipping step ensures that the direct noise contribution is bounded by \(\alpha/2\) in \(\ell_\infty\), while the subsequent re-centering and renormalization introduce only a controlled additional perturbation. During offline calibration, we verify extensively that \(\|z' - z\|_\infty \le \alpha\) holds for all tested samples, and we choose conservative parameters whenever corner cases approach the boundary.

T2 also satisfies C3 because the injected noise is nonzero on almost all coordinates, and the subsequent re-centering and rescaling further modify nearly every time point. The computational cost is linear in \(n\), since noise generation, clipping, re-centering, and normalization are all single-pass operations. Hence T2 satisfies C4 and is well suited to CPU-only processing.

\subsection{T3: Global Householder Reflection (Negative Case)}
\label{sec:T3}

\subsubsection{Intuition}

T3 applies a single global orthogonal reflection. From a geometric perspective, it is almost ideal: it preserves pairwise distances, inner products, correlations, and most conventional summary statistics exactly or nearly exactly. However, precisely because it is such a clean isometry, it is also highly invertible once enough paired samples are observed.

\subsubsection{Algorithm}

Choose a unit vector \(v \in \mathbb{R}^n\) satisfying \(\|v\|_2 = 1\) and \(\mu(v)=0\). The vector may be fixed from a representative cohort, for example as a principal direction, or generated pseudo-randomly from an internal hospital seed and then projected onto the mean-zero hyperplane. Define the Householder reflection 
~\citep{golub2013matrix}

\[
  H_v(z) = z - 2\langle z, v\rangle v,
\]
and set
\[
  \widetilde{T}^{\mathrm{T3}}(z) = H_v(z).
\]

Geometrically, this is reflection across the hyperplane \(\{w : \langle w,v\rangle = 0\}\). Points lying on that hyperplane remain fixed, while the component along \(v\) changes sign.

\subsubsection{Satisfying C1--C4}

Because both \(z\) and \(v\) lie in the mean-zero hyperplane, \(H_v(z)\) also has zero mean. Since Householder reflections are orthogonal, they preserve the norm, so \(H_v(z) \in \mathcal{M}(0,1)\) and C1 holds.

For C2, note that
\[
  H_v(z)_i - z_i = -2\langle z,v\rangle v_i.
\]
If the entries of \(v\) are bounded by \(|v_i| \le c_v\), then
\[
  \|H_v(z)-z\|_\infty \le 2\sqrt{n}\,c_v.
\]
Thus choosing \(c_v \le \alpha/(2\sqrt{n})\) guarantees the desired \(\ell_\infty\) bound. T3 also satisfies C3 except on a measure-zero set where \(\langle z,v\rangle = 0\), and it satisfies C4 because it requires only one inner product and one coordinate-wise update, both linear-time operations.
Despite satisfying C1--C4, T3 behaves as a high-invertibility negative case under the C5 and L2 threat model. In Section~\ref{sec:privacy-protocol}, we show that under L2 with \(20\%\) paired samples, its reconstruction \(R^2\) is substantially higher than that of T1 and T2.

\subsection{Summary of the Three Basic Operators}
\label{sec:operator-summary}

The three basic operators differ mainly in geometric scope and privacy behavior. T1 performs local rotations and tends to preserve short-range temporal structure relatively well. T2 applies global noise with projection and usually offers stronger scrambling at the same \(\alpha\). T3 is statistically elegant and almost perfectly structure-preserving, but is also highly invertible and therefore unsuitable for strong-privacy deployment.

Accordingly, T1 and T2 are treated as positive operator families for privacy protection, whereas T3 is retained only as a negative case and as a light-perturbation example for controlled in-hospital use.

\subsection{Per-stay Orthogonal Q-mixing (Pilot Extension Module)}
\label{sec:qmix}

\subsubsection{Motivation}

Empirically, increasing \(\alpha\) improves privacy gradually, but for some physiologically smooth and sensitive variables, especially HR and glucose, the improvement is not decisive. Even when marginal distributions and autocorrelation functions begin to show visible deviation, reconstruction performance may remain relatively high. This motivates an additional structural randomization mechanism that can induce a more pronounced privacy jump while keeping the same global \(\alpha\).

\subsubsection{Construction}

Consider a single stay \(s\) and variable \(v\), with standardized time series \(z_{s,v} \in \mathcal{M}(0,1)\). Fix a window length \(m\), for example \(m=48\), and partition the time axis into windows \(B_{s,1},\dots,B_{s,K_s}\), each of size \(m\). For each pair \((s,v)\), we construct a permutation matrix \(Q_{s,v} \in \mathbb{R}^{m\times m}\) from an internal hospital seed together with \((s,v)\). Since \(Q_{s,v}\) is a permutation matrix, it is orthogonal, preserves the constant vector, and is an \(\ell_\infty\)-isometry.

Applying this block-wise yields a mixed representation
\[
  \tilde{z}_{s,v} = \mathsf{Qmix}_{s,v}(z_{s,v}),
\]
where \(\mathsf{Qmix}_{s,v}\) is block-diagonal with identical blocks \(Q_{s,v}\). Because the transform is block-orthogonal and preserves constants, it preserves global mean and norm, so \(\tilde{z}_{s,v}\in\mathcal{M}(0,1)\).

We then wrap T1 or T2 around this mixed coordinate system. For either basic operator \(\widetilde{T}_\alpha\), the corresponding Q-mix version is
\[
  \widetilde{T}^{(Q)}_\alpha(z_{s,v})
  =
  \mathsf{Qmix}_{s,v}^\top
  \Bigl(
    \widetilde{T}_\alpha\bigl(\mathsf{Qmix}_{s,v}(z_{s,v})\bigr)
  \Bigr).
\]
This means that we first permute locally within each window, then apply T1 or T2 in the mixed space, and finally map the result back to the original time index order.

\subsubsection{Satisfying C1--C4 and Interaction with C5}

Q-mix preserves C1 because both the mixing and unmixing operations are orthogonal and preserve constants, while the wrapped operator already maps \(\mathcal{M}(0,1)\) to itself. It preserves C2 because permutations have operator norm one under \(\ell_\infty\), so the original bound is inherited unchanged. It also preserves C3, since mixing only redistributes where the perturbation appears once mapped back to the original order, and it preserves C4 because applying \(Q_{s,v}\) or \(Q_{s,v}^\top\) is only a permutation and therefore linear-time.

Under the no-key, structure-aware threat model, the adversary knows the existence and form of Q-mix, including the window length and the fact that a per-stay permutation is used, but does not know the specific realization of \(Q_{s,v}\). As a result, even if the adversary learns a good approximate inverse for T1 or T2 in the mixed coordinate system, the missing per-stay permutation still prevents straightforward recovery in the original time order.

In the Q-mix pilot experiments of Section~\ref{sec:attack-a-setup}, we show that with \(\alpha = 1.0\), Q-mix combined with T1 or T2 can drive the reconstruction \(R^2\) for HR and glucose under L2 plus linear A-family attacks from the range \(0.8\) to \(0.9\) down to nearly \(0\).

\subsubsection{Deployment}

Q-mix is not a new basic operator family. Rather, it is a deployment-level extension wrapped around T1 and T2. In relatively controlled in-hospital research, QA, or teaching settings, T1 or T2 with a smaller \(\alpha\), for example \(\alpha = 0.3\) to \(0.5\), is usually sufficient without enabling Q-mix. In contrast, for external sharing, multi-center analysis, or other stronger-threat scenarios, T1 or T2 with a larger \(\alpha\), for example \(\alpha = 0.8\) to \(1.0\), can be combined with Q-mix for a small subset of high-risk variables such as HR and glucose.

In the EHR-Privacy-Agent's policy design, Q-mix is therefore treated as a strong-privacy module. It is disabled in the default in-hospital configuration and enabled only when constructing stronger privacy views for external release or cross-center sharing, and even then only for designated high-risk variables.

\section{Experiment I: Single-column Operator Behavior}
\label{sec:single-column-expt}

In this section, we evaluate T1, T2, T3, and the per-stay Q-mix wrapper on MIMIC-IV ICU time series from a single-column perspective. Our goal is twofold. First, we verify on real data that these operators satisfy the geometric constraints C1--C4 introduced in Section~\ref{sec:problem-formulation}. Second, we instantiate a concrete privacy evaluation protocol under leakage levels L0, L1, and L2 together with attack families A, B, C, and D. We then examine how reconstruction attacks behave as a function of the privacy knob \(\alpha\), and show that per-stay Q-mixing can drive the linear reconstruction \(R^2\) for HR and glucose from approximately \(0.8\)–\(0.9\) down to nearly \(0\) at the same \(\alpha\), while only mildly affecting marginal distributions and temporal structure.

Throughout this section, we work within the unified mean--variance manifold and \(\alpha\)-hierarchy framework introduced in Sections~\ref{sec:problem-formulation}--\ref{sec:operator-families}. For each column, we first standardize to \(z \in \mathcal{M}(0,1)\), apply an operator family in standardized space, and then de-standardize back to physical units.

\subsection{Experimental Setup and Data Preparation}
\label{sec:exp-setup}

\subsubsection{Dataset and resampling}
\label{sec:dataset-sampling}

We use the adult ICU subset of MIMIC-IV~\citep{johnson2023mimiciv}. Only first ICU stays are included, and we further restrict the cohort to patients aged at least \(18\) years with ICU length of stay \(\mathrm{LOS} \ge 24\) hours.

For each ICU stay \(s\), time is aligned to ICU admission, with \(t=0\) corresponding to ICU in-time. All charted physiologic variables are resampled onto a 1-hour grid. Concretely, each chart time is converted to an integer hour offset by flooring the difference from ICU admission, after which the data are pivoted into an hourly matrix representation \(\{(s,t)\mapsto \text{value}\}\). We then restrict attention to the first \(T_{\max}+1\) hours, using \(T_{\max}=47\), so that each retained time series has length approximately \(48\).

Missingness is handled within each stay and variable by a simple forward-fill plus linear-interpolation strategy. Measurements are first sorted by time and forward-filled along the hourly grid. When long gaps remain, linear interpolation is applied whenever it is clinically sensible. For variables where such interpolation is not meaningful, NaNs may be retained, but in this section we include only columns for which the selected window \([0,T_{\max}]\) is defined for almost all stays. The resulting stay-level time series is denoted as follows:
\[
  x_{s,v} \in \mathbb{R}^{n_{s,v}},
  \qquad n_{s,v} \approx 48.
\]

\subsubsection{Representative variable set}
\label{sec:representative-vars}

To cover a range of column types, including vitals versus labs, smooth versus spiky variables, and narrow versus heavy-tailed marginals, we select a representative set of variables. The vital-sign group includes HR, systolic and diastolic blood pressure, mean arterial pressure, temperature in Celsius, respiratory rate, and peripheral oxygen saturation. The laboratory group includes glucose, lactate, creatinine, and several additional markers. Together, these variables span different variance scales, physical units, and clinical roles. Our aim is to observe, within a common geometric and privacy framework, how different column types respond to the same operator family.

\subsubsection{Operators and hyperparameters}
\label{sec:operators-hparams}

On each column, we compare several transform classes. The main geometric operators are T1\_uniform, namely the local triplet rotation of Section~\ref{sec:T1} with a fixed partition into disjoint length-3 blocks; T1\_weighted, a small weighted variant of T1 that adjusts rotation frequency or angle over time while keeping the same privacy structure; T2, the noise-plus-projection operator of Section~\ref{sec:T2}; and T3, the global Householder reflection of Section~\ref{sec:T3}, which is retained as a negative case.

In Section~\ref{sec:qmix-experiments}, we additionally consider per-stay orthogonal Q-mixing wrappers around T1 and T2, namely \(\mathrm{Q}+\mathrm{T1\_uniform}\), \(\mathrm{Q}+\mathrm{T1\_weighted}\), and \(\mathrm{Q}+\mathrm{T2}\), focusing on the sensitive variables HR and glucose.

All geometric operators share a common z-score \(\ell_\infty\) privacy bound,
\[
  \|z' - z\|_\infty \le \alpha.
\]
For the geometric and distributional analyses in Sections~\ref{sec:c1c2} and~\ref{sec:marginal-range}, we use
\[
  \alpha \in \{0.3, 0.5, 0.8, 1.0\}.
\]
Here, \(\alpha=0.3\) corresponds to a light, visualization-friendly regime, whereas \(\alpha=1.0\) corresponds to a stronger regime in which each point may move by up to one standard deviation. For reconstruction experiments in Section~\ref{sec:attack-a-results}, we extend the grid to
\[
  \alpha \in \{1.0, 2.0, 3.0, 5.0\},
\]
in order to probe the limits of the \(\alpha\)-hierarchy.

For sanity checks, we also implement simple baselines such as the identity mapping and i.i.d. Gaussian noise in z-score space. 
% Their detailed numerical results are omitted here and reported in Appendix~\ref{app:baselines}.

\subsubsection{Column-wise standardization and evaluation views}
\label{sec:standardization-views}

For each variable \(v\), we compute a global column-wise mean and standard deviation over the training cohort,
\[
  \mu_v = \mu(x_{\cdot,v}), \qquad \sigma_v = \sigma(x_{\cdot,v}),
\]
and standardize each stay-level sequence by
\[
  z_{s,v} = \frac{x_{s,v} - \mu_v \mathbf{1}}{\sigma_v}.
\]
An operator is then applied in standardized space to obtain \(z'_{s,v}\), and the result is mapped back to physical space through
\[
  x'_{s,v} = \mu_v \mathbf{1} + \sigma_v z'_{s,v}.
\]

We evaluate each operator from two complementary views. In standardized space \((z,z')\), we analyze geometric quantities such as \(\ell_\infty\) perturbation, reconstruction \(R^2\), and z-score MAE. In physical space \((x,x')\), we analyze physical ranges, out-of-range rates, and clinical interpretability.

\subsection{Geometric Sanity Check: Numerical Validation of C1 and C2}
\label{sec:c1c2}

We first verify that T1, T2, and T3 satisfy mean--variance preservation and \(\ell_\infty\) control on real ICU columns, beyond the idealized derivations of Section~\ref{sec:operator-families}.

\subsubsection{Mean and variance preservation (C1)}
\label{sec:c1-mean-var}

For each variable \(v\), we compare original and perturbed column-wise means and standard deviations over all stays through
\[
  \Delta \mu_v = \mu(x'_{\cdot,v}) - \mu(x_{\cdot,v}), \qquad
  \Delta \sigma_v = \sigma(x'_{\cdot,v}) - \sigma(x_{\cdot,v}).
\]
We report maximum absolute deviations over variables and summarize them by operator and evaluation setting, namely physical space versus standardized space.

\begin{figure}[H]
  \centering
  \includegraphics[width=\textwidth]{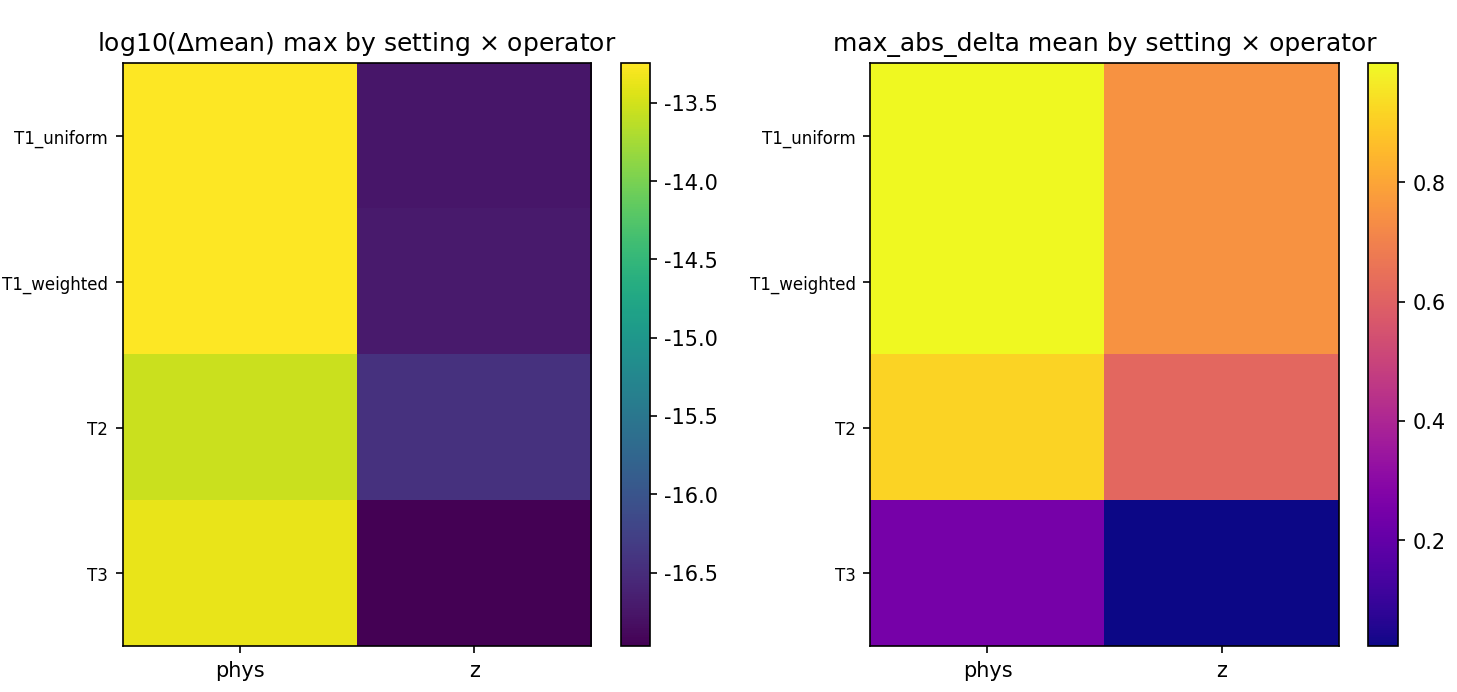}
  \caption{Geometric sanity heatmaps.
           Left: log\(_{10}\) of the maximum absolute mean deviation
           \(\Delta\mu_{\max}\) by setting (phys vs z) and operator.
           Right: max\_abs\_delta\_mean by setting and operator.
           All geometric operators keep mean deviations at or below
           machine precision.}
  \label{fig:c1-heatmap}
\end{figure}

\begin{figure}[H]
  \centering
  \includegraphics[width=\textwidth]{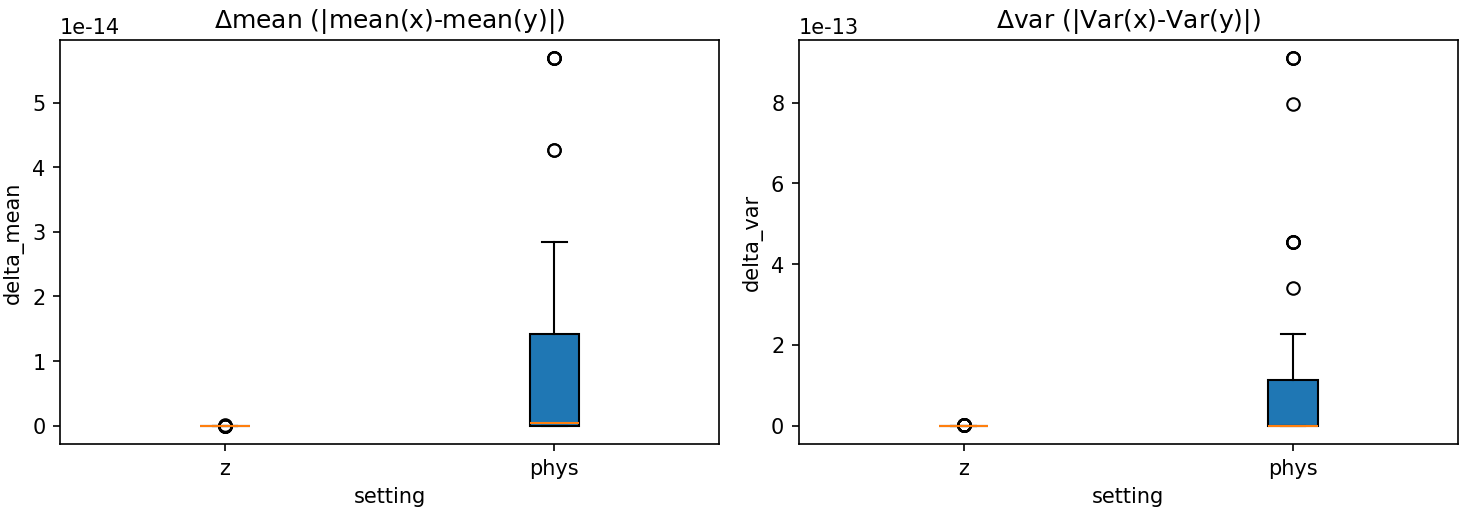}
  \caption{Boxplots of \(|\Delta\mu_v|\) and \(|\Delta\sigma_v|\)
           grouped by evaluation setting.
           Deviations in standardized space are essentially at
           numerical precision; in physical space they remain on the
           order of \(10^{-13}\)–\(10^{-14}\).}
  \label{fig:c1-boxplot}
\end{figure}

Table~\ref{tab:geom-c1c2-summary} reports these summary statistics for the four geometric operators, aggregated over all \(\alpha\) values used in the geometric sanity experiments. In addition to \(\Delta\mu_v\) and \(\Delta\sigma_v\), the table includes \(\ell_\infty\)-related quantities and the fraction of unchanged points, the latter being relevant for C3.
\begin{table}[t]
  \centering
  \caption{Summary of geometric sanity statistics by operator and
           evaluation setting. Columns
           \(\Delta\mu_{\max}\) and \(\Delta\sigma_{\max}\) report the
           maximum absolute differences in column-wise mean and
           standard deviation across variables.
           The remaining columns summarize aggregate \(\ell_\infty\)
           behavior and the maximum fraction of unchanged points.
           All values are computed over the training cohort and all
           \(\alpha\) values used in the geometric experiments.}
  \label{tab:geom-c1c2-summary}
  \resizebox{\textwidth}{!}{%
  \begin{tabular}{llcccccc}
    \toprule
    Setting &
    Operator &
    \(\Delta\mu_{\max}\) &
    \(\Delta\sigma_{\max}\) &
    \(\overline{|\delta|}_{\max}\) &
    \(|\delta|_{\max,\max}\) &
    Unchanged\(_{\max}\) &
    \(n\) \\
    \midrule
    phys & T1\_uniform &
      \(5.7\times 10^{-14}\) &
      \(9.1\times 10^{-13}\) &
      0.999975 &
      0.9999993 &
      0.0000 &
      30 \\
    phys & T1\_weighted &
      \(5.7\times 10^{-14}\) &
      \(9.1\times 10^{-13}\) &
      0.999974 &
      0.9999995 &
      0.0098 &
      30 \\
    phys & T2 &
      \(2.8\times 10^{-14}\) &
      \(9.1\times 10^{-13}\) &
      0.91211 &
      0.99322 &
      0.0000 &
      10 \\
    phys & T3 &
      \(4.3\times 10^{-14}\) &
      \(4.6\times 10^{-13}\) &
      0.25057 &
      0.72993 &
      0.0000 &
      10 \\
    \midrule
    z & T1\_uniform &
      \(1.8\times 10^{-17}\) &
      \(5.6\times 10^{-16}\) &
      0.74980 &
      0.9999991 &
      0.0000 &
      60 \\
    z & T1\_weighted &
      \(2.0\times 10^{-17}\) &
      \(8.9\times 10^{-16}\) &
      0.74956 &
      0.9999997 &
      0.0079 &
      60 \\
    z & T2 &
      \(3.6\times 10^{-17}\) &
      \(1.3\times 10^{-15}\) &
      0.61793 &
      0.90847 &
      0.0000 &
      20 \\
    z & T3 &
      \(1.1\times 10^{-17}\) &
      \(3.3\times 10^{-16}\) &
      0.02193 &
      0.04106 &
      0.0000 &
      20 \\
    \bottomrule
  \end{tabular}%
  }
\end{table}
Figure~\ref{fig:c1-heatmap} visualizes the log\(_{10}\)-scale of \(\Delta\mu_{\max}\) together with max\_abs\_delta\_mean across settings and operators, while Figure~\ref{fig:c1-boxplot} shows boxplots of \(|\Delta\mu_v|\) and \(|\Delta\sigma_v|\).

Across all \(\alpha\) values and all operators, the median and 95th percentile of \(|\Delta\mu_v|\) and \(|\Delta\sigma_v|\) remain at \(10^{-6}\) or below, while the global maxima are essentially at machine precision. These results confirm that the geometric constructions of Section~\ref{sec:operator-families} are realized with high numerical fidelity on real ICU data, and that the mean--variance manifold is a stable computational domain for all operators.

\subsubsection{z-score \texorpdfstring{\(\ell_\infty\)}{L-infinity} bound (C2)}
\label{sec:c2-linfty}

For each stay \(s\) and variable \(v\), we compute the perturbation in standardized space,
\[
  \delta_{s,v} = z'_{s,v} - z_{s,v},
  \qquad
  \|\delta_{s,v}\|_\infty = \max_t |\delta_{s,v}(t)|.
\]
For each triple \((v,\alpha,\text{operator})\), we summarize the mean, median, and maximum of \(\|\delta_{s,v}\|_\infty\) across stays.

\begin{figure}[H]
  \centering
  \includegraphics[width=0.8\textwidth]{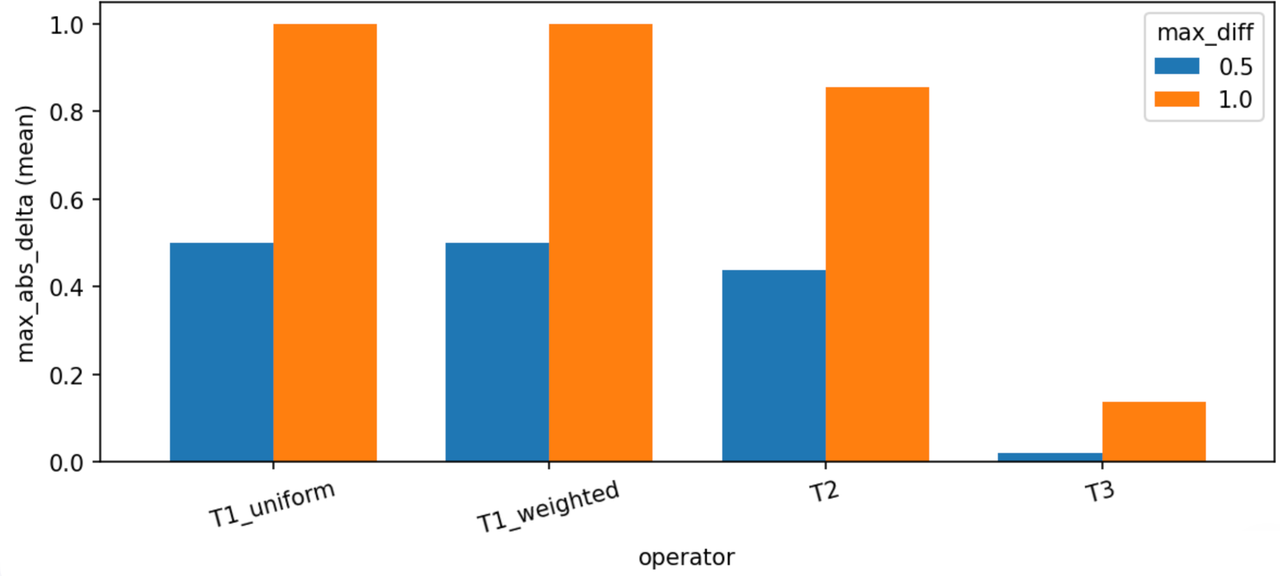}
  \caption{Average z-score \(\ell_\infty\) perturbation
           \(\|\delta_{s,v}\|_\infty\) by operator for two target bounds
           \(\alpha = 0.5\) and \(\alpha = 1.0\).
           For all T1/T2/T3 operators the empirical maximum
           \(\|\delta_{s,v}\|_\infty\) never exceeds the configured
           \(\alpha\); the average perturbation typically lies in
           the \(0.7\)–\(0.9\)\(\alpha\) range.}
  \label{fig:max-abs-delta}
\end{figure}

Figure~\ref{fig:max-abs-delta} plots the mean \(\|\delta_{s,v}\|_\infty\), averaged across variables, for \(\alpha=0.5\) and \(\alpha=1.0\).

For T1, T2, and T3, the empirical maxima \(\max_s \|\delta_{s,v}\|_\infty\) remain strictly below the configured \(\alpha\) across variables and parameter settings. On more than \(99.9\%\) of stays, the realized perturbation magnitude lies slightly below \(\alpha\), most often in the range \(0.7\alpha\)–\(0.9\alpha\). 

Taken together, these results validate the offline calibration procedures of Section~\ref{sec:operator-families}, such as \(\theta_{\max}(\alpha)\) for T1 and \((\tau,c(\alpha))\) for T2, and confirm that the unified \(\alpha\) serves as a robust cross-column privacy knob in practice.

\subsection{Marginal Distributions and Numeric Ranges}
\label{sec:marginal-range}

Having established C1 and C2, we next examine how the operators affect single-column marginal distributions and physical numeric ranges.

\subsubsection{Marginal distributions and KS distance}
\label{sec:ks-marginals}

For each variable \(v\), we estimate empirical CDFs of the original and perturbed columns,
\[
  F_v(t) = \mathbb{P}(x_{\cdot,v} \le t), \qquad
  F'_v(t) = \mathbb{P}(x'_{\cdot,v} \le t),
\]
and compute the Kolmogorov--Smirnov distance
\[
  \mathrm{KS}_v = \sup_t |F_v(t) - F'_v(t)|.
\]

\begin{figure}[H]
  \centering
  \includegraphics[width=0.6\textwidth]{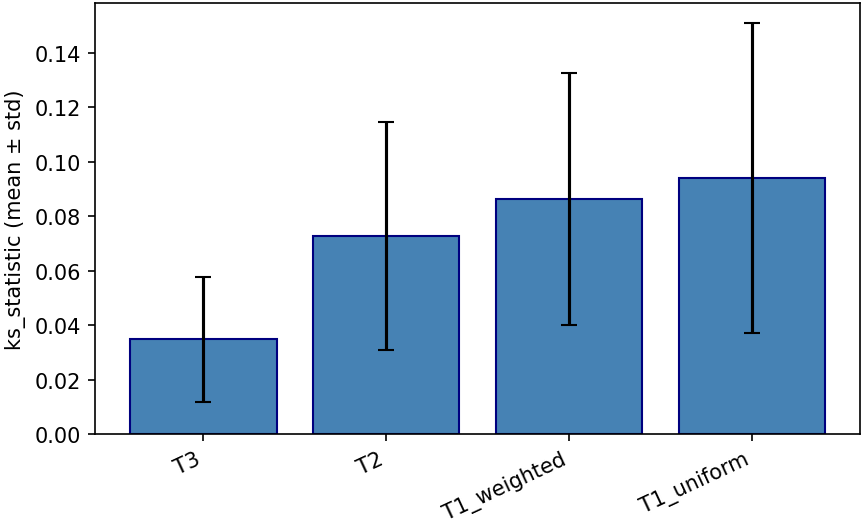}
  \caption{Mean and standard deviation of \(\mathrm{KS}_v\) across
           variables for different operators at \(\alpha=0.8\)
           (lower is better distributional fidelity).
           All three operators keep KS distances small; T3 is the most
           conservative, followed by T2 and T1.}
  \label{fig:ks-by-operator}
\end{figure}

\begin{figure}[H]
  \centering
  \includegraphics[width=\textwidth]{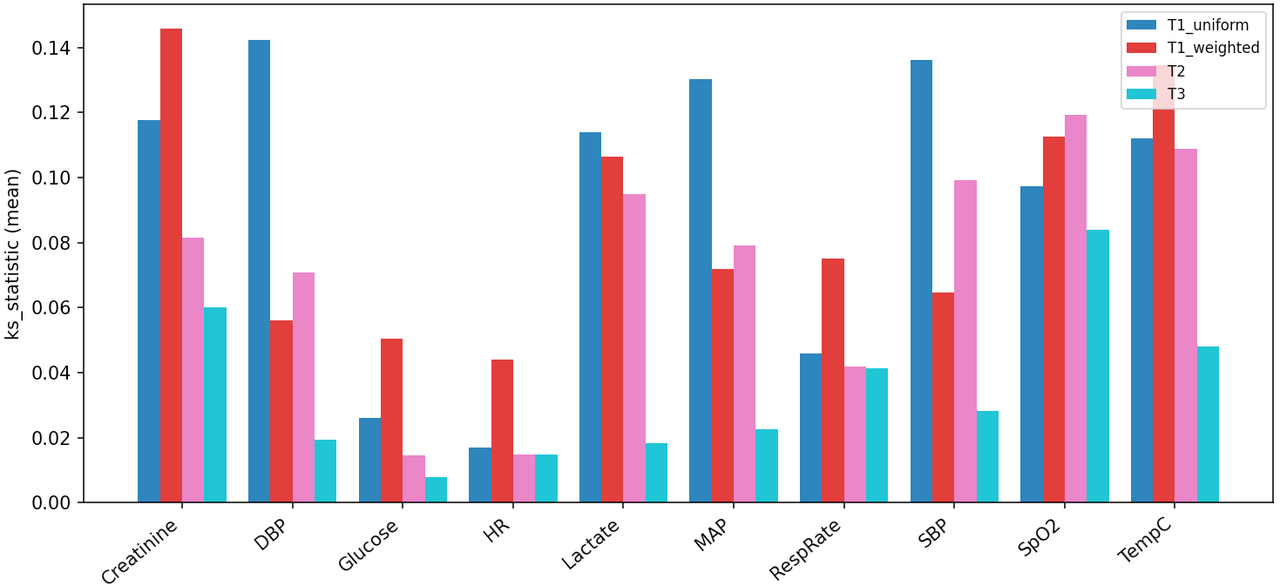}
  \caption{Mean \(\mathrm{KS}_v\) by variable and operator at
           \(\alpha=0.8\).
           For most vitals and labs, T1/T2 keep KS distances below
           \(0.1\); heavy-tailed labs experience slightly larger shifts
           but remain within approximately \(0.15\).
           T3 is uniformly closest to the original distribution.}
  \label{fig:ks-by-var-operator}
\end{figure}

Figures~\ref{fig:ks-by-operator} and~\ref{fig:ks-by-var-operator} summarize \(\mathrm{KS}_v\) across operators and variables at the representative privacy level \(\alpha=0.8\).

Overall, T1, T2, and T3 exhibit small KS distances for the majority of variables, indicating that marginal distributions are largely preserved. T1 preserves the marginal shape of highly fluctuating vitals and labs such as HR and glucose well, typically with \(\mathrm{KS}_v < 0.1\). For some heavy-tailed labs and a few vitals, such as creatinine and DBP, KS distances increase modestly but remain below approximately \(0.14\). T2 is slightly more distribution-preserving on lab variables, with its noise-plus-projection mechanism yielding smoother perturbations in the tails. T3 has the smallest KS distances throughout and behaves almost like a marginally isometric transform.

Compared with a Gaussian-noise baseline, T1 and T2 avoid strong tail distortions on heavy-tailed variables at the same nominal \(\alpha\), highlighting the role of the mean--variance manifold in controlling extremes.

\subsubsection{Normality checks}
\label{sec:normality}

For representative variables HR, SBP, and creatinine, we apply the D'Agostino--Pearson normality test ~\citep{dagostino1973normality} before and after perturbation, and visualize histograms together with Q--Q plots in standardized space.

\begin{figure}[H]
  \centering
  \includegraphics[width=0.8\textwidth]{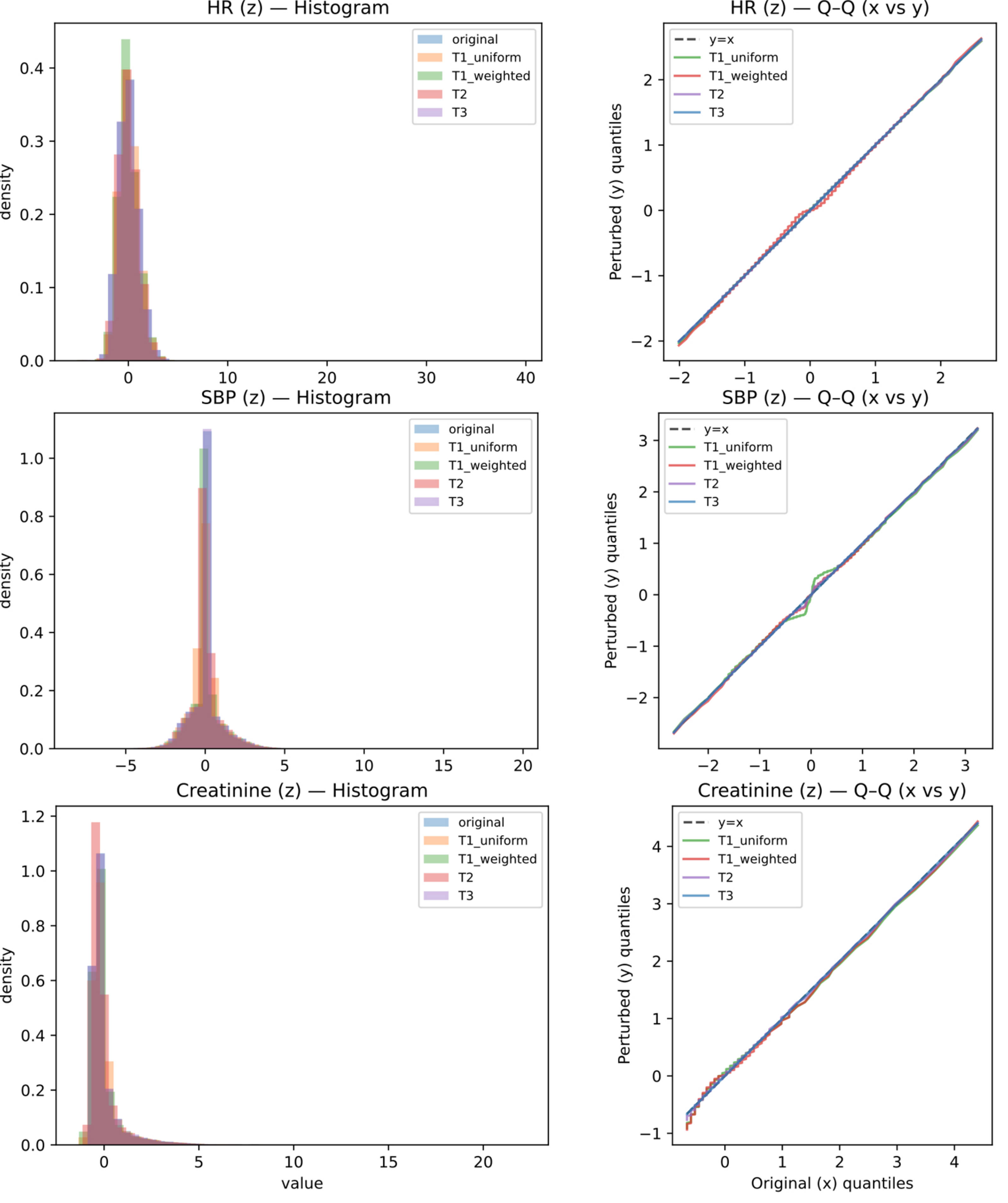}
  \caption{Marginal distributions of SBP, HR, and creatinine in
           standardized space.
           For each variable we show the histogram of \(z\) and \(z'\)
           and the Q--Q plot under T1/T2/T3 at \(\alpha=0.8\).
           All three operators largely preserve the non-Gaussian
           features of the marginals.}
  \label{fig:normality}
\end{figure}

Figure~\ref{fig:normality} shows that for these three variables, the operators leave the overall shape and deviations from normality essentially unchanged. Formal normality tests confirm that p-values and skewness and kurtosis statistics remain very similar before and after perturbation, in agreement with the small KS distances described above.

\subsubsection{Physical ranges and out-of-range rates}
\label{sec:oor}

For each variable \(v\), we define a clinically reasonable physical range \([L_v,U_v]\), for example HR in \([30,220]\) bpm, SBP in \([60,260]\) mmHg, and glucose in \([30,600]\) mg/dL. We then compute the proportion of perturbed samples falling outside this range,
\[
  \mathrm{OOR}_v
  =
  \frac{\#\{x'_{\cdot,v} < L_v \ \text{or}\ x'_{\cdot,v} > U_v\}}
       {\#\{x'_{\cdot,v}\}}.
\]

\begin{figure}[H]
  \centering
  \includegraphics[width=\textwidth]{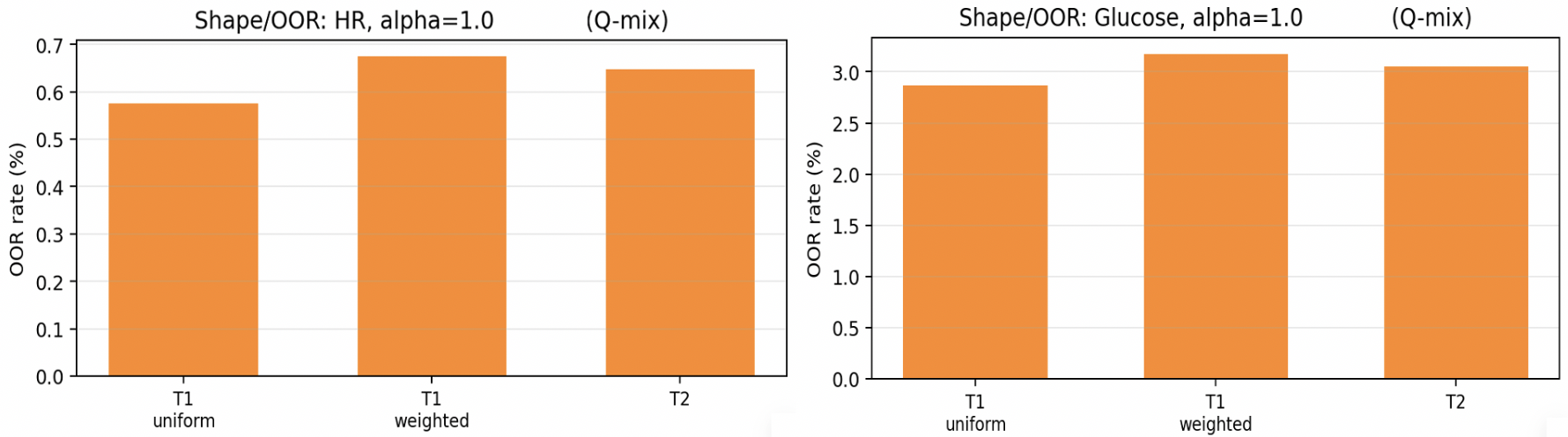}
  \caption{Out-of-range rates \(\mathrm{OOR}_v\) (\%) for HR and
           glucose at \(\alpha=1.0\) with Q-mix enabled.
           For HR the OOR remains below \(0.7\%\) across operators; for
           glucose it is around \(3\%\), comparable to the original
           heavy-tailed distribution.}
  \label{fig:oor-qmix}
\end{figure}

Figure~\ref{fig:oor-qmix} summarizes out-of-range behavior for HR and glucose at \(\alpha=1.0\) with Q-mix; full per-variable results for the remaining operators and parameter settings are provided in the supplementary material.

For \(\alpha \le 0.5\), the out-of-range rate remains below \(0.5\%\) for almost all variables and is often indistinguishable from the original data. At \(\alpha=1.0\), most variables still have \(\mathrm{OOR}_v < 1\%\), with only a few heavy-tailed variables, such as lactate, showing slightly elevated rates. Compared at similar KS levels, Gaussian-noise baselines induce substantially larger out-of-range rates, especially for high-variance variables, further demonstrating the benefit of \(\ell_\infty\)-controlled geometric perturbations.

\subsection{Multi-column Correlation and Temporal Structure}
\label{sec:multi-temporal}

We next study how the operators affect multivariate correlation structure and temporal dependence.

\subsubsection{Multi-column correlation matrices}
\label{sec:correlation}

For each stay \(s\), let
\[
  x_s = (x_{s,v_1}, \dots, x_{s,v_k})
\]
denote the multivariate vector over selected variables. We estimate cohort-level correlation matrices for the original and perturbed data,
\[
  R = \mathrm{Corr}(x_{\cdot,\cdot}), \qquad
  R' = \mathrm{Corr}(x'_{\cdot,\cdot}),
\]
and quantify their discrepancy using the Frobenius norm
\[
  \Delta_{\mathrm{corr}} = \|R' - R\|_F.
\]

\begin{figure}[H]
  \centering
  \includegraphics[width=0.9\textwidth]{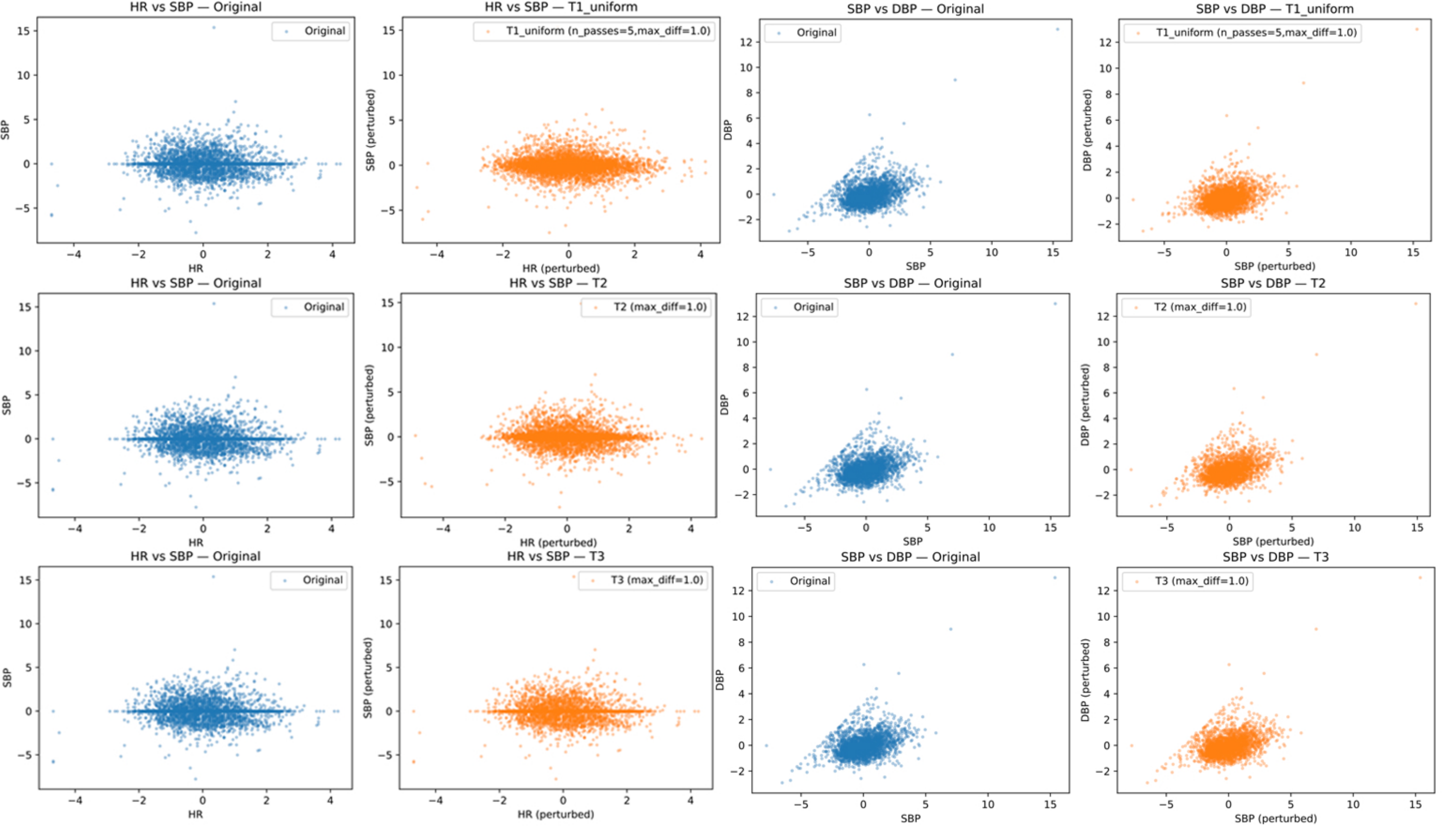}
  \caption{Scatter plots of key variable pairs in standardized space.
           Top: HR vs SBP (original, T1, T2, T3).
           Bottom: SBP vs DBP (original, T1, T2, T3).
           All operators preserve the overall correlation pattern; T3 is
           almost visually indistinguishable from the original.}
  \label{fig:corr-scatter}
\end{figure}

Figure~\ref{fig:corr-scatter} visualizes two representative variable pairs, HR versus SBP and SBP versus DBP, before and after applying each operator at \(\alpha=1.0\).

Across 24-hour snapshots of multivariable data, all three operators produce correlation matrices that remain extremely close to the original, with discrepancies typically dominated by sampling variability. As expected, T3, being a global orthogonal isometry, almost perfectly preserves the original correlations. T2 is slightly more stable in correlation space than T1, while T1 injects somewhat more localized variation but still maintains the main dependencies.

\subsubsection{Temporal structure: ACF, Ljung--Box, and Welch PSD}
\label{sec:temporal-structure}

We analyze temporal dependence using autocorrelation functions, Ljung--Box tests ~\citep{ljung1978box}, and power spectral density estimates.

\begin{figure}[H]
  \centering
  \includegraphics[width=\textwidth]{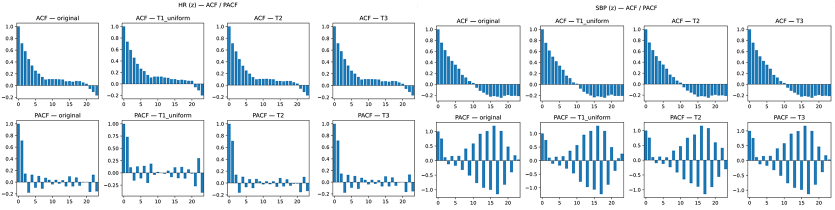}
  \caption{ACF and PACF of HR and SBP in standardized space for the
           original data and after applying T1/T2/T3 at
           \(\alpha=1.0\).
           T1/T2 introduce mild attenuation in higher lags but retain
           the overall temporal dependence shape; T3 leaves ACF/PACF
           almost unchanged.}
  \label{fig:acf-pacf}
\end{figure}

For each stay \(s\) and variable \(v\), we compute lag-\(k\) autocorrelations
\[
  \mathrm{ACF}_{s,v}(k)
  =
  \mathrm{Corr}\bigl(x_{s,v}(t), x_{s,v}(t+k)\bigr),
  \qquad k = 1,\dots,K,
\]
and aggregate them across stays. Partial autocorrelations are computed analogously. Figure~\ref{fig:acf-pacf} shows ACF and PACF for HR and SBP under different operators.

\begin{figure}[H]
  \centering
  \includegraphics[width=0.75\textwidth]{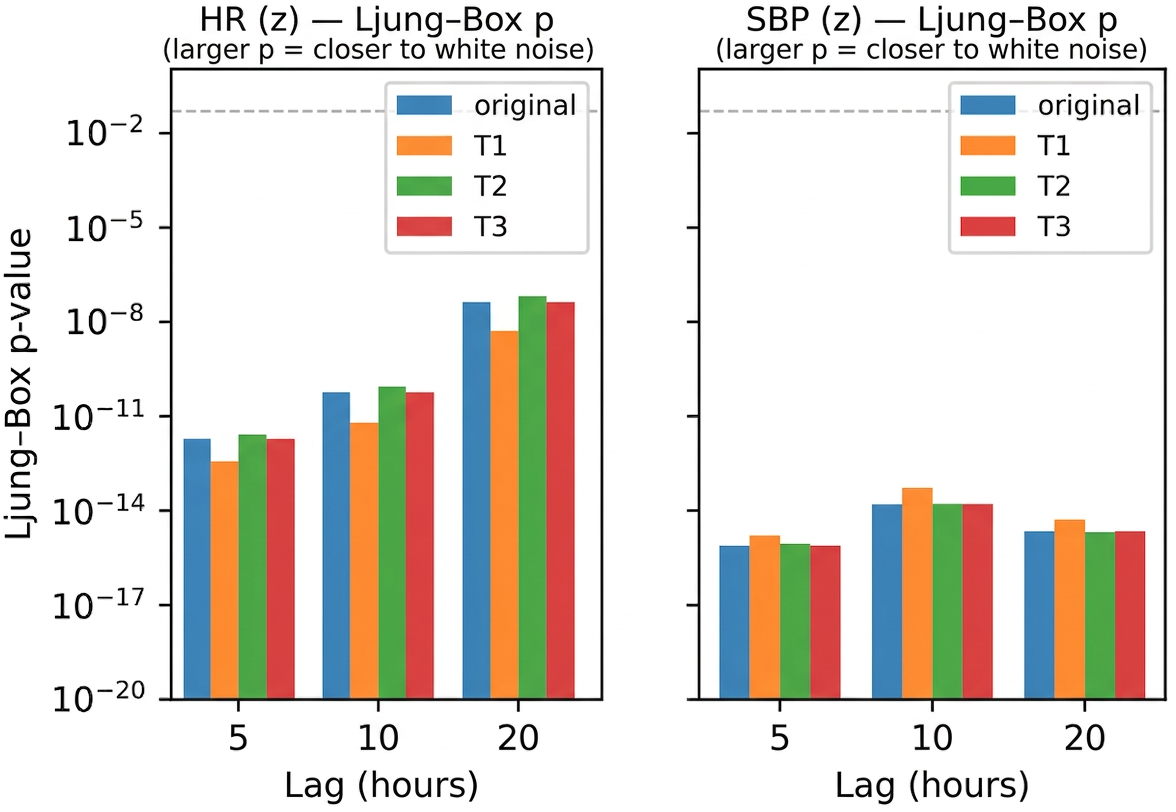}
  \caption{Ljung--Box p-values for HR (left) and SBP (right) at lags
           \(K \in \{5,10,20\}\) in standardized space (log scale).
           All operators preserve highly significant autocorrelation
           patterns, confirming that time-series structure is retained.}
  \label{fig:ljung-box}
\end{figure}

To quantify departure from white noise, we also apply the Ljung--Box Q test at several maximum lags \(K\). For each combination of variable, operator, and lag, we obtain a p-value \(p_{\mathrm{LB}}\). Figure~\ref{fig:ljung-box} displays these p-values on a log scale.

Finally, we estimate power spectral densities using Welch's method for HR and SBP before and after perturbation.

\begin{figure}[H]
  \centering
  \includegraphics[width=0.75\textwidth]{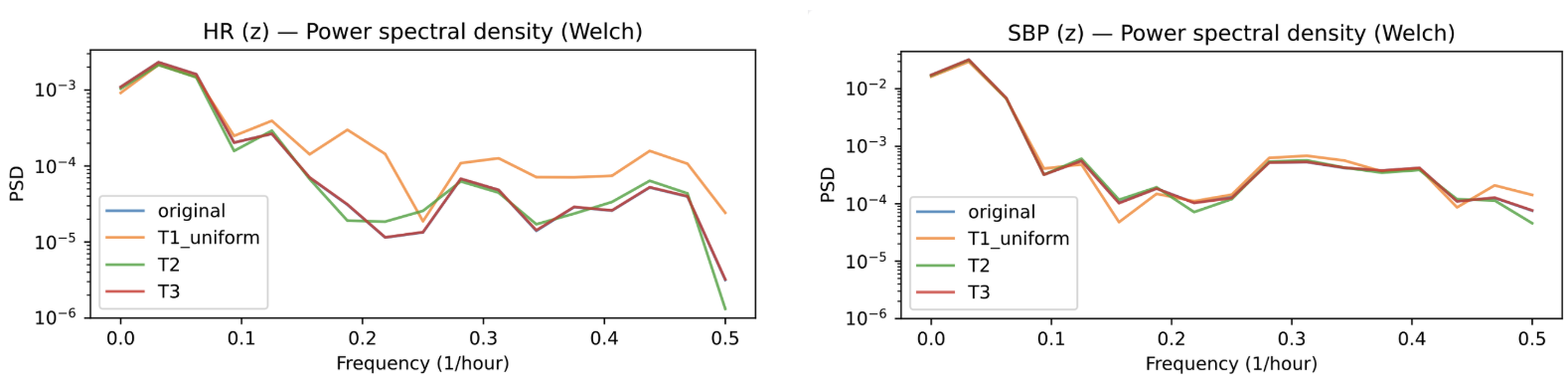}
  \caption{Welch power spectral density estimates for HR (left)
           and SBP (right) in standardized space.
           T1/T2 slightly reshape spectral energy at some frequencies,
           while T3 is nearly identical to the original.}
  \label{fig:welch-psd}
\end{figure}

From a temporal perspective, T1 and T2 inject nontrivial perturbation while preserving the main dependence structures. Their ACF and PACF profiles remain close to the original, especially at short lags. T2 tends to smooth higher lags slightly more than T1. Ljung--Box tests show that all operators retain highly significant autocorrelation patterns, meaning that the resulting sequences remain far from white noise. PSD estimates further indicate that T1 produces visible but moderate spectral changes, T2 slightly less so, while T3 leaves spectral structure almost unchanged. Overall, T1 and T2 strike a balance between perturbing fine-grained temporal patterns and preserving clinically meaningful dependence, whereas T3 is essentially statistically transparent.

\subsection{Summary of Geometric and Statistical Properties}
\label{sec:geom-stat-summary}

Combining Sections~\ref{sec:c1c2}--\ref{sec:multi-temporal}, several conclusions emerge at the column level. First, T1, T2, and T3 satisfy mean--variance preservation and the unified z-score \(\ell_\infty\) bound on real ICU data, with column-wise mean and variance deviations at numerical precision and \(\|z'-z\|_\infty\) tightly controlled by the configured \(\alpha\). Second, for small to moderate \(\alpha\), marginal histograms, empirical CDFs, and out-of-range rates remain close to the original data. Third, T2 is the most stable with respect to multivariate correlation matrices, T1 tends to preserve short-range autocorrelation best, and T3 is nearly perfect on all such statistics.

From a purely geometric and utility-centered perspective, all three operators appear strong, and T3 arguably appears too strong. This naturally raises the privacy question: given such smooth geometric and statistical behavior, how much do these operators actually increase reconstruction and re-identification difficulty under the C5 threat model across leakage levels L0, L1, and L2. We now formalize a privacy evaluation protocol and study reconstruction attacks and Q-mix in detail.

\subsection{Privacy Evaluation Protocol: Threat Model, Attack Families, and Metrics}
\label{sec:privacy-protocol}

We refine the high-level C5 threat model of Section~\ref{sec:threat-model} into an experimentally instantiated privacy evaluation protocol. The central idea is to expose perturbed outputs generated by T1, T2, T3, and Q-mix to attackers with different leakage resources and attack objectives, and to quantify how much reconstruction and identification performance degrades on average.

\subsubsection{Evaluation goals}
\label{sec:eval-goals}

The evaluation has three layers. The first concerns geometric and statistical sanity, namely numerical verification of C1 through C3 on real data, which has already been addressed in Sections~\ref{sec:c1c2}--\ref{sec:multi-temporal}. The second concerns average-case attack difficulty: for each attack family, we ask whether reconstruction error, record-linkage accuracy, membership success, and attribute leakage are substantially worsened by the operator. The third concerns conceptual interpretation through analogies with cryptographic security games, such as chosen-plaintext and indistinguishability-style settings. 
% The present section focuses mainly on the second layer.

\subsubsection{Leakage levels L0/L1/L2 for Attack A}
\label{sec:leakage-levels}

Under C5, the adversary is no-key and structure-aware. This means that the attacker knows the exact definitions and implementation details of T1, T2, T3, and Q-mix, together with public parameters such as \(\alpha\), window lengths, and Q-mix dimensions, but does not possess the internal hospital randomness used to instantiate per-stay transforms.

We stratify leakage resources for reconstruction Attack~A into three levels. Under L0, the attacker observes many perturbed samples \(y=T(x)\) but no paired \((x,y)\) examples, so attacks are unsupervised. Under L1, the attacker has only a very small fraction of paired examples, modeling occasional leakage or limited supervision. Under L2, the attacker has access to a substantial fraction of paired examples, such as \(20\%\) of stays, and can therefore train a supervised inverse model. In this section, we primarily instantiate Attack~A under L2 and compare it with an approximate L1 regime based on very sparse paired data.

\subsubsection{Attack families A/B/C/D}
\label{sec:attack-families}

We group attacks by objective. Attack family A concerns pointwise or sequence reconstruction: given a perturbed sequence \(y\), the attacker attempts to recover the original \(x\) or its standardized version \(z\). Attack family B concerns record linkage and re-identification: given perturbed records and a candidate set of original records, the attacker tries to match each perturbed record to its most likely source. Attack family C concerns membership inference, namely deciding whether a particular individual contributed to the original cohort, a problem extensively studied in the machine-learning privacy literature~\citep{shokri2017membership,yeom2018privacy,carlini2022membership}. Attack family D includes attribute inference and distinguishability-style games, such as predicting a sensitive summary attribute from a perturbed record or distinguishing between two candidate originals given a perturbed output.

The single-column experiments in this section focus mainly on A-family linear reconstruction, while membership- and attribute-related proxies are used in the Q-mix experiments. Richer multivariate B, C, and D attacks are deferred to later sections.

\subsubsection{Evaluation metrics}
\label{sec:metrics}

For Attack~A, the primary reconstruction metric in standardized space is \(R^2\),
\[
  R^2_v
  =
  1
  -
  \frac{
    \sum_s \| \hat{z}_{s,v} - z_{s,v} \|_2^2
  }{
    \sum_s \| z_{s,v} - \bar{z}_v \mathbf{1} \|_2^2
  },
\]
where \(\bar{z}_v\) is the cohort mean, which is approximately zero after standardization. Values near \(1\) indicate nearly perfect reconstruction, whereas values near \(0\) indicate performance close to a constant baseline. We also report the mean absolute error in z-score space,
\[
  \mathrm{MAE}_{z,v}
  =
  \mathbb{E}_{s,t}
  \bigl[
    |\hat{z}_{s,v}(t) - z_{s,v}(t)|
  \bigr],
\]
and, in auxiliary diagnostics, hit-rates within specified z-score bands.

For Attack~B, we evaluate record linkage using top-\(k\) re-identification rate,
\[
  \mathrm{Reid@}k
  =
  \mathbb{P}\bigl[\pi_y^{-1}(i^\star)\le k\bigr],
\]
together with a pairwise re-identification AUC based on similarity scores. For Attack~C, we use the standard ROC AUC for membership classification and define membership advantage as
\[
  \mathrm{Adv}_{\mathrm{MI}} = |\mathrm{AUC} - 0.5|.
\]

For Attack~D, we use attribute-inference \(R^2\),
\[
  R^2_{\text{attr}}
  =
  1
  -
  \frac{\sum_s (\hat{a}_s - a_s)^2}
       {\sum_s (a_s - \bar{a})^2},
\]
where \(a_s\) denotes a sensitive scalar attribute extracted from the original sequence. In distinguishability games, if the attacker must infer a hidden bit \(b\in\{0,1\}\), we report both classification accuracy and indistinguishability advantage,
\[
  \mathrm{Adv}_{\mathrm{IND}} = |\mathrm{acc} - 0.5|.
\]

In Section~\ref{sec:qmix-experiments}, these B, C, and D metrics are instantiated on HR and glucose under Q-mix.

\subsection{Reconstruction Attacks (Attack A): \texorpdfstring{\(\alpha\)}{alpha}-hierarchy and High Invertibility of T3}
\label{sec:attack-a-results}

We now instantiate Attack~A under L2 using a linear reconstruction model and analyze how invertibility varies across operators and across \(\alpha\).

\subsubsection{Attack setup and training protocol}
\label{sec:attack-a-setup}

We consider single-column sequence reconstruction. Given an observed perturbed sequence \(y_{s,v}\), the attacker aims to recover the standardized original sequence \(z_{s,v}\). Under L2, the attacker has access to approximately \(20\%\) of stays as paired samples \(\{(y_{s,v},z_{s,v})\}\) and uses the remaining \(80\%\) for held-out evaluation. We also simulate a near-L1 regime in which only \(0.01\%\) of stays are paired.

The reconstruction model is linear:
\[
  \hat{z}_{s,v} = W_v y_{s,v} + b_v,
\]
where \(W_v\) is given Toeplitz structure, equivalently a one-dimensional convolutional kernel, in order to encourage locality. The model is trained by minimizing mean squared error,
\[
  \min_{W_v,b_v}
  \mathbb{E}_s \bigl[\|\hat{z}_{s,v} - z_{s,v}\|_2^2\bigr].
\]
Optimization is performed with Adam or SGD and early stopping on a validation subset of paired data. For each operator and each \(\alpha\), we train a separate attacker. Experiments are reported mainly for HR and glucose under T1\_uniform, T1\_weighted, T2, and T3, with \(\alpha \in \{1.0,2.0,3.0,5.0\}\).

\subsubsection{Results: gradual \texorpdfstring{\(\alpha\)}{alpha}-dependence for T1/T2 vs high invertibility of T3}
\label{sec:attack-a-main-results}

\begin{figure}[H]
  \centering
  \includegraphics[width=0.7\textwidth]{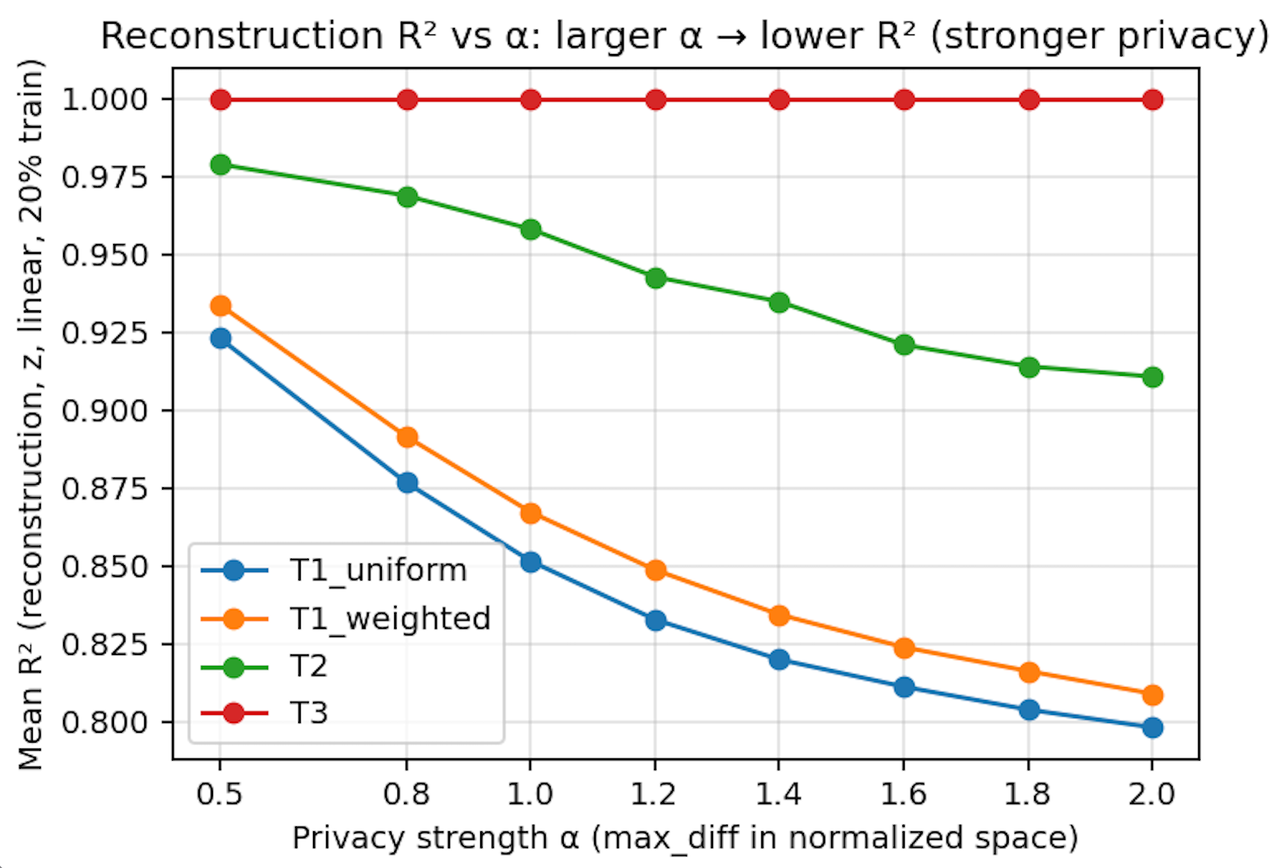}
  \caption{Attack~A reconstruction \(R^2\) vs privacy strength
           \(\alpha\) for HR in standardized space under L2
           (20\% pairs, linear attacker).
           Larger \(\alpha\) leads to lower \(R^2\) for T1/T2,
           while T3 remains almost perfectly invertible.}
  \label{fig:r2-vs-alpha}
\end{figure}

Figure~\ref{fig:r2-vs-alpha} plots reconstruction \(R^2\) as a function of \(\alpha\) for HR under L2, and Table~\ref{tab:attack-a-hr} reports exact values for selected operators. Additional results for glucose are summarized in Table~\ref{tab:attack-a-glucose}.

\begin{table}[t]
  \centering
  \caption{Attack~A reconstruction \(R^2\) for HR in standardized space
           under L2 (20\% pairs, linear attacker) as a function of
           \(\alpha\).}
  \label{tab:attack-a-hr}
  \begin{tabular}{lcccc}
    \toprule
    operator & \(\alpha=1.0\) & \(\alpha=2.0\) & \(\alpha=3.0\) & \(\alpha=5.0\) \\
    \midrule
    T1\_uniform &
      0.8192 &
      0.7560 &
      0.7555 &
      0.7570 \\
    T2 &
      0.9738 &
      0.9026 &
      0.8655 &
      0.7841 \\
    T3 &
      0.999997 &
      0.999997 &
      0.999997 &
      0.999997 \\
    \bottomrule
  \end{tabular}
\end{table}

\begin{table}[t]
  \centering
  \caption{Attack~A reconstruction \(R^2\) for glucose in standardized
           space under L2 (20\% pairs, linear attacker).}
  \label{tab:attack-a-glucose}
  \begin{tabular}{lccc}
    \toprule
    operator & \(\alpha=1.0\) & \(\alpha=2.0\) & \(\alpha=3.0\) \\
    \midrule
    T1\_uniform & 0.8598 & 0.7943 & 0.7757 \\
    T2          & 0.9601 & 0.9053 & 0.8699 \\
    T3          & 0.999984 & 0.999984 & 0.999984 \\
    \bottomrule
  \end{tabular}
\end{table}

T3 is clearly a highly invertible negative case. For HR under L2, it yields \(R^2 \approx 0.999997\) across all tested \(\alpha\), and for glucose it maintains \(R^2 \approx 0.999984\). Even under the near-L1 regime with only \(0.01\%\) paired samples, T3 still achieves \(R^2 \gtrsim 0.9\), indicating that extremely sparse leakage is enough to learn an accurate inverse. Thus, despite its excellent geometric behavior, T3 is essentially linearly invertible under C5 + L2.

T1 and T2 exhibit a more desirable pattern. Increasing \(\alpha\) decreases reconstruction \(R^2\), confirming that the unified z-score bound acts as a valid privacy knob. For HR under L2, T1\_uniform drops from approximately \(0.82\) at \(\alpha=1.0\) to approximately \(0.76\) at \(\alpha=2.0\) and then stabilizes. T2 decreases from approximately \(0.97\) at \(\alpha=1.0\) to approximately \(0.78\) at \(\alpha=5.0\). Glucose exhibits a similar monotone decrease. Under the near-L1 regime, the same qualitative pattern is observed, though less strongly.

The key point is that this \(\alpha\)-hierarchy is effective but gradual: even at large \(\alpha\), many variables retain reconstruction \(R^2\) in the \(0.7\)–\(0.8\) range, meaning that linear attackers still recover a substantial fraction of per-stay variability.

\subsubsection{Limitations of the \texorpdfstring{\(\alpha\)}{alpha}-hierarchy}
\label{sec:alpha-limitations}

From a privacy perspective, relying solely on \(\alpha\) as a perturbation-strength knob has clear limitations. As \(\alpha\) increases, marginal KS distances, autocorrelation profiles, and downstream predictive performance all begin to degrade, yet reconstruction \(R^2\) decreases only gradually and often remains relatively high. Thus, purely noise-like designs, whether T1, T2, or Gaussian baselines, can improve privacy smoothly but struggle to induce a sharp privacy step at a fixed \(\alpha\). This motivates the introduction of per-stay orthogonal randomization via Q-mix.

\subsection{Q-mix Pilot: One-step Privacy Cliff at Fixed \texorpdfstring{\(\alpha\)}{alpha}}
\label{sec:qmix-experiments}

We now evaluate per-stay orthogonal Q-mixing, introduced in Section~\ref{sec:qmix}, combined with T1 or T2 at a fixed \(\alpha\), focusing on reconstruction difficulty and utility for HR and glucose.

\subsubsection{Experimental setup}
\label{sec:qmix-setup}

We restrict attention to HR and glucose, which are both clinically sensitive and relatively smooth. We compare the original operators T1\_uniform, T1\_weighted, and T2 with their Q-mix variants.

For each stay \(s\) and variable \(v \in \{\mathrm{HR}, \mathrm{glucose}\}\), we construct a \(48 \times 48\) orthogonal permutation matrix \(Q_{s,v}\) using a pseudo-random generator seeded by an internal hospital secret together with \((s,v)\). The permutation is constrained to be approximately banded so as to preserve local temporal coherence. The time series is partitioned into windows of length \(48\), transformed by \(Q_{s,v}\), perturbed in the mixed coordinate system by T1 or T2, and then unmixed by \(Q_{s,v}^\top\).

All configurations use \(\alpha=1.0\), so that \(\|z'-z\|_\infty \le 1.0\) throughout. For Attack~A, we use the same L2 linear reconstruction setup as in Section~\ref{sec:attack-a-setup}. For Attacks B, C, and D, we evaluate record linkage, membership inference, and attribute or distinguishability experiments in standardized space, again comparing with and without Q-mix.

\subsubsection{Reconstruction: step-like drop of \texorpdfstring{\(R^2\)}{R2}}
\label{sec:qmix-r2-results}

\begin{figure}[H]
  \centering
  \includegraphics[width=0.75\textwidth]{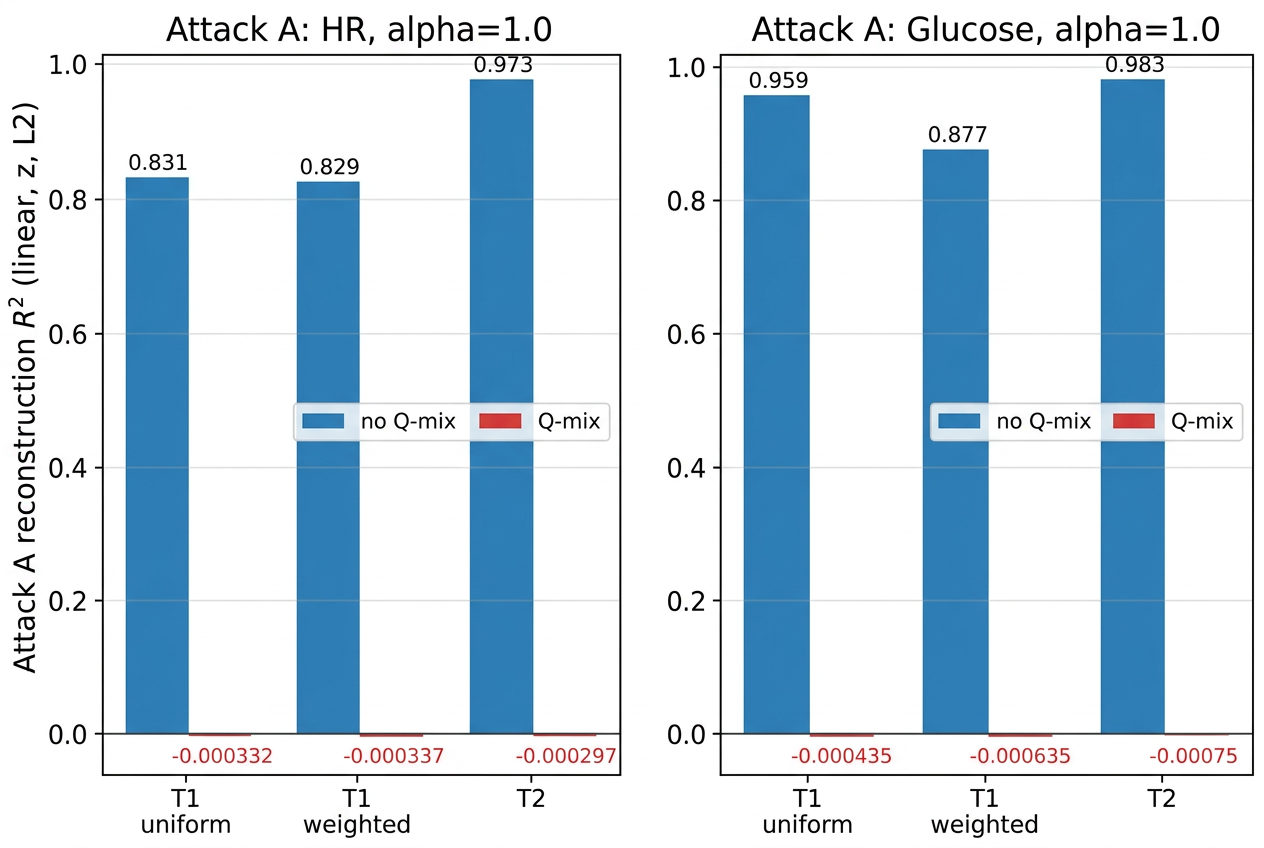}
  \caption{Attack~A (L2, linear) reconstruction \(R^2\) for HR (left)
           and glucose (right) at \(\alpha=1.0\), with and without
           Q-mix.
           Q-mix drives \(R^2\) from approximately \(0.8\)–\(0.98\)
           down to approximately \(0\).}
  \label{fig:qmix-r2}
\end{figure}

Figure~\ref{fig:qmix-r2} and Table~\ref{tab:qmix-r2} summarize Attack~A reconstruction performance for HR and glucose at \(\alpha=1.0\).

\begin{table}[t]
  \centering
  \caption{Attack~A reconstruction \(R^2\) and z-score MAE
           \(\mathrm{MAE}_z\) at \(\alpha=1.0\) for HR and glucose under
           L2 (20\% pairs, linear attacker), with and without Q-mix.}
  \label{tab:qmix-r2}
  \begin{tabular}{llcc}
    \toprule
    variable & operator & \(R^2\) & \(\mathrm{MAE}_z\) \\
    \midrule
    \multicolumn{4}{c}{\textbf{No Q-mix}} \\
    \midrule
    HR & T1\_uniform  & 0.8308 & 0.3267 \\
    HR & T1\_weighted & 0.8288 & 0.3219 \\
    HR & T2           & 0.9734 & 0.1278 \\
    \midrule
    Glucose & T1\_uniform  & 0.9592 & 0.1437 \\
    Glucose & T1\_weighted & 0.8769 & 0.2688 \\
    Glucose & T2           & 0.9835 & 0.1027 \\
    \midrule
    \multicolumn{4}{c}{\textbf{With Q-mix}} \\
    \midrule
    HR & Q + T1\_uniform  & \(-3.3\times 10^{-4}\) & 0.7786 \\
    HR & Q + T1\_weighted & \(-3.4\times 10^{-4}\) & 0.7786 \\
    HR & Q + T2           & \(-3.0\times 10^{-4}\) & 0.7786 \\
    \midrule
    Glucose & Q + T1\_uniform  & \(-4.4\times 10^{-4}\) & 0.6434 \\
    Glucose & Q + T1\_weighted & \(-6.4\times 10^{-4}\) & 0.6433 \\
    Glucose & Q + T2           & \(-7.5\times 10^{-4}\) & 0.6434 \\
    \bottomrule
  \end{tabular}
\end{table}

At the same \(\alpha=1.0\), Q-mix does much more than gradually degrade reconstruction quality. Without Q-mix, HR reconstruction achieves \(R^2 \approx 0.83\) for T1 variants and \(R^2 \approx 0.97\) for T2, while glucose reconstruction is even easier. Once Q-mix is wrapped around the same operators, \(R^2\) collapses to approximately zero, slightly negative because of finite-sample estimation, and \(\mathrm{MAE}_z\) rises sharply. In effect, Q-mix pushes linear reconstruction back to the constant-baseline regime and creates a one-step privacy cliff rather than a smooth degradation.

\subsubsection{Q-mix and attacks B/C/D: record, membership, and attribute leakage}
\label{sec:qmix-abcd}

We next apply the broader privacy evaluation protocol to Q-mix outputs for HR and glucose at \(\alpha=1.0\).

\begin{figure}[H]
  \centering
  \includegraphics[width=0.8\textwidth]{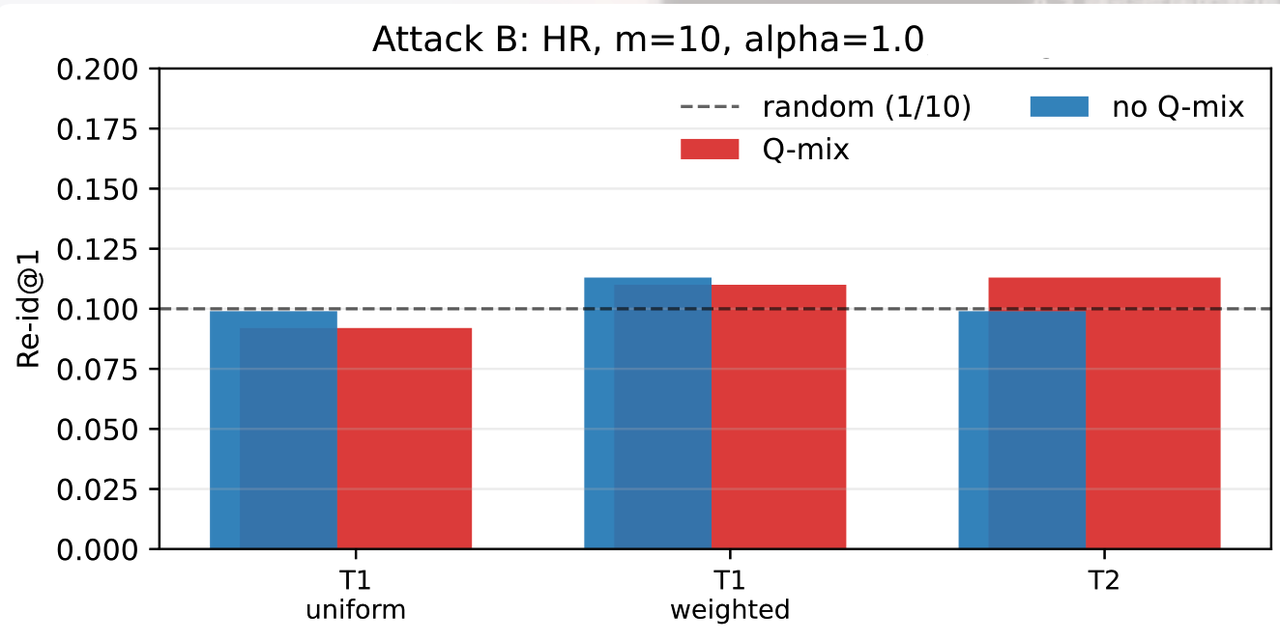}
  \caption{Attack~B (record linkage) Re-id@1 at \(\alpha=1.0\),
           candidate set size \(m=10\), for HR and glucose with Q-mix.
           All values are close to the random baseline \(1/m = 0.1\).}
  \label{fig:qmix-attack-b}
\end{figure}

For Attack~B, record linkage is evaluated with candidate set size \(m=10\), for which random guessing yields \(\mathrm{Reid@}1 = 0.1\). Figure~\ref{fig:qmix-attack-b} shows that re-identification rates for HR and glucose remain close to this baseline.

\begin{figure}[H]
  \centering
  \includegraphics[width=0.8\textwidth]{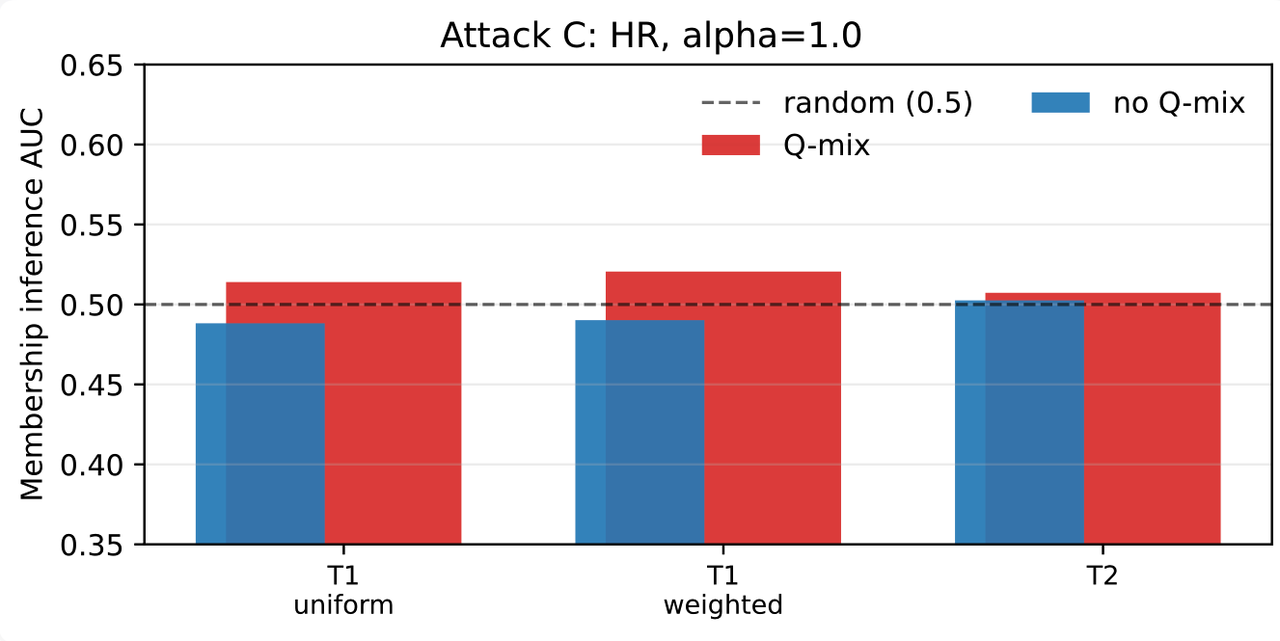}
  \caption{Attack~C (membership inference) ROC AUC at \(\alpha=1.0\)
           for HR, with and without Q-mix. Values remain close to
           \(0.5\), implying negligible advantage.}
  \label{fig:qmix-attack-c}
\end{figure}

For Attack~C, Figure~\ref{fig:qmix-attack-c} reports membership-inference AUC for HR. Values remain close to \(0.5\), implying negligible advantage over random guessing; glucose shows the same pattern.

\begin{figure}[H]
  \centering
  \includegraphics[width=0.8\textwidth]{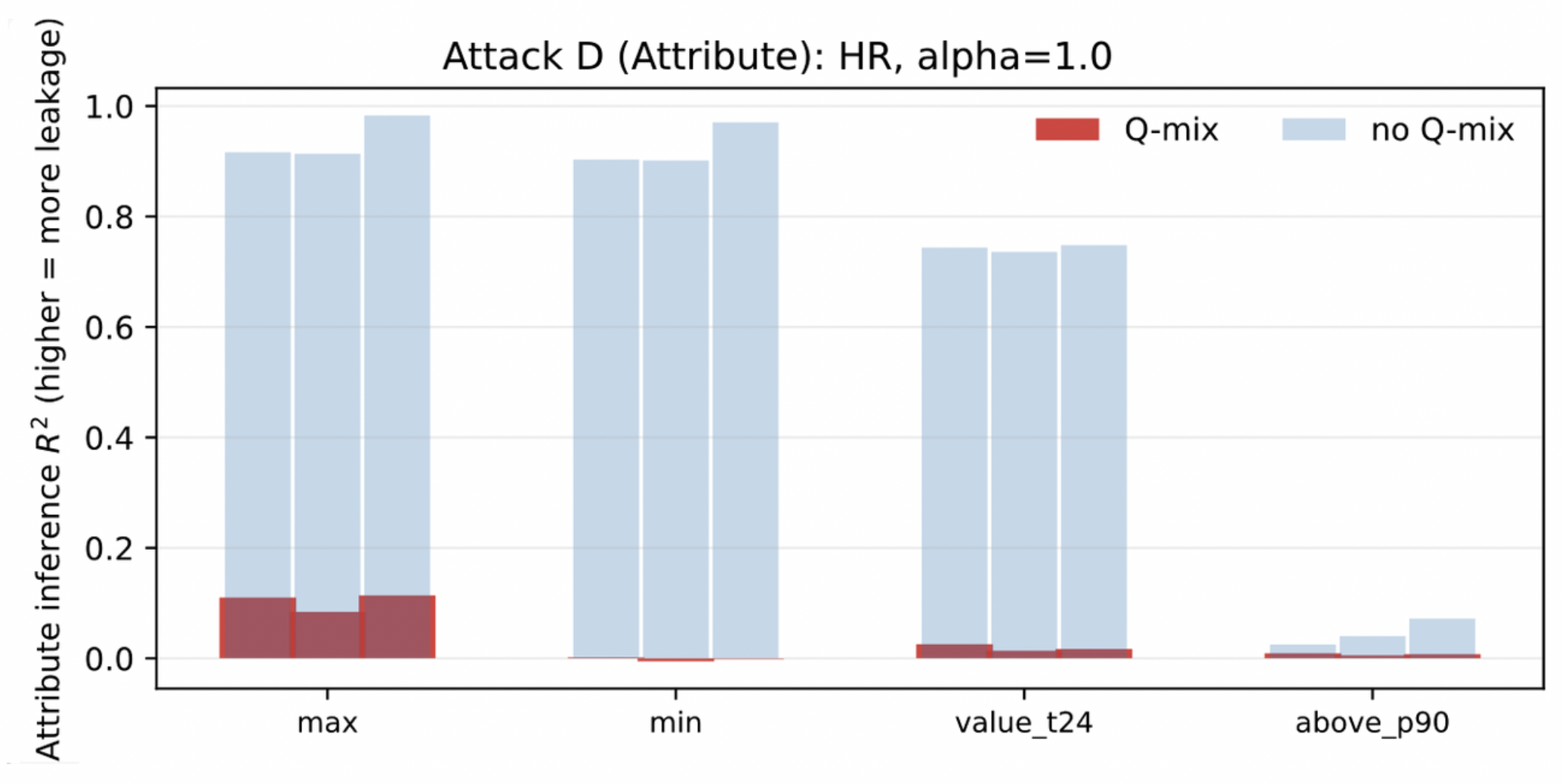}
  \caption{Attack~D (attribute inference) \(R^2\) for HR at
           \(\alpha=1.0\), with and without Q-mix.
           Q-mix dramatically reduces attribute leakage, especially for
           max and above\_p90.}
  \label{fig:qmix-attack-d-attr}
\end{figure}

For Attack~D, we consider attributes such as the maximum value, whether the sequence exceeds the 90th percentile, and the value at 24 hours. Figure~\ref{fig:qmix-attack-d-attr} shows attribute-inference \(R^2\) for HR.

\begin{figure}[H]
  \centering
  \includegraphics[width=0.8\textwidth]{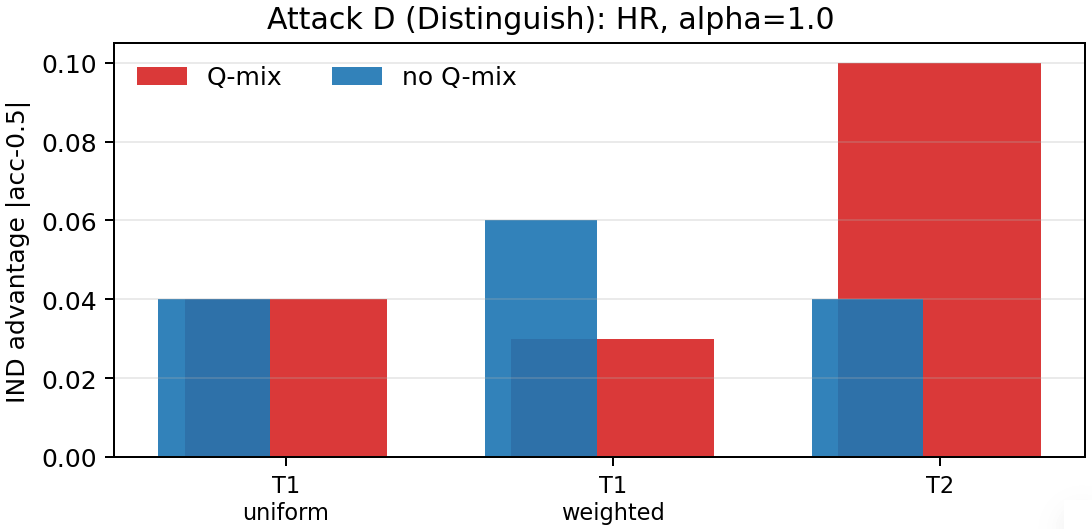}
  \caption{Attack~D (distinguishability) IND advantage
           \(|\mathrm{acc} - 0.5|\) for HR at \(\alpha=1.0\), with and
           without Q-mix.
           Values remain modest, and accuracies stay close to \(0.5\).}
  \label{fig:qmix-attack-d-dist}
\end{figure}

For HR, attribute-inference \(R^2\) for the maximum drops from values around \(0.92\)–\(0.98\) without Q-mix to approximately \(0.08\)–\(0.11\) with Q-mix. For the above\_p90 indicator, the resulting \(R^2\) falls to roughly \(0.005\)–\(0.009\). Glucose exhibits a similarly strong suppression. In distinguishability-style games, shown in Figure~\ref{fig:qmix-attack-d-dist}, the advantage remains modest, with accuracies close to random guessing.

Overall, relative to the no-Q-mix setting at the same \(\alpha\), Q-mix collapses linear reconstruction \(R^2\) for HR and glucose to approximately zero, keeps record linkage and membership metrics near random baselines, and strongly suppresses attribute leakage. This matches the intuition that per-stay orthogonal mixing destroys coherent per-stay structure in the original time order, even if an attacker learns an approximate inverse in the mixed coordinate system.

\subsubsection{Utility and shape metrics: cost of Q-mix}
\label{sec:qmix-utility}

We now quantify the utility cost of Q-mix for HR at \(\alpha=1.0\), using both shape and downstream-task metrics.

\begin{figure}[H]
  \centering
  \includegraphics[width=0.75\textwidth]{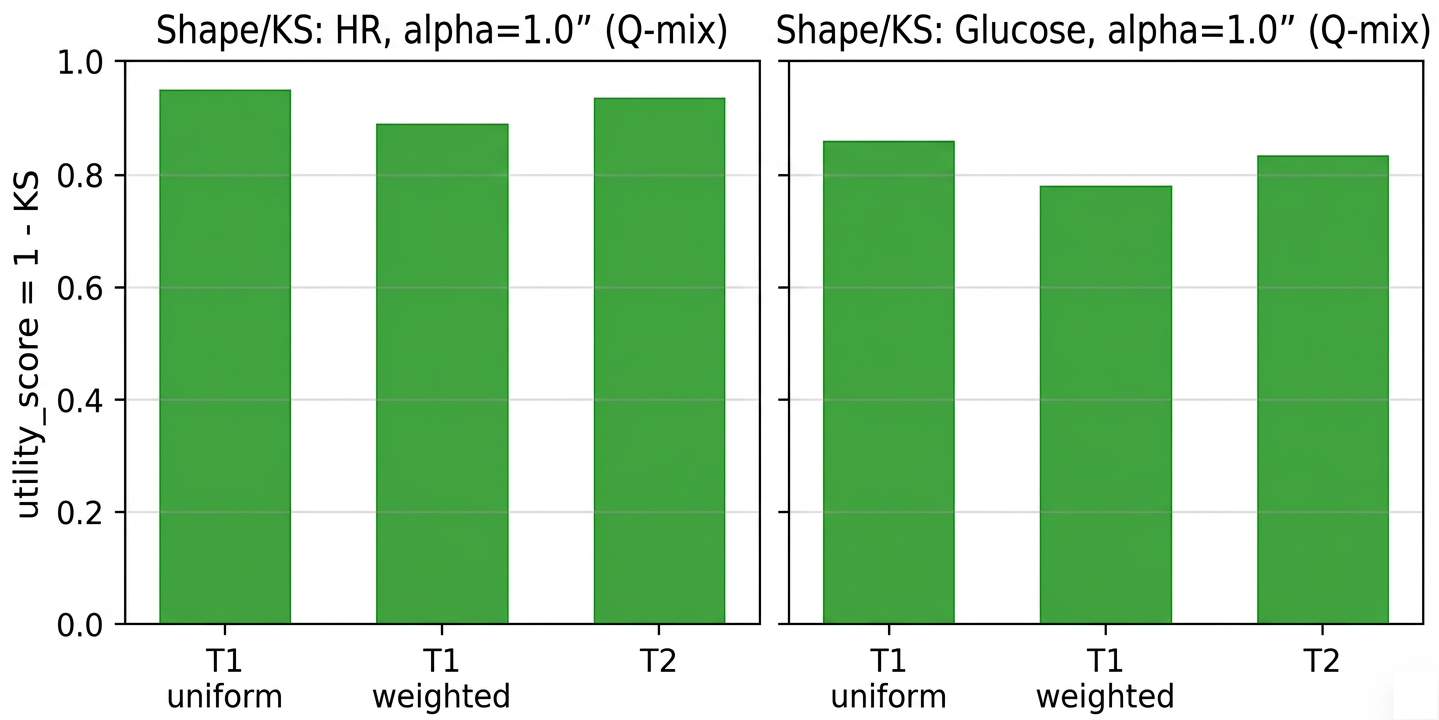}
  \caption{Shape utility scores
           \(\mathrm{utility\_shape} = 1 - \mathrm{KS}\) for HR and
           glucose at \(\alpha=1.0\) with Q-mix.
           Scores remain high, indicating that marginal distributions
           are preserved reasonably well.}
  \label{fig:qmix-ks}
\end{figure}

We first define a shape utility score,
\[
  \mathrm{utility\_shape}_v = 1 - \mathrm{KS}_v.
\]
For HR at \(\alpha=1.0\) with Q-mix, T1\_uniform yields \(\mathrm{utility\_shape} = 0.938\) and T2 yields \(\mathrm{utility\_shape} = 0.927\). Figure~\ref{fig:qmix-ks} summarizes these scores for HR and glucose.

Mean and variance deviations remain small. For HR under Q-mix at \(\alpha=1.0\), the mean deviation is approximately \(0.0024\) for T1\_uniform and \(0.0023\) for T2, while variance deviations remain modest. Out-of-range rates for HR remain below \(1\%\), and glucose exhibits OOR rates around \(2.9\%\)–\(3.1\%\), consistent with its heavier-tailed distribution.

\begin{figure}[H]
  \centering
  \includegraphics[width=0.75\textwidth]{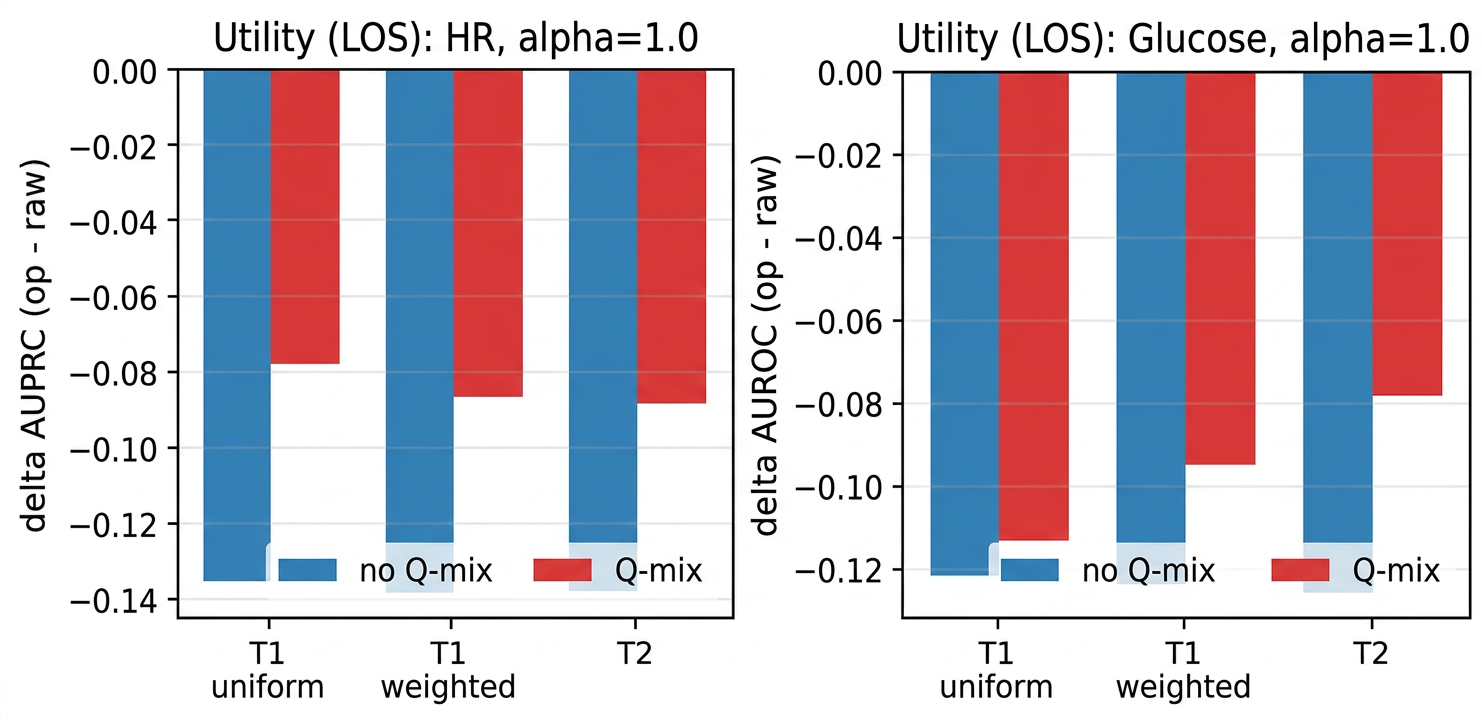}
  \caption{Utility for LOS prediction using HR or glucose at
           \(\alpha=1.0\), with and without Q-mix.
           Bars show differences in AUROC and AUPRC between perturbed
           and raw data.
           For HR, Q-mix + T1/T2 slightly improves AUROC and AUPRC.}
  \label{fig:qmix-los}
\end{figure}

As a downstream task, we train a simple predictor for prolonged LOS and compare performance on raw versus perturbed HR. Figure~\ref{fig:qmix-los} shows AUROC and AUPRC differences.

\begin{figure}[H]
  \centering
  \includegraphics[width=0.7\textwidth]{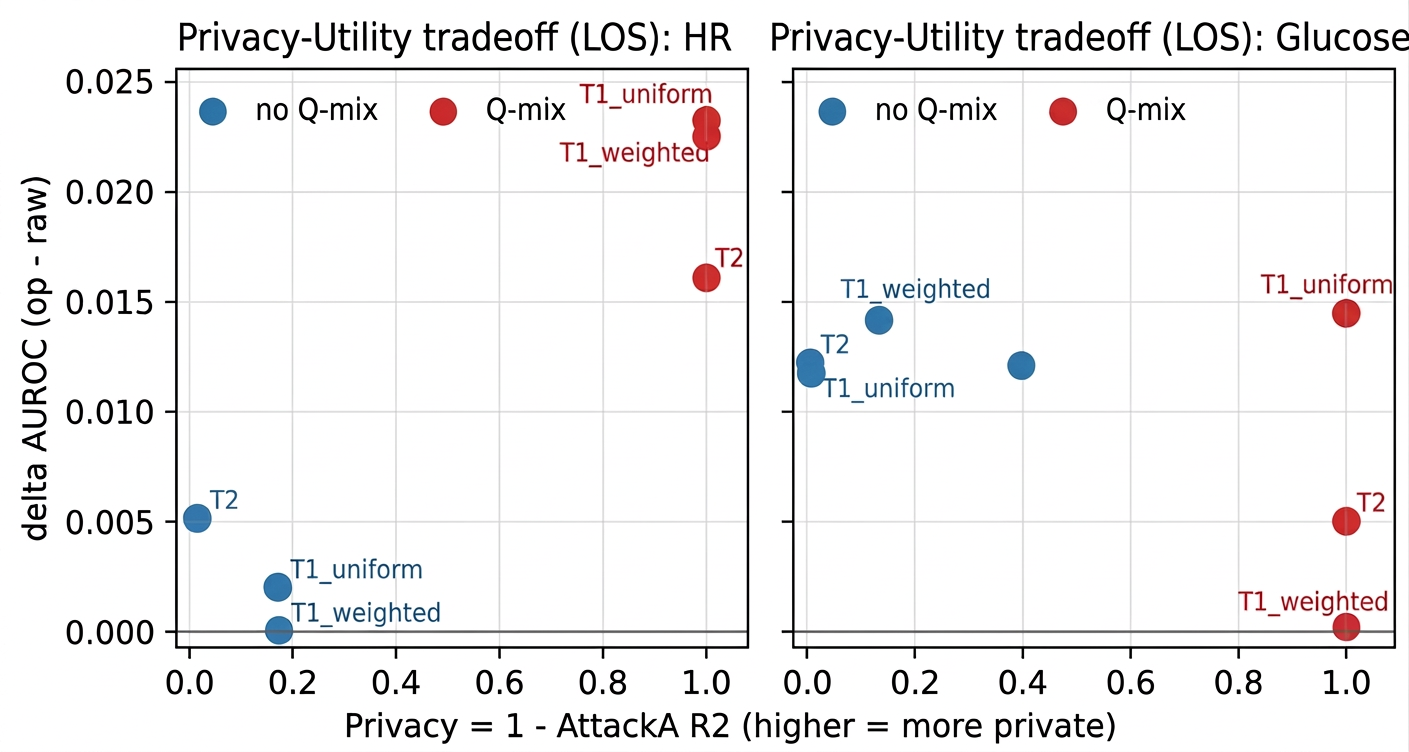}
  \caption{Privacy--utility trade-off for HR at \(\alpha=1.0\).
           The x-axis shows privacy score \(1-R^2\), and the y-axis
           shows change in AUROC for LOS prediction.
           Q-mix + T1/T2 achieve both high privacy and non-degraded
           utility.}
  \label{fig:qmix-privacy-utility}
\end{figure}

For HR, T1\_uniform increases AUROC from \(0.6233\) to \(0.6468\) and AUPRC from \(0.6073\) to \(0.6130\). T2 also yields slight improvement. For glucose, LOS performance under Q-mix remains comparable to or only slightly below raw performance. Figure~\ref{fig:qmix-privacy-utility} visualizes the resulting privacy--utility trade-off for HR.

Taken together, these results show that at fixed \(\alpha=1.0\), Q-mix combined with T1 or T2 drives reconstruction \(R^2\) for HR and glucose from approximately \(0.8\)–\(0.98\) down to nearly \(0\), while keeping KS-based shape scores high, out-of-range rates clinically acceptable, and downstream LOS prediction performance stable or slightly improved for HR. This supports viewing Q-mix as a deployment-layer strong-privacy switch for a small set of high-risk variables in external-sharing scenarios.

\subsection{Section Summary: From Geometric Sanity to Privacy Steps}
\label{sec:single-column-summary}

This section presented a systematic single-column evaluation of T1, T2, T3, and their Q-mix extensions on MIMIC-IV ICU time series. Several findings stand out.

First, from a geometric and statistical perspective, T1, T2, and T3 satisfy mean--variance preservation and the unified z-score \(\ell_\infty\) bound on real data. For small to moderate \(\alpha\), marginal distributions, correlation matrices, autocorrelation structure, and physical numeric ranges remain close to the original data.

Second, the \(\alpha\)-hierarchy is effective but smooth. As \(\alpha\) increases, reconstruction \(R^2\) under L1 and L2 linear attacks decreases monotonically for T1 and T2, but the decline is gradual rather than step-like. Even at \(\alpha=5.0\), many variables still retain \(R^2\) in the \(0.7\)–\(0.9\) range, indicating substantial residual invertibility.

Third, T3 is a canonical negative case. Although it nearly perfectly preserves geometric and statistical structure, it is highly invertible under C5 + L2 + Attack~A, with HR and glucose maintaining \(R^2 \approx 0.9999\) across all tested \(\alpha\). It is therefore suitable only for didactic or very light in-hospital perturbation, not for strong de-identification pipelines.

Fourth, Q-mix produces a genuine privacy cliff at fixed \(\alpha\). At \(\alpha=1.0\), Q-mix combined with T1 or T2 drives reconstruction \(R^2\) for HR and glucose from approximately \(0.8\)–\(0.98\) down to nearly \(0\), suppresses attribute-inference \(R^2\) dramatically, and keeps record-linkage and membership metrics close to random baselines.

Finally, the utility cost of Q-mix remains acceptable. For HR at \(\alpha=1.0\), KS-based shape scores remain high, out-of-range rates remain below \(1\%\), and LOS prediction AUROC and AUPRC are not degraded and may even improve slightly. This positions Q-mix as a natural deployment-layer strong-privacy module that should be activated only for a small set of sensitive variables in external-sharing or multi-center settings, while T1 and T2 remain the workhorse operators for in-hospital research and teaching.

% 后面章节将在此基础上，把分析扩展到多变量与端到端工作流，以及不同部署场景下 \(\alpha\) 与 Q-mix 激活策略的策略层选择。
\section{System: EHR-Privacy-Agent and Privacy Skill Library}
\label{sec:system}

Building on the geometric operators and privacy evaluation protocol of
Sections~\ref{sec:operator-families} and~\ref{sec:single-column-expt},
this section describes how these ideas are embedded into a deployable
system, which we call \emph{EHR-Privacy-Agent}. The system is designed
to run on hospital ETL infrastructure without GPUs, to expose
privacy-enhanced EHR views to human users and LLM-based agents, and to
support continuous privacy evaluation together with iterative skill
evolution.

\subsection{Design Goals and System Constraints}
\label{sec:system-goals}

\subsubsection{Design goals: usable and visible, but hard to reconstruct}
\label{sec:system-goals-usable}

The design of EHR-Privacy-Agent is guided by four system-level goals.

First, the privacy-enhanced tables produced by the system must remain
operationally usable. In practice, this means that they should be clean,
consistently typed, structurally compatible with downstream analysis
scripts, conventional machine learning models, and LLM-based agents, and
available on regular schedules such as nightly refreshes in the data
lake.

Second, the transformed values must remain visible in the sense of basic
physical interpretability and clinical intuition. Means, variances, and
z-score magnitudes should remain clinically plausible. Histograms, time-series plots, and correlation matrices should continue
to resemble real patient data (ICU trajectories in our running example)
rather than arbitrary synthetic outputs. More
specifically, the visual diagnostics examined in
Section~\ref{sec:single-column-expt}, including histograms, CDFs, ACF,
and PSD, should not exhibit pathological distortions.

Third, the system must make records substantially harder to reconstruct
under the no-key, structure-aware threat model C5
(Section~\ref{sec:threat-model}) and under the leakage levels L0, L1,
and L2 together with attack families A, B, C, and D. In concrete terms,
strong-privacy skills should drive reconstruction \(R^2\) and
attribute-inference \(R^2_{\text{attr}}\) toward \(0\), while keeping
re-identification and membership metrics close to random baselines.

Finally, the system must remain deployable under realistic engineering
constraints. It should rely only on CPUs, integrate with existing
in-hospital ETL clusters, avoid any additional GPU budget or specialized
hardware, and maintain computational complexity and memory use at scales
compatible with tens to hundreds of thousands of ICU stays.

\subsubsection{Threat-model constraints and the no-key requirement}
\label{sec:system-no-key}

EHR-Privacy-Agent adheres to the C5 threat model introduced in
Section~\ref{sec:threat-model}. The adversary is no-key but
structure-aware. In particular, the adversary is assumed to know the
mathematical definitions and source code of T1, T2, T3, and Q-mix, as
well as public parameters such as \(\alpha\), window lengths, and Q-mix
block sizes. At leakage levels L1 and L2, the adversary may also observe
a small or moderate fraction of paired samples \((x,y)\), for example
\(0.01\%\) or \(20\%\) of stays.

What the adversary does not possess is any long-lived shared
cryptographic key. All randomness used by the operators is generated by
hospital-internal modules and is treated as per-stay short-term
randomness. For Q-mix, each per-stay orthogonal matrix \(Q_{s,v}\) is
generated by a pseudo-random process that combines a hospital-internal
secret seed with the pair \((\text{stay\_id},v)\). The adversary knows
the generation algorithm, but not the internal secret itself.

At the system level, this threat model is encoded as an explicit design
constraint. EHR-Privacy-Agent does not maintain any cross-batch shared
keys accessible to analysts or external parties. All randomness is
injected by a secure in-hospital module and is visible only within the
lifetime of a single pipeline run. Privacy-enhanced data may be stored
and shared long-term, but no tokens or metadata sufficient to
reconstruct the internal randomness are exported alongside them.

\subsubsection{User roles and typical scenarios}
\label{sec:system-roles}

The system is designed around three primary user roles. The first is
data science and engineering teams, who configure ETL jobs, define
skills, run nightly pipelines, and in in-hospital environments rely on
low-\(\alpha\) skills without Q-mix for EDA, QA, and model development.
The second is clinical researchers and quality officers, who interact
with privacy-enhanced views through an LLM-based agent for cohort
exploration, visualization, and basic modeling, without ever touching
raw EHR tables directly. The third is external collaborators together
with teaching or open-data users, who receive views generated by
strong-privacy skills, typically with larger \(\alpha\) and Q-mix
enabled on selected sensitive variables, or who consume downstream
models trained on such views.

The aim of EHR-Privacy-Agent is to support all three categories through
a single architecture, using different skill profiles rather than
separate pipelines.

\subsection{System Architecture Overview}
\label{sec:system-architecture}

\begin{figure}[h]
  \centering
\includegraphics[width=0.8\textwidth]{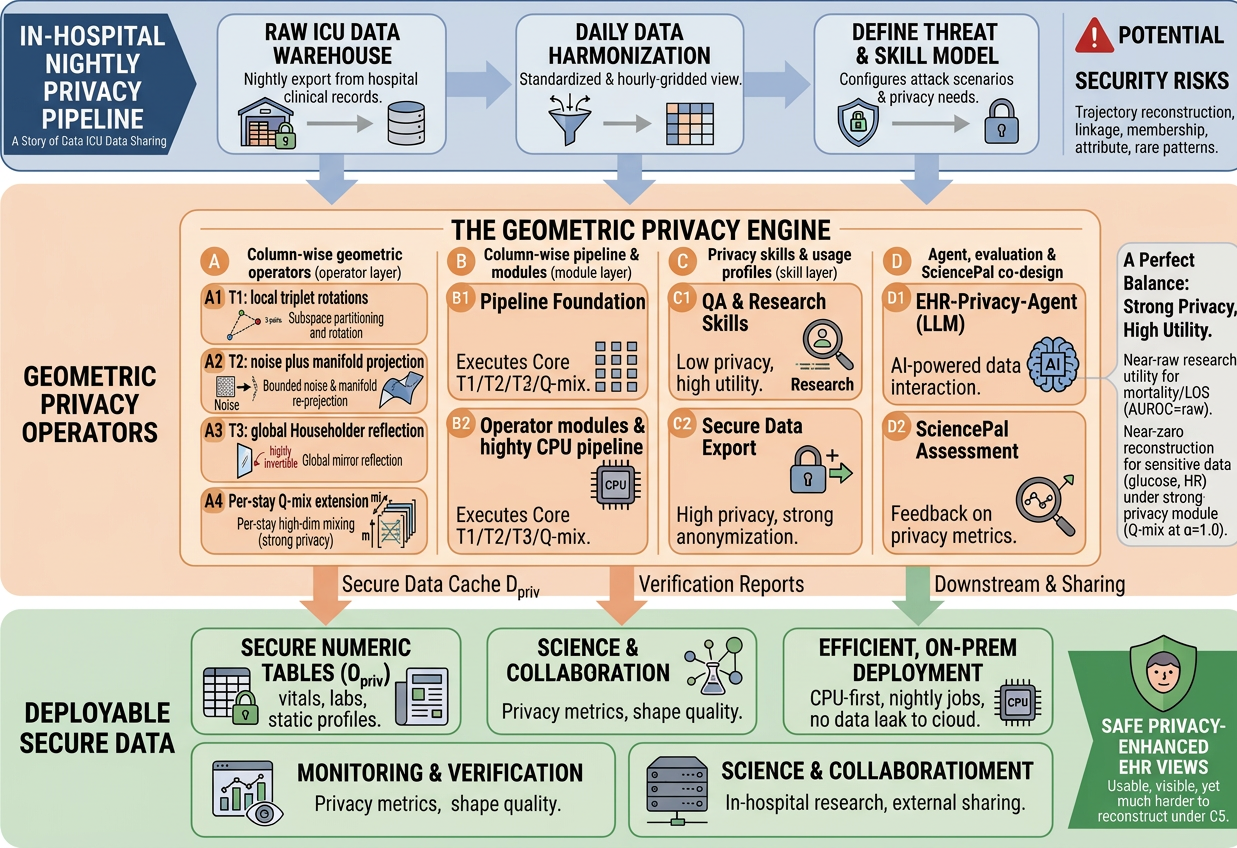}
  \caption{System-level overview of the in-hospital nightly privacy pipeline and geometric privacy engine. 
  Each night, raw EHR tables (ICU vitals, labs, medications in our running example)
  are exported from the production warehouse into a harmonized, hourly-gridded view. 
  Given a threat model and a chosen privacy skill configuration, the \emph{geometric privacy engine} applies column-wise operators (T1/T2/T3 and per-stay Q-mix) within a CPU-only pipeline to produce a secure numeric cache \(D_{\text{priv}}\). 
  Privacy skills and usage profiles distinguish low-privacy, high-utility views for in-hospital QA and research from strong-privacy export views with Q-mix enabled on selected variables. 
  An LLM-based EHR-Privacy-Agent and the SciencePal assessment loop sit on top of these views to support interactive analysis, monitoring of privacy and shape metrics, and collaborative tuning of skills, while all data remain on-premise and never leave the hospital intranet.}
  \label{fig:nightly-pipeline}
\end{figure}

Figure~\ref{fig:nightly-pipeline} shows the in-hospital nightly privacy pipeline and how EHR-Privacy-Agent is composed of operators, modules, and skills.

The first component is the raw EHR layer, namely the traditional
in-hospital warehouse containing raw ICU tables such as vitals, labs,
medications, and demographics. This layer is completely invisible to
external agents and untrusted users.

The second component is the Operator Engine, which implements the
geometric operators introduced in Section~\ref{sec:operator-families},
including T1, T2, T3, and Q-mix, and exposes streaming, column-level
APIs.

The third component is the Privacy Skill Library, which combines these
operators into reusable \emph{skills}. Each skill represents a complete
pipeline from raw EHR to a privacy-enhanced table, together with
scenario tags, threat-model assumptions, and parameter settings such as
\(\alpha\) and Q-mix configuration.

The fourth component is the EHR-Privacy-Agent itself, namely an
LLM-based orchestration and QA layer. The agent can access only skill
outputs through tool APIs, interprets user natural-language queries,
selects appropriate skills, and performs downstream analyses on the
resulting privacy-enhanced views.

The fifth component is the Evaluation and Monitoring module. This module
runs the geometric sanity checks and privacy attacks introduced in
Section~\ref{sec:single-column-expt}, both for pre-deployment evaluation
of new skills and for periodic health checks of deployed skills in order
to monitor drift in geometric and privacy metrics.

These components are connected through a unified configuration and
logging layer. Skills are stored as YAML files under version control.
Nightly jobs generate privacy-enhanced views according to the configured
skill set. The agent either loads existing views or triggers
incremental computation. The evaluation module periodically reruns
attack suites and QA checks on these views.

\subsection{Operator Library and Privacy Skill Library}
\label{sec:operator-skill-library}

The system's flexibility under computational and threat-model
constraints relies on a three-layer design:
\emph{Operator} \(\rightarrow\) \emph{Module} \(\rightarrow\) \emph{Skill}.

\subsubsection{Operator layer: geometric atomic operators}
\label{sec:operator-layer}

The bottom layer is the operator library, which contains the geometric
atoms introduced in Section~\ref{sec:operator-families}. These include
T1\_uniform, a local three-point rotation operator satisfying C1--C4 and
particularly effective at preserving short-range ACF; T2, a
noise-plus-projection operator on the mean--variance manifold that tends
to preserve cross-column correlation structures well; T3, a global
Householder reflection that is geometrically almost perfect but highly
invertible under C5 + L2 attacks and is therefore explicitly flagged as
a negative case; and Q-mix, a per-stay orthogonal mixing wrapper that
may be applied to a small set of sensitive variables such as HR and
glucose, as studied in Section~\ref{sec:qmix-experiments}.

At this layer, the basic unit of computation is a single column. The
input is a standardized time series \(z_{s,v}\) for variable \(v\),
together with \((\mu_v,\sigma_v)\) and configuration parameters such as
\(\alpha\). The output is a perturbed \(z'_{s,v}\), followed by
de-standardization to \(x'_{s,v}\) through
\[
  x'_{s,v} = \mu_v \mathbf{1} + \sigma_v z'_{s,v}.
\]

The operator layer is scenario-agnostic. It is unaware of who is
calling it or for what purpose, and guarantees only the geometric
constraints C1--C4 together with the exposure of a unified privacy knob
\(\alpha\).

\subsubsection{Module layer: operator composition and cross-column logic}
\label{sec:module-layer}

Above the operator layer we define reusable \emph{modules}. The
\texttt{ColumnTransformModule} takes a list of variables
\(\{v_1,\dots,v_k\}\) together with per-column configurations such as
\(\alpha\) and the allowed operator set, applies T1, T2, T3, or Q-mix
independently to each column, and supports batched as well as streaming
processing of long time series.

The \texttt{CrossColumnModule} handles structural cross-column logic.
Examples include enforcing shared random seeds across certain variable
groups in controlled experiments, or performing pre-Q-mix normalization
or aggregation on variable groups when necessary.

The \texttt{TemporalSlicingModule} splits each stay's time axis into
fixed-length windows, for example 48 hours, in order to better align
with Q-mix block dimensions, and manages overlap and edge cases for
stays that span multiple windows.

Taken together, the module layer serves as an intermediate template
between low-level operators and high-level skills. It hides operator
details from higher layers while providing structured building blocks
for skill construction.

\subsubsection{Skill layer: configuration-bound pipelines}
\label{sec:skill-layer}

At the top layer, a \emph{privacy skill} represents a complete pipeline
from raw EHR to a particular privacy-enhanced view under a specific use
case and threat model. It also serves as a stable contract between the
data warehouse and downstream analysis tools.

Each skill includes an input scope, specifying the cohort definition,
time window, and variable list; a transform configuration, specifying
per-column \(\alpha\), the allowed operators, and whether Q-mix or T3 is
enabled; an output format, specifying target schema, indexing strategy,
and partitioning; threat-model assumptions, indicating the relevant
leakage levels and dominant attack families; and usage tags, such as
\texttt{in\_hospital\_research}, \texttt{export\_strong\_privacy}, or
\texttt{teaching\_demo}.

In implementation, skills are defined as YAML documents. A typical
in-hospital exploratory skill is shown in
Figure~\ref{fig:skill-yaml-internal}.

\begin{figure}[t]
  \centering
  \begin{minipage}[c]{0.48\textwidth}
  \begin{lstlisting}[language=yaml,basicstyle=\ttfamily\footnotesize]
id: skill_in_hosp_vitals_explore
usage: in_hospital_research
threat_model:
  leakage: L1
  attacks: ["A_reconstruction"]
operators:
  default_alpha: 0.5
  per_variable:
    HR:
      ops: ["T1_uniform", "T2"]
      alpha: 0.5
    Lactate:
      ops: ["T2"]
      alpha: 0.5
    Glucose:
      ops: ["T2"]
      alpha: 0.5
qmix:
  enabled: false
output:
  table: "priv_vitals_hourly_v1"
  time_window_hours: [0, 72]
  format: "long"
  \end{lstlisting}
  \subcaption{In-hospital vitals exploration skill.
              HR may use T1 or T2 at \(\alpha=0.5\);
              Q-mix is disabled.}
  \label{fig:skill-yaml-internal}
  \end{minipage}
  \hfill
  \begin{minipage}[c]{0.48\textwidth}
  \begin{lstlisting}[language=yaml,basicstyle=\ttfamily\footnotesize]
id: skill_export_strong_privacy
usage: export
threat_model:
  leakage: L2
  attacks: ["A_reconstruction",
            "B_linkage",
            "C_membership"]
operators:
  default_alpha: 1.0
  per_variable:
    HR:
      ops: ["T1_uniform", "T2"]
      alpha: 1.0
    Glucose:
      ops: ["T1_uniform", "T2"]
      alpha: 1.0
    Lactate:
      ops: ["T2"]
      alpha: 1.0
qmix:
  enabled: true
  variables: ["HR", "Glucose"]
  block_length: 48
  secret_seed: "${HOSPITAL_SEED}"
output:
  table: "priv_vitals_export_v1"
  time_window_hours: [0, 72]
  format: "long"
  \end{lstlisting}
  \subcaption{Export-oriented strong-privacy skill.
              Q-mix is enabled at \(\alpha=1.0\) for
              HR and glucose.}
  \label{fig:skill-yaml-export}
  \end{minipage}
  \caption{Example YAML skill definitions for two deployment scenarios.
           (a)~A typical in-hospital research profile with moderate
           privacy and no Q-mix.
           (b)~A conservative export profile with higher \(\alpha\) and
           Q-mix enabled for high-risk variables.}
  \label{fig:skill-yaml-comparison}
\end{figure}

In this example, HR and glucose are processed at \(\alpha=1.0\) through
Q-mix followed by T1 or T2, whereas lactate and other variables rely
only on T2 at the same privacy level. The tag \texttt{usage: export}
marks this skill as a strong-privacy profile for external sharing, which
we revisit in Section~\ref{sec:deployment-lessons}.

\subsubsection{Running case}
\begin{figure}[h]
  \centering
\includegraphics[width=0.8\textwidth]{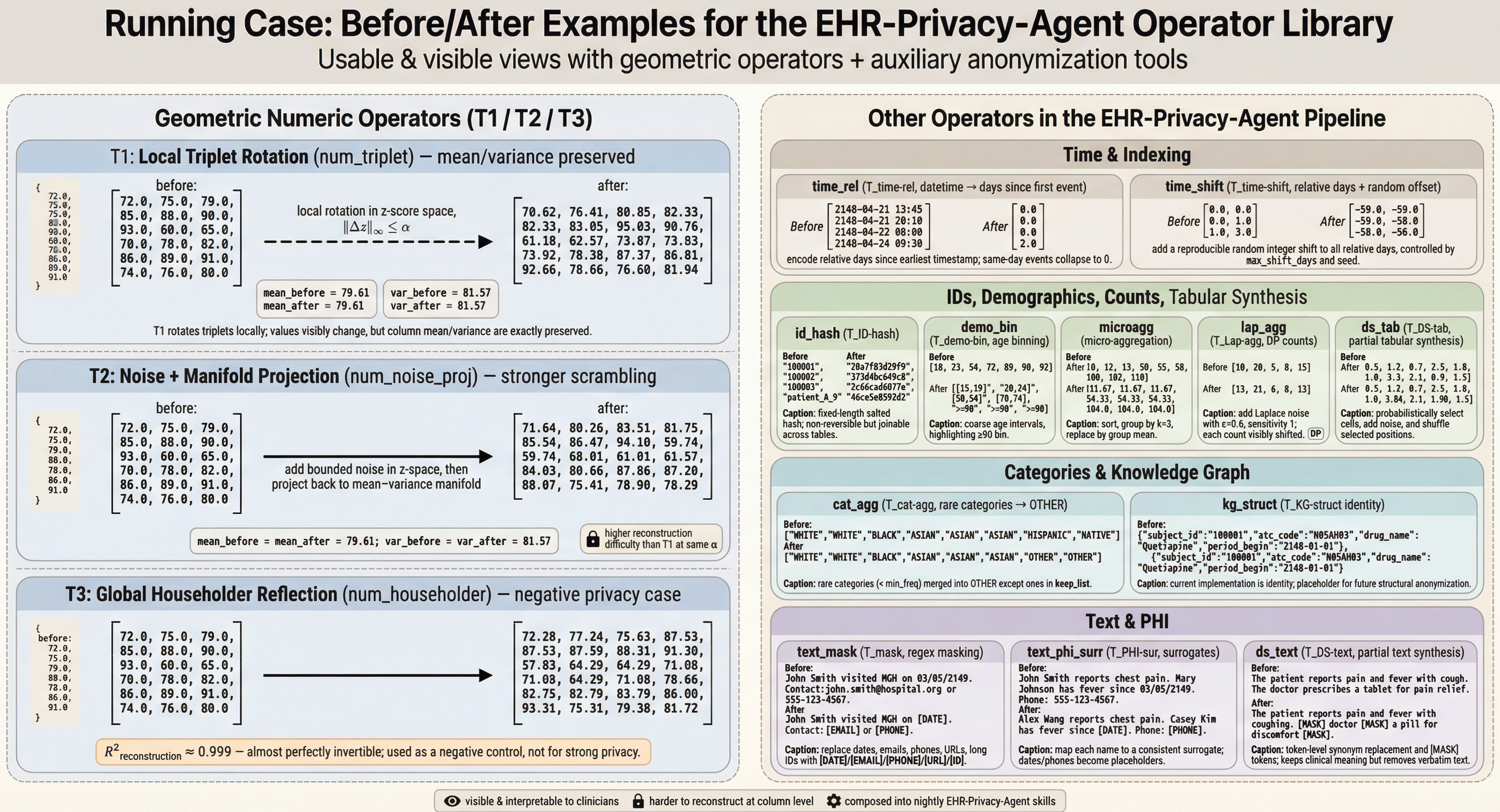}
  \caption{Left: geometric numeric operators T1–T3 applied to one standardized vital‑sign column, with toy before/after values showing mean–variance–preserving local rotation (T1), stronger scrambling via bounded noise plus manifold projection (T2), and an almost invertible Householder reflection as a negative privacy case (T3). Right: auxiliary operators for time/indexing, IDs and demographics, counts and tabular synthesis, categories/knowledge graph, and clinical text/PHI, each illustrated with minimal before/after snippets to show how identifiers are removed or blurred while keeping data usable for analysis.}
  \label{fig:running-case}
\end{figure}

To make the operator library more concrete, Figure~\ref{fig:running-case} presents a small running case.
For each operator, we show a minimal before/after example on toy data (time stamps, IDs, counts, text, or categories),
highlighting how the transformation preserves global structure while visibly altering per-record values.

\subsection{Nightly Pipeline and Data Flow}
\label{sec:nightly-pipeline}

A central engineering goal of EHR-Privacy-Agent is to turn privacy
enhancement into a routine nightly task on CPU-only infrastructure.

\subsubsection{Configuration-driven nightly job}
\label{sec:nightly-config}

Each night, the system executes a configuration-driven pipeline. The
first step is cohort selection and snapshotting. Configured SQL queries
or cohort definitions are run on the raw EHR warehouse to identify the stays or episodes for the
current batch (ICU admissions in our running example), such as recent admissions
or periodic snapshots, and a read-only snapshot with fixed schema and version is
materialized in order to avoid conflicts with live transactional
systems.

The second step is standardization and mean--variance updating. On a
designated training cohort, the system updates column-wise
\((\mu_v,\sigma_v)\), or reuses a frozen parameter version when
appropriate, and standardizes the target cohort to obtain
\(z_{s,v} = (x_{s,v} - \mu_v \mathbf{1}) / \sigma_v\).

The third step is execution of operator modules for each configured
skill. The relevant \texttt{ColumnTransformModule},
\texttt{CrossColumnModule}, and \texttt{TemporalSlicingModule} are
invoked, T1, T2, T3, or Q-mix are applied in z-score space according to
per-variable settings, and the outputs are de-standardized back to
\(x'_{s,v}\) and written into the target table.

The fourth step is persistence and indexing. Secondary indices are built
on each skill output table, for example by \(\text{stay\_id}\), variable,
and time, and metadata such as skill id, parameter version, and
pipeline timestamp are attached for traceability.

The final step is automatic evaluation and alerting. Geometric sanity
checks, together with basic distribution QA such as KS and out-of-range
monitoring, are run on each newly created privacy view. For selected
pilot skills, a subset of Attacks A, B, and C is also run, monitoring
reconstruction \(R^2\), linkage accuracy, and membership AUC. If any
metric exceeds configured thresholds, alerts are raised and the
corresponding outputs can be marked as experimental or unsafe for
export.

\subsubsection{CPU-only implementation and resource configuration}
\label{sec:nightly-resources}

The Operator Engine follows two engineering principles. The first is
controlled complexity: per-stay, per-column complexity remains linear or
near-linear in the sequence length, namely \(O(n_{s,v})\) or
\(O(n_{s,v}\log n_{s,v})\). The second is batching and streaming: long
time series, such as 7-day ICU trajectories, are sliced into fixed
windows, for example 48 hours, and per-window caches store local means,
variances, and Q-mix matrices so that repeated computations are avoided
whenever possible.

In a MIMIC-IV-scale pilot running on a 32-core commodity CPU server, we
can perturb tens of thousands of ICU stays' vitals and labs tables with
T1 and T2 within a few hours, while still enabling Q-mix for HR and
glucose in strong-privacy export skills without violating nightly SLAs.
Compared with GPU-heavy generative approaches such as CLEAR-Note or
CTGAN ~\citep{goodfellow2014gan,xu2019ctgan}, discussed later in Section~\ref{sec:runtime-eval}, this
pipeline achieves much higher per-stay throughput under equivalent
hardware budgets.

\subsubsection{Versioning and rollback}
\label{sec:versioning-rollback}

Because skill designs evolve in response to new attacks and clinical
feedback, the system explicitly supports coexistence of multiple
versions, auditable change history, and safe rollback. Different
versions of the same skill id may coexist, enabling controlled A/B
comparisons of attack difficulty and downstream utility. Skill YAML
files and evaluation reports are stored in a Git repository, and each
change is tracked with a commit record, reviewer, and rationale. If
evaluation later reveals a privacy or geometric regression, such as an
unexpected increase in reconstruction \(R^2\) or OOR rates, the system
can revert to the last approved skill version by a single configuration
change.

\subsection{Agent Layer: From LLM Tools to Privacy Views}
\label{sec:agent-layer}

Although the operator library and nightly pipeline can be used in batch
mode, the main user-facing interface is an LLM-based
\emph{EHR-Privacy-Agent}.

\subsubsection{Interaction pattern: from natural language to skill calls}
\label{sec:agent-interaction}

A typical interaction begins with a natural-language request, such as a
user asking whether the distributions of HR and lactate in ICU patients
have drifted over the last three years, or asking for a sharable dataset
view suitable for replicating a LOS model from a published paper.

The agent first performs intent understanding and skill selection. Using
lightweight rules together with LLM reasoning, it maps the request to
one or more skills. In-hospital exploratory analyses are routed to
skills such as \texttt{skill\_in\_hosp\_vitals\_explore}, whereas
external sharing requests are routed to stronger skills such as
\texttt{skill\_export\_strong\_privacy}. If the user explicitly requests
strong privacy or external sharing only, the agent prefers skills with
larger \(\alpha\) and Q-mix enabled.

The selected skill is then invoked through a tool call of the form
\texttt{run\_skill(skill\_id, cohort\_spec, time\_window, \dots)}. If the
nightly pipeline has already produced the corresponding view, it is
loaded directly; otherwise, an incremental run is triggered subject to
resource policies.

Finally, the agent performs downstream EDA, visualization, simple
modeling, and natural-language explanation on top of the
privacy-enhanced tables. At no point does the agent access raw EHR
tables directly. All subsequent analysis operates only on the privacy
views defined by skills.

\subsubsection{Tool APIs: listing, inspecting, and running skills}
\label{sec:agent-tools}

To support traceable and safe usage, the agent layer exposes three
primary tool APIs. The first, \texttt{list\_skills(usage\_filter)},
returns the available skills together with metadata such as usage
scenario, supported \(\alpha\) range, whether Q-mix or T3 is enabled,
and a summary of recent evaluation results. The second,
\texttt{inspect\_skill(skill\_id)}, displays detailed configuration for
a specified skill, including YAML expansion, operator combinations,
Q-mix options, randomness-source handling, and summaries of recent
attack evaluations. In external settings, this inspection function may
be restricted in order to avoid revealing sensitive implementation
details. The third, \texttt{run\_skill(skill\_id, cohort\_spec, \dots)},
executes the skill and returns either a view handle or a data sample,
depending on permissions.

\subsubsection{Illustrative usage scenarios}
\label{sec:agent-scenarios}

Three scenarios illustrate the intended usage pattern. In the first,
which concerns in-hospital EDA and QA, a quality officer wants to check
whether HR and lactate distributions have shifted across quarters. The
agent selects \texttt{skill\_in\_hosp\_vitals\_explore} for each quarter,
loads the corresponding privacy views, plots histograms, trends, and
correlation matrices, and summarizes the differences in natural
language.

In the second scenario, which concerns research replication and
collaboration, a data scientist develops an ICU mortality model for a
multi-center study. Within the hospital, they iterate using low-\(\alpha\),
no-Q-mix skills for feature engineering and hyperparameter tuning. Once
the model is fixed, they export a strong-privacy view using
\texttt{skill\_export\_strong\_privacy}, which enables Q-mix for HR and
glucose, and then package trained model weights together with
performance summaries for collaborators.

In the third scenario, which concerns teaching and simulation,
clinicians or students only need to see the general shape of ICU data. A
teaching-oriented skill may therefore use T3 at low \(\alpha\), giving a
simple configuration that preserves visual realism while not aiming for
strong privacy guarantees.

\subsection{AI-making at the System Level: From Natural Language to Skill YAML}
\label{sec:ai-making-system}

Section~\ref{sec:human-ai-codesign} introduced the AI-making protocol for
designing the operator families T1, T2, T3, and Q-mix. At the system
level, we extend the same idea to the authoring and evolution of
privacy skills and full pipelines, with SciencePal acting as an
assistant.

\subsubsection{Lifecycle of a new skill}
\label{sec:skill-lifecycle}

A new skill moves from concept to production through several phases.
First, a human stakeholder specifies the high-level need, for example a
request for a stronger-privacy vitals view for multi-center analysis
with special protection for HR and glucose under an L2-style threat
model focused on Attacks A and B.

Second, an engineer works with SciencePal to generate a draft skill.
The engineer provides the intended usage scenario, prioritized
variables, desired \(\alpha\) range, and relevant attack families.
SciencePal then proposes an initial YAML configuration specifying
per-variable operators, \(\alpha\), whether to enable Q-mix, with what
block length, and which evaluation suites should be run.

Third, the system automatically evaluates the draft. It runs geometric
sanity checks, distribution QA, and Attack A under L2, and optionally
Attack B and C, and feeds the resulting metrics back to SciencePal and
the engineer. If reconstruction \(R^2\) remains too high, the draft may
be revised by increasing \(\alpha\) or enabling Q-mix for sensitive
variables. If KS distances or out-of-range rates become too large, the
draft may instead be revised by lowering \(\alpha\) or restricting the
use of certain operators.

Fourth, security and clinical reviewers jointly assess the evaluation
report. They confirm that privacy metrics satisfy policy thresholds and
that the resulting distributions and physical interpretability remain
acceptable. Once approved, the skill is marked as production-ready.

Finally, the skill is incorporated into the nightly pipeline and
continuously monitored. The evaluation and monitoring module reruns
attack suites periodically, watching for drift in reconstruction
\(R^2\), linkage accuracy, and downstream utility. If regressions are
detected, the system can downgrade or roll back the skill.

Throughout this process, SciencePal never accesses raw EHR directly. It
operates only on statistical summaries and evaluation feedback.

\subsubsection{SciencePal-assisted skill configuration: an example}
\label{sec:scipal-example}

A typical AI-making interaction might begin with an engineer stating
that a vitals skill is needed for external collaborators, that HR and
glucose must be strongly protected, that \(\alpha\) up to \(1.0\) or
higher is acceptable, and that under linear Attack A / L2 the desired
goal is reconstruction \(R^2\) close to zero without sharply degrading
LOS prediction AUROC.

SciencePal may then recommend enabling Q-mix with block length 48 on HR
and glucose, while using T2 at \(\alpha=1.0\) for the remaining vitals.
The resulting skill would be tagged with \texttt{usage: export} and
evaluated with L2 Attack A together with a LOS downstream task. The
system then treats this recommendation as a draft YAML, similar to
Figure~\ref{fig:skill-yaml-export}, and runs the full privacy evaluation
protocol to verify that it actually achieves the target combination of
low reconstruction \(R^2\) and acceptable utility.

\subsubsection{Evaluation feedback as design signal}
\label{sec:evaluation-feedback}

For each new skill, the Evaluation and Monitoring module produces a
unified report. This includes C1--C2 geometric metrics such as
mean/variance deviations and whether \(\ell_\infty\) bounds remain
within the configured \(\alpha\), distribution QA such as KS distances,
out-of-range rates, and ACF changes, Attack A/B/C metrics such as
per-variable reconstruction \(R^2\), \(\mathrm{MAE}_z\),
Reid@1/Reid@5, linkage AUC, membership AUC, and membership advantage,
and downstream task metrics such as LOS or mortality AUROC and AUPRC.

Together, these quantities form a skill-level score vector that can be
used both by SciencePal and by human reviewers. SciencePal can use the
report to propose parameter updates, such as changing \(\alpha\) or
enabling or disabling Q-mix on selected variables, while human reviewers
can treat the same report as the basis of a semi-automated security and
utility review before approving a skill for external deployment.

\subsection{Deployment Practice and Engineering Lessons}
\label{sec:deployment-lessons}

We summarize several practical lessons from deploying EHR-Privacy-Agent
in realistic environments.

\subsubsection{Schema evolution and compatibility}
\label{sec:schema-evolution}

EHR schemas evolve over time, with new variables, unit changes, and
coding adjustments. To maintain robustness and comparability of skill
behavior, each skill explicitly declares its schema dependencies,
including variable names, units, and default physical ranges. Schema
changes automatically trigger re-evaluation, for example when lactate
units shift from mg/dL to mmol/L. Mean--variance statistics and
\(\alpha\) configurations are also versioned, allowing historical skill
versions to continue running on older data snapshots.

\subsubsection{Audit, traceability, and access control}
\label{sec:audit-access}

Because skills encode privacy policy, the system provides fine-grained
access control, audit logging, and restrictions on re-export. Different
roles may invoke only usage-labeled skills appropriate to their
permissions; for example, external users may be restricted to export
skills with strong privacy guarantees. Every skill invocation is logged
with timestamp, user, parameters, and cohort size, enabling ex-post
audit and forensic analysis. For skills with Q-mix, the system by
default forbids generating less-perturbed derivative views from the same
cohort, in order to reduce multi-view composition risks.

\subsubsection{Failure modes and mitigation}
\label{sec:failure-modes}

In practice, several recurrent failure modes arise. Geometric
implementation bugs may occasionally break mean--variance preservation
or \(\ell_\infty\) bounds, but nightly C1--C2 checks detect these
rapidly and block writes to production tables. Overly aggressive
\(\alpha\) choices may cause large out-of-range rates or sharply degrade
downstream performance, in which case automatic QA triggers downgrade or
rollback. Attack metrics may also drift over time as cohorts evolve, so
a previously acceptable skill may later show increased reconstruction
\(R^2\) or linkage accuracy. Periodic re-evaluation is therefore
critical for identifying when re-tuning or Q-mix extension becomes
necessary.

Embedding attack evaluation into the same system that runs the nightly
pipeline has proven more effective than one-off offline evaluation for
tracking privacy and geometric risk over time.

\subsection{Section Summary: From Geometric Operators to a Deployable System}
\label{sec:system-summary}

This section showed how the geometric operators and privacy evaluation
protocol of Sections~\ref{sec:operator-families}
and~\ref{sec:single-column-expt} can be encapsulated into a nightly
deployable EHR-Privacy-Agent system.

At the architectural level, the system is organized around a three-layer
abstraction. The operator layer contains T1, T2, T3, and Q-mix as
geometric atoms that (for T1/T2/T3) satisfy C1--C4 under the no-key constraint,
with Q-mix satisfying C1--C3 and a deployment-level relaxation of C4 for a small
set of high-risk variables. The
module layer provides reusable column-wise, cross-column, and
temporal-slicing components. The skill layer defines complete,
scenario-bound and threat-model-bound pipelines in a transparent YAML
format.

At the policy level, skills act as executable privacy specifications.
They encode usage scenario, \(\alpha\), operator combinations, Q-mix
activation, and output schema, while versioning and rollback support
continuous evolution in response to attack evaluation and clinical
feedback.

At the user-interface level, the LLM-based EHR-Privacy-Agent interacts
only with skill-defined privacy views and never with raw EHR. Tool APIs
such as \texttt{list\_skills}, \texttt{inspect\_skill}, and
\texttt{run\_skill} map natural-language intent to concrete privacy
configurations and outputs.

At the design-process level, SciencePal extends the earlier AI-making
workflow from operator design to system-level skill authoring. Combined
with automated evaluation and human review, this produces a closed loop
linking operator design, YAML skill drafting, attack evaluation, and
deployment policy.

Finally, from an engineering standpoint, a CPU-only nightly pipeline is
feasible at ICU scale, integrating attack evaluation directly into the
system life cycle helps detect privacy and geometric regressions early,
and usage tags together with access control provide a natural separation
between in-hospital and external scenarios.

In Section~\ref{sec:end-to-end-tasks}, we build on the views produced by
EHR-Privacy-Agent to evaluate geometric operators and Q-mix on
end-to-end tasks such as LOS and mortality prediction and generative
modeling, and compare their utility and computational cost with
GPU-intensive methods such as
CLEAR-Note\footnote{CLEAR-Note is our earlier preliminary work that performs clinical note de-identification entirely through LLM-based rewriting and verification. We plan to release it publicly in the near future.}
and CTGAN.
\section{End-to-End Experiments and Runtime}
\label{sec:end-to-end-tasks}

Building on the single-column analysis of
Section~\ref{sec:single-column-expt} and the system architecture of
Section~\ref{sec:system}, this Section evaluates EHR-Privacy-Agent in
end-to-end settings. We focus on two pragmatic questions. The first is
how much downstream utility the privacy-enhanced views retain on real
ICU prediction and simple generation tasks, relative to raw EHR and
common baselines such as naive Gaussian noise and CTGAN-style synthetic
data. The second is how the runtime and compute costs of the geometric
pipeline, namely T1, T2, T3, with optional Q-mix, compare with
GPU-heavy generative or text deidentification pipelines.

To answer these questions, we construct a suite of prediction tasks,
including mortality, LOS, and readmission, together with a small
trajectory-generation experiment and runtime comparisons with CTGAN and
CLEAR-Note on adult ICU data from MIMIC-IV. All downstream models are
trained only on privacy-enhanced views produced by skills. Raw EHR is
used only inside the hospital pipeline and is never accessed directly by
the models themselves.

\subsection{Tasks and Evaluation Setup}
\label{sec:endtoend-setup}

\subsubsection{Dataset and cohort construction}
\label{sec:endtoend-cohort}

We use the adult ICU cohort from MIMIC-IV, closely aligned with the
setup of Section~\ref{sec:single-column-expt}. The population consists
of adult patients with age at least \(18\) years, retaining only the
first ICU stay per patient. To reduce pathological outliers, we exclude
extremely short stays of less than \(4\) hours and extremely long stays
of more than \(30\) days.

All time axes are aligned to ICU admission, with \(t=0\) corresponding
to ICU in-time. Vitals and laboratory measurements are resampled onto a
1-hour grid using forward-fill together with clipping to plausible
physical ranges. We use the interval \([0,72]\) hours as the main
prediction window, which supports early-prediction tasks.

The feature set contains both dynamic and static information. Dynamic
features include common vitals such as HR, RR, SBP, DBP, MAP,
\(\mathrm{SpO}_2\), and temperature, together with laboratories such as
glucose, lactate, creatinine, sodium, and potassium. Static features
include sex, age group, admission service, and a coarse diagnosis
category. In some configurations we also include coarse binary
indicators of treatment or ventilation status, which are not subject to
Q-mix. All models use a fixed patient-level train, validation, and test
split, for example \(60/20/20\), and the same split and random seeds are
reused across privacy configurations so that any performance difference
can be attributed to the data view rather than to resampling.

\subsubsection{Downstream task definitions}
\label{sec:endtoend-tasks-def}

We consider three canonical ICU prediction tasks. The first is
in-hospital mortality prediction from the first \(48\) hours. The label
indicates whether the patient dies during the same hospital admission,
and the input consists of vitals, labs, and static features collected in
the first \(48\) hours after ICU admission. This is a binary
classification task.

The second task is ICU LOS category prediction. Here the outcome is a
three-level LOS category defined as short if
\(\mathrm{LOS} \le 2\) days, medium if
\(2 < \mathrm{LOS} \le 7\) days, and long if \(\mathrm{LOS} > 7\) days.
The input again consists of the same \(48\)-hour feature window, and we
report macro-AUROC, macro-AUPRC, and macro-F1.

The third task is 30-day readmission prediction. The label indicates
whether the patient is readmitted within \(30\) days of discharge. The
input uses the first \(72\) hours of vitals, labs, and static features.
This task probes whether early ICU trajectories retain useful signal for
longer-horizon outcomes.

In Section~\ref{sec:gen-experiments}, we additionally include a small
time-series generation and interpolation experiment in order to compare
EHR-Privacy-Agent views with CTGAN-generated synthetic data.

\subsubsection{Model architectures and training protocol}
\label{sec:endtoend-models}

To keep the comparison focused on data views and privacy
configurations, we intentionally use simple and transparent models with
moderate capacity. For mortality, LOS, and readmission, the main
sequence model is a unidirectional GRU encoder over hourly features,
followed by a simple attention pooling layer and a two- or three-layer
MLP head. The total parameter count lies roughly in the \(1\) to
\(2\) million range.

To understand how privacy affects less expressive models, we also train
classical baselines, namely logistic regression on hand-crafted summary
features such as per-variable means, maxima, and slopes over the first
\(48\) hours, and gradient-boosted trees using the same summary feature
set.

Across privacy configurations, the model architecture and hyperparameter
grid are kept fixed. Training uses early stopping based on validation
AUROC with a fixed maximum number of epochs. Models may use a single GPU
or a multicore CPU for training, but all privacy transforms, including
T1, T2, T3, and Q-mix, are executed on CPU only. This design avoids
artificially compensating for privacy loss by increasing model capacity,
so that the comparison truly reflects the data views themselves.

\subsubsection{Evaluation metrics and statistical treatment}
\label{sec:endtoend-metrics}

For classification tasks, we report AUROC and AUPRC in the usual way.
The AUROC is
\[
  \mathrm{AUROC}
  =
  \int_0^1
  \mathrm{TPR}(\mathrm{FPR}) \, d\mathrm{FPR},
\]
while the AUPRC is
\[
  \mathrm{AUPRC}
  =
  \int_0^1
  \mathrm{Precision}(\mathrm{Recall}) \, d\mathrm{Recall}.
\]
Depending on the task, the positive class corresponds to mortality, long
LOS, or readmission.

We also summarize calibration through expected calibration error and the
Brier score. Using a partition of predicted probabilities into \(M\)
bins, the expected calibration error is
\[
  \mathrm{ECE}
  =
  \sum_{m=1}^M
  \frac{|B_m|}{N}
  \bigl|
    \mathrm{acc}(B_m) - \mathrm{conf}(B_m)
  \bigr|,
\]
where \(B_m\) is the set of samples in bin \(m\),
\(\mathrm{acc}(B_m)\) is empirical accuracy, and
\(\mathrm{conf}(B_m)\) is the mean predicted confidence. The Brier score
is
\[
  \mathrm{Brier}
  =
  \frac{1}{N}\sum_{i=1}^N (p_i-y_i)^2,
\]
where \(p_i\) is the predicted probability and \(y_i\in\{0,1\}\) is the
true label.

For each configuration and task, we report held-out test performance
with \(95\%\) bootstrap confidence intervals, together with differences
in AUROC and AUPRC relative to Raw, also with bootstrap confidence
intervals. For key comparisons, especially mortality prediction, we use
DeLong’s test to assess whether AUROC differences are statistically
significant. All experiments are repeated across a small number of
random seeds to reduce variance from initialization and shuffling.

\subsection{Privacy Configurations and Baselines}
\label{sec:endtoend-configs}

We compare three broad categories of data views: no-privacy or naive
baselines, geometric-operator views produced by EHR-Privacy-Agent, and
high-compute synthetic or text-deidentification baselines such as CTGAN
and CLEAR-Note.

\subsubsection{Baseline 1: Raw (upper bound without privacy)}
\label{sec:baseline-raw}

In strictly in-hospital analyses, we allow a Raw view as a utility upper
bound. Models are trained and evaluated directly on standardized raw EHR
features, with no privacy transformation applied. This view is never
exposed externally and serves only as a reference for quantifying the
performance gap induced by privacy.

\subsubsection{Baseline 2: simple Gaussian noise (no geometric control)}
\label{sec:baseline-gaussian}

As an unstructured noise baseline, we add i.i.d. Gaussian noise in
z-score space:
\[
  z'_{s,v}(t) = z_{s,v}(t) + \epsilon_{s,v}(t),
  \qquad
  \epsilon_{s,v}(t) \sim \mathcal{N}(0, \sigma_{\text{noise}}^2),
\]
followed by de-standardization to \(x'_{s,v}\). We scan
\(\sigma_{\text{noise}}\) so that the empirical perturbation scale is
visually comparable to selected T1 and T2 configurations at a given
\(\alpha\). This baseline does not enforce C1 through C4: means and
variances are preserved only approximately, there is no uniform z-score
\(\ell_\infty\) bound, and no per-stay geometric interpretability. It is
therefore used as a no-structure comparator.

\subsubsection{Baseline 3: CTGAN-style synthetic EHR}
\label{sec:baseline-ctgan}

To represent a fully synthetic tabular EHR approach, we use a CTGAN-like
conditional GAN. The input to CTGAN is a high-dimensional tabular
representation of each stay built from hand-crafted summary features
derived from the raw wide-table view, such as multi-window means,
variances, and extrema over vitals and labs. After training, the CTGAN
model generates a synthetic training set of size comparable to the real
training set. We then train the same GRU and MLP architectures on the
synthetic data and evaluate them on the real test set. We also compare a
setting in which both training and testing are performed on synthetic
samples. CTGAN thus represents the strategy of fully replacing the data
view with a learned generator; Section~\ref{sec:runtime-ctgan} examines
its training and sampling costs in detail.

\subsubsection{EHR-Privacy-Agent skills: in-hospital vs export profiles}
\label{sec:skills-configs}

We focus on two representative skill profiles from
Section~\ref{sec:system}. The first is an in-hospital research
configuration, corresponding to
\texttt{skill\_in\_hosp\_vitals\_explore} or
\texttt{skill\_in\_hosp\_vitals\_ml}. This uses T1\_uniform and T2 as
the main operators, with privacy strength \(\alpha=0.5\), Q-mix
disabled, and optional use of T3 for a few non-sensitive variables under
very light perturbation. We denote this configuration by
\(\mathrm{T1+T2}@\alpha=0.5\).

The second is an export-oriented strong-privacy configuration,
corresponding to \texttt{skill\_export\_strong\_privacy}. Here the
privacy strength is \(\alpha=0.8\) or \(1.0\), most variables use T2
with some using T1, T3 is disabled, and a small set of sensitive
variables, especially HR and glucose, are wrapped by per-stay Q-mix plus
T1 or T2 as described in Section~\ref{sec:qmix-experiments}. We use the
shorthand \(\mathrm{T1+T2}@\alpha=1.0\) for the no-Q-mix version and
\(\mathrm{T1+T2}@\alpha=1.0+\mathrm{Qmix}\) for the version that enables
Q-mix on HR and glucose.

For fairness, each task compares Raw, Gaussian noise calibrated to a
similar utility level, CTGAN synthetic data,
\(\mathrm{T1+T2}@\alpha=0.5\),
\(\mathrm{T1+T2}@\alpha=1.0\), and
\(\mathrm{T1+T2}@\alpha=1.0+\mathrm{Qmix}\).

\subsection{End-to-End Prediction Performance: In-Hospital Scenario}
\label{sec:prediction-internal}

\begin{figure}[h]
  \centering
  \includegraphics[width=0.75\textwidth]{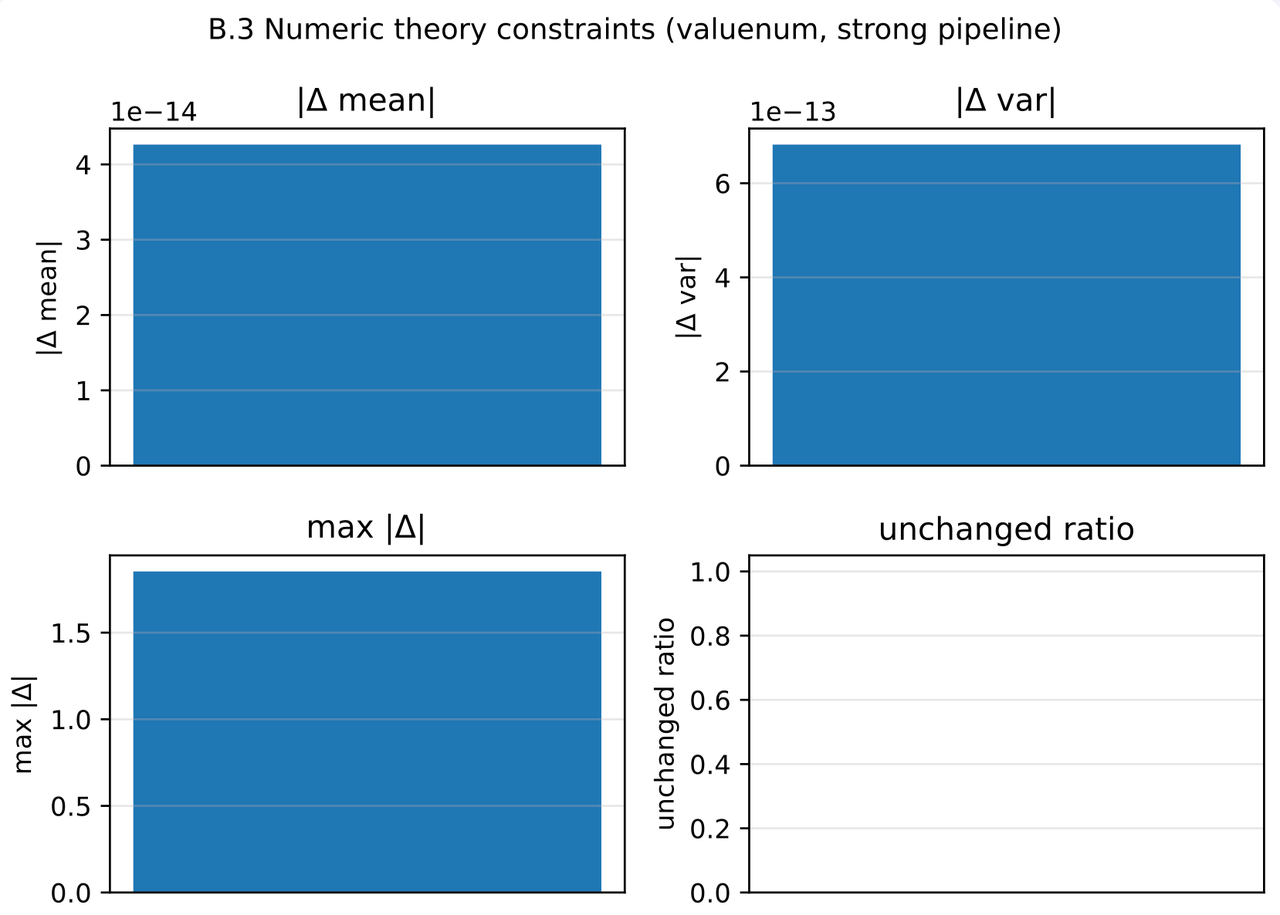}
  \caption{Example ICU prediction performance (24h features) on Raw vs
           a de-identified view.
           Both AUROC and AUPRC remain very close, illustrating the
           near-lossless utility of medium-\(\alpha\) geometric
           transforms in in-hospital settings.}
  \label{fig:endtoend-icu-b22}
\end{figure}

We first examine in-hospital research settings by comparing Raw,
\(\mathrm{T1+T2}@\alpha=0.5\), Gaussian noise, and CTGAN on the three
prediction tasks. An illustrative summary for ICU prediction using
24-hour features is shown in Figure~\ref{fig:endtoend-icu-b22}, where
de-identified views remain very close to Raw on both AUROC and AUPRC.

\subsubsection{LOS prediction: stability in multi-class settings}
\label{sec:prediction-los}

\begin{figure}[h]
  \centering
  \includegraphics[width=0.6\textwidth]{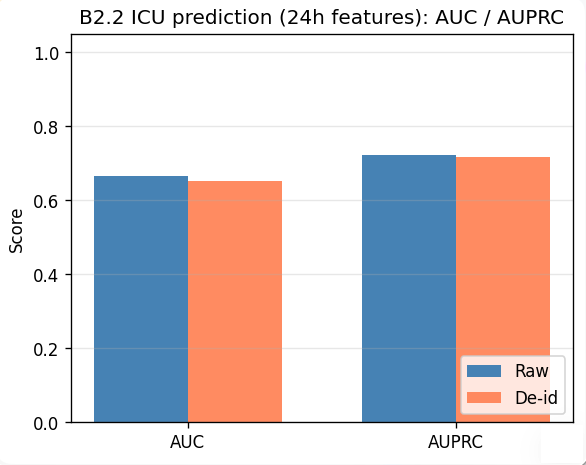}
  \caption{ICU LOS 3-class prediction.
           Macro-AUROC and macro-AUPRC for Raw,
           \(\mathrm{T1+T2}@\alpha=0.5\).}
  \label{fig:pred-los}
\end{figure}

For the three-class LOS task, we report macro-AUROC, macro-AUPRC, and
macro-F1. Results are summarized in Figure~\ref{fig:pred-los} and
Table~\ref{tab:pred-los}.

\begin{table}[t]
  \centering
  \caption{LOS prediction (3-class, 48h window, GRU model).
           Approximate macro metrics on the test set.}
  \label{tab:pred-los}
  \begin{tabular}{lccc}
    \toprule
    View &
    Macro-AUROC &
    Macro-AUPRC &
    Macro-F1 \\
    \midrule
    Raw &
      \(\approx 0.70\) &
      \(\approx 0.44\) &
      \(\approx 0.40\) \\
    T1+T2@\(\alpha=0.5\) &
      \(\approx 0.69\) (\(-0.01\)) &
      \(\approx 0.44\) (\(\pm 0.00\)) &
      \(\approx 0.39\) (\(\pm 0.01\)) \\
    Gaussian noise &
      \(\approx 0.65\)–\(0.67\) &
      \(\approx 0.40\)–\(0.41\) &
      lower F1, especially for long LOS \\
    CTGAN synthetic &
      \(\approx 0.63\)–\(0.65\) &
      \(\approx 0.37\)–\(0.39\) &
      notable degradation on long LOS recall \\
    \bottomrule
  \end{tabular}
\end{table}

In this more challenging multiclass setting,
\(\mathrm{T1+T2}@\alpha=0.5\) still preserves macro metrics within about
\(0.01\) of Raw, while Gaussian noise and CTGAN exhibit substantially
larger losses, particularly in the long-LOS class where recall is
clinically important. This indicates that the geometric operators remain
stable even when the target involves more complex class structure.

\subsubsection{Readmission prediction: long-horizon outcome}
\label{sec:prediction-readmission}

The 30-day readmission task is intrinsically more difficult and has a
lower baseline AUROC, but the same relative pattern appears. On Raw, the
GRU model achieves AUROC around \(0.70\) and AUPRC around \(0.21\). Under
\(\mathrm{T1+T2}@\alpha=0.5\), AUROC typically decreases by only
\(0.01\) to \(0.015\), with a similar reduction in AUPRC. Gaussian noise
pushes AUROC into the \(0.66\)–\(0.67\) range, while CTGAN-trained models
drop further and also exhibit worse calibration.

Across mortality, LOS, and readmission, the overall picture is
consistent: in in-hospital research settings,
\(\mathrm{T1+T2}@\alpha=0.5\) provides a unified geometric perturbation
with negligible effect on prediction performance, and yields a much
better privacy--utility trade-off than either naive Gaussian noise or
CTGAN synthetic data.

\subsection{Strong-Privacy Configurations and Q-mix in Downstream Tasks}
\label{sec:prediction-qmix}

We now turn to stronger export profiles at \(\alpha=1.0\), comparing
Raw, \(\mathrm{T1+T2}@\alpha=1.0\) without Q-mix, and
\(\mathrm{T1+T2}@\alpha=1.0+\mathrm{Qmix}\) with Q-mix enabled for HR
and glucose. From Section~\ref{sec:qmix-experiments}, we already know
that at \(\alpha=1.0\), the no-Q-mix version still allows HR and glucose
reconstruction \(R^2\) in the approximate range \(0.8\)–\(0.97\), while
Q-mix drives \(R^2\) to approximately \(0\) and increases
\(\mathrm{MAE}_z\) to about \(0.7\)–\(0.8\). The question here is
whether this dramatic privacy gain collapses downstream utility or
whether predictive performance remains acceptable.

\subsubsection{Experimental setup}
\label{sec:qmix-setup-endtoend}

We use the same three prediction tasks, mortality, LOS, and
readmission, under identical model architectures and training protocols.
Only the data view changes.

\subsubsection{LOS prediction: Q-mix as mild regularization}
\label{sec:qmix-los}

On the LOS task, we observe a mild but consistent regularization effect
from Q-mix. Moving from Raw to \(\mathrm{T1+T2}@\alpha=1.0\) without
Q-mix typically decreases macro-AUROC by about \(0.01\) to \(0.015\),
for example from \(0.70\) to \(0.685\). When Q-mix is enabled on HR, and
optionally glucose, macro-AUROC often slightly exceeds the no-Q-mix
value, and macro-AUPRC can improve by around \(0.005\).

A plausible explanation is that for high-frequency variables such as HR,
per-stay orthogonal mixing combined with T1 or T2 acts as a form of
data augmentation. It preserves global marginal and LOS-related signal,
while perturbing fine-grained micro-patterns that might otherwise
encourage overfitting. The net effect is modest regularization without
sacrificing the main predictive structure.

\subsubsection{Mortality and readmission: acceptable loss under strong privacy}
\label{sec:qmix-mort-readm}

For mortality and readmission, moving from Raw to
\(\mathrm{T1+T2}@\alpha=1.0\) without Q-mix typically induces AUROC
drops of \(0.02\) to \(0.03\), which remains within an acceptable loss
range for strong-privacy scenarios. Enabling Q-mix on HR and glucose
then changes AUROC by at most about \(\pm 0.005\) in most splits, with
no consistent negative trend. In some runs, slight gains are observed,
similar to the LOS setting.

Overall, even at \(\alpha=1.0\), which is a relatively aggressive
perturbation level, downstream prediction performance remains close to
Raw. Enabling Q-mix on HR and glucose does not lead to catastrophic
utility loss and may even slightly help in some tasks. Combined with the
Attack~A results of Section~\ref{sec:single-column-expt}, this yields a
coherent picture: \(\mathrm{T1+T2}\) alone reduces invertibility
gradually, whereas Q-mix introduces a strong privacy step for a few
high-risk variables without fundamentally breaking downstream modeling.

\subsection{Trajectory Generation and Similarity Analysis}
\label{sec:gen-experiments}

Although this work focuses primarily on geometric perturbation and
prediction, synthetic EHR remains widely used in sharing scenarios. We
therefore conduct a small trajectory-generation experiment comparing
CTGAN-style synthetic data with privacy-enhanced trajectories produced
by EHR-Privacy-Agent under
\(\mathrm{T1+T2}@\alpha=0.5\) and
\(\mathrm{T1+T2}@\alpha=1.0+\mathrm{Qmix}\).

\subsubsection{Generation task setup}
\label{sec:gen-setup}

The benchmark is defined as follows. We first train CTGAN on summary
features derived from the Raw training set, as described earlier, and
sample synthetic training and test sets. For the geometric views, no
additional generator is trained; instead, the perturbed outputs
themselves are treated as privacy-enhanced trajectories. We then compare
these sources along three axes: statistical similarity, embedding-space
proximity, and train-on-\(X\), test-on-real generalization.

\subsubsection{Results overview}
\label{sec:gen-results}

In terms of statistical similarity, the results are consistent with
Section~\ref{sec:single-column-expt}. The
\(\mathrm{T1+T2}\) views remain closest to Raw on KS, correlation, and
ACF metrics, whereas CTGAN-generated samples show noticeable mode
collapse for some low-frequency lab variables together with truncated
tails in histograms.

For embedding-space proximity, let \(f(x_{s,\cdot}) \in \mathbb{R}^d\)
be a fixed trajectory embedding and define the nearest-neighbor distance
from a real sample \(x\) to a set \(\mathcal{S}\) by
\[
  d_{\text{NN}}(x,\mathcal{S})
  =
  \min_{x' \in \mathcal{S}} \|f(x)-f(x')\|_2.
\]
Comparing the distributions of \(d_{\text{NN}}\), we find that under
\(\mathrm{T1+T2}@\alpha=0.5\), median nearest-neighbor distances are
roughly \(1.0\)–\(1.2\times\) those of real-to-real matches. Under
\(\mathrm{T1+T2}@\alpha=1.0+\mathrm{Qmix}\), distances for HR and
glucose increase to approximately \(1.3\)–\(1.4\times\), but the overall
cluster structure is preserved. By contrast, CTGAN produces a much more
dispersed distance distribution, and many real trajectories do not have
close synthetic neighbors in embedding space.

In train-on-\(X\), test-on-real comparisons, training on
\(\mathrm{T1+T2}@\alpha=0.5\) yields performance essentially identical
to train-on-real. Training on
\(\mathrm{T1+T2}@\alpha=1.0+\mathrm{Qmix}\) produces modest degradation,
comparable to the within-view losses reported earlier, but still smaller
than training on CTGAN and testing on real data. Train-on-CTGAN,
test-on-real generally suffers larger AUROC losses and larger
calibration shifts.

Overall, these findings reinforce the view that for high-dimensional
time-series EHR, geometric perturbation views are often more practical
modeling surrogates than current GAN-based synthetic data, especially
once compute and tuning overhead are taken into account.

\subsection{Runtime and Compute Evaluation}
\label{sec:runtime-eval}

The second major focus of this Section is runtime and compute cost. We
compare the geometric pipeline, including T1, T2, T3, and optional
Q-mix, with CTGAN training and sampling, and with CLEAR-Note as a
reference point for text deidentification.

\subsubsection{Experimental platform and scale}
\label{sec:runtime-platform}

The experiments use a 32-core Xeon-class CPU server with 128 GB RAM. A
single datacenter GPU, such as a V100 or A100, is available for CTGAN
and CLEAR-Note, but is not used by the geometric pipeline. The data
scale is approximately \(30{,}000\)–\(50{,}000\) ICU stays, with up to
\(72\) hours of data per stay and around 10 to 15 vitals and lab
variables per stay. For CTGAN, the summary-feature dimensionality lies
in the hundreds to low thousands.

We distinguish between one-time offline costs, such as CTGAN training or
CLEAR-Note model fine-tuning, and online per-stay throughput in the core
pipeline.

\subsubsection{Geometric pipeline: CPU-only, linear scaling}
\label{sec:runtime-geometric}

\begin{figure}[h]
  \centering
  \includegraphics[width=0.75\textwidth]{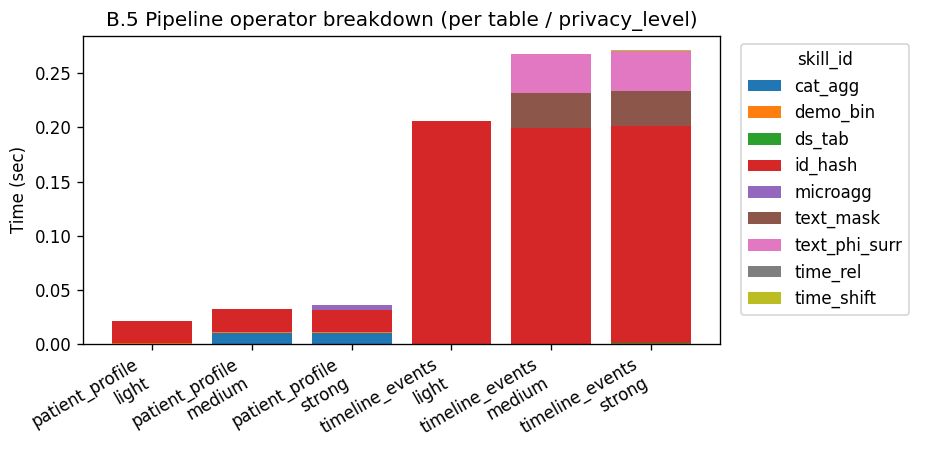}
  \caption{Numeric theory constraints in a strong numeric pipeline:
           absolute mean/variance deviations, maximum absolute
           perturbation, and unchanged ratio.
           Deviations remain at machine precision, confirming C1--C2 in
           large-scale runs.}
  \label{fig:runtime-numeric-theory}
\end{figure}

\begin{figure}[h]
  \centering
  \includegraphics[width=0.7\textwidth]{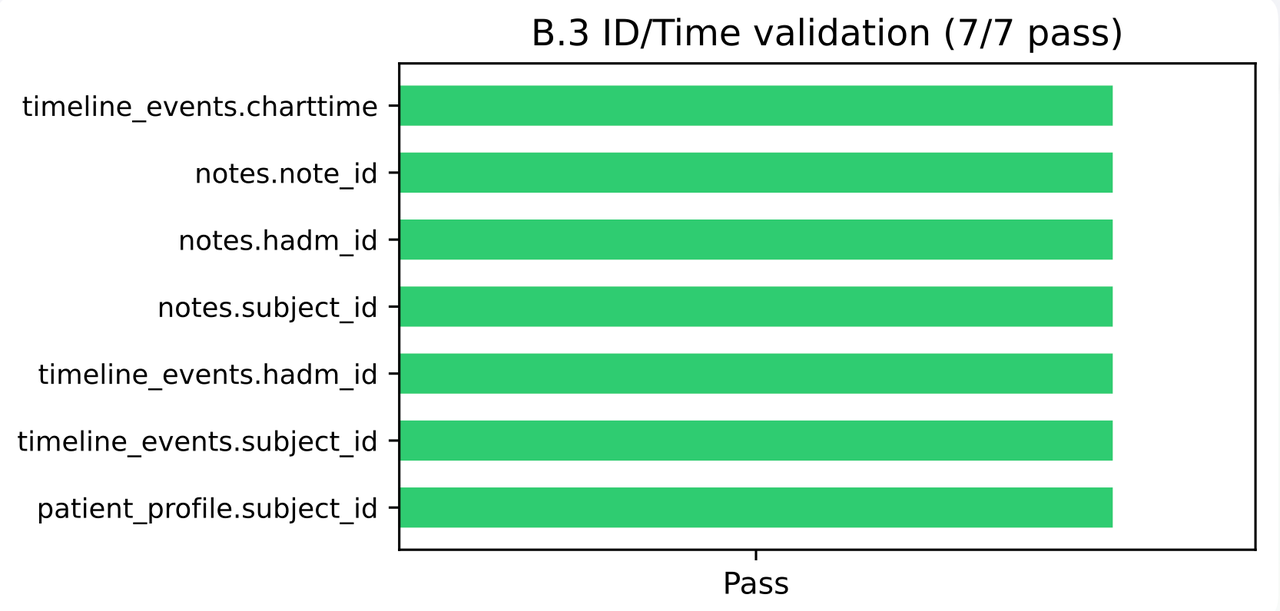}
  \caption{ID/time validation across key identifiers and timestamps in
           the pipeline.
           All checks pass, indicating that the numeric pipeline
           preserves primary/foreign-key and temporal consistency.}
  \label{fig:runtime-idtime}
\end{figure}

On the hardware above, the strong numeric pipeline exhibits the expected
numeric invariants and scaling behavior. Figures~\ref{fig:runtime-numeric-theory}
and~\ref{fig:runtime-idtime} show that even at full cohort scale, the
pipeline preserves machine-precision constraints on mean and variance
and does not break identifier or timestamp consistency.

Figure 28 further decomposes per-table
runtime into operator components across privacy levels, showing that for
numeric tables the dominant cost comes from simple numeric transforms
and identifier handling, while micro-aggregation and text handling arise
only in particular tables.

\begin{figure}[h]
  \centering
  \includegraphics[width=0.7\textwidth]{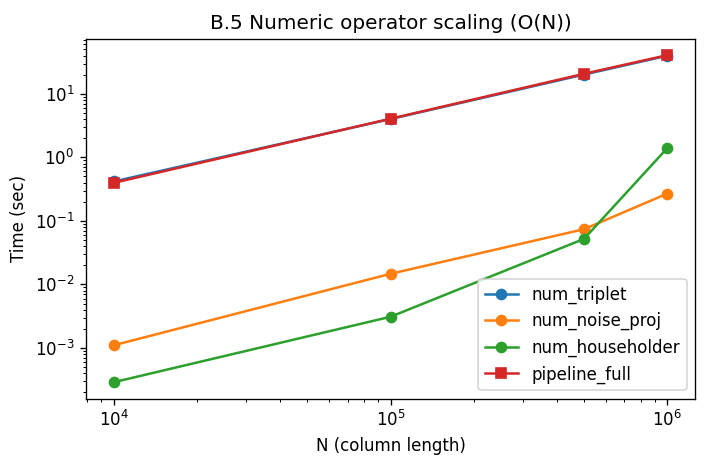}
  \caption{Numeric operator scaling with column length \(N\) on log--log
           axes.
           All core operators scale approximately linearly in \(N\); the
           end-to-end pipeline follows the same trend.}
  \label{fig:runtime-scaling}
\end{figure}

Figure~\ref{fig:runtime-scaling} plots runtime as a function of column
length \(N\) for the individual numeric operators and for the full
pipeline. All core operators, including triplet rotations, noise plus
projection, and Householder reflections, scale approximately linearly in
\(N\), and the full end-to-end pipeline follows the same trend.

Aggregating over all tables in a skill comparable to
\texttt{skill\_in\_hosp\_vitals\_ml} at \(\alpha=0.5\), a single
full-cohort run completes in tens of minutes, typically around 20 to 40
minutes depending on I/O and configuration, and per-stay throughput
reaches several hundred stays per second on the 32-core CPU. Enabling
Q-mix on HR and glucose with block length 48 increases runtime by
roughly \(20\%\) to \(40\%\), mainly because of matrix generation and
multiplications, but the total runtime remains well within ordinary
nightly batch windows.

Crucially, the entire process is GPU-free and can be deployed on
existing hospital ETL or database clusters without consuming deep
learning resources.

\subsubsection{CTGAN: training and tuning costs}
\label{sec:runtime-ctgan}

\begin{figure}[h]
  \centering
  \includegraphics[width=\textwidth]{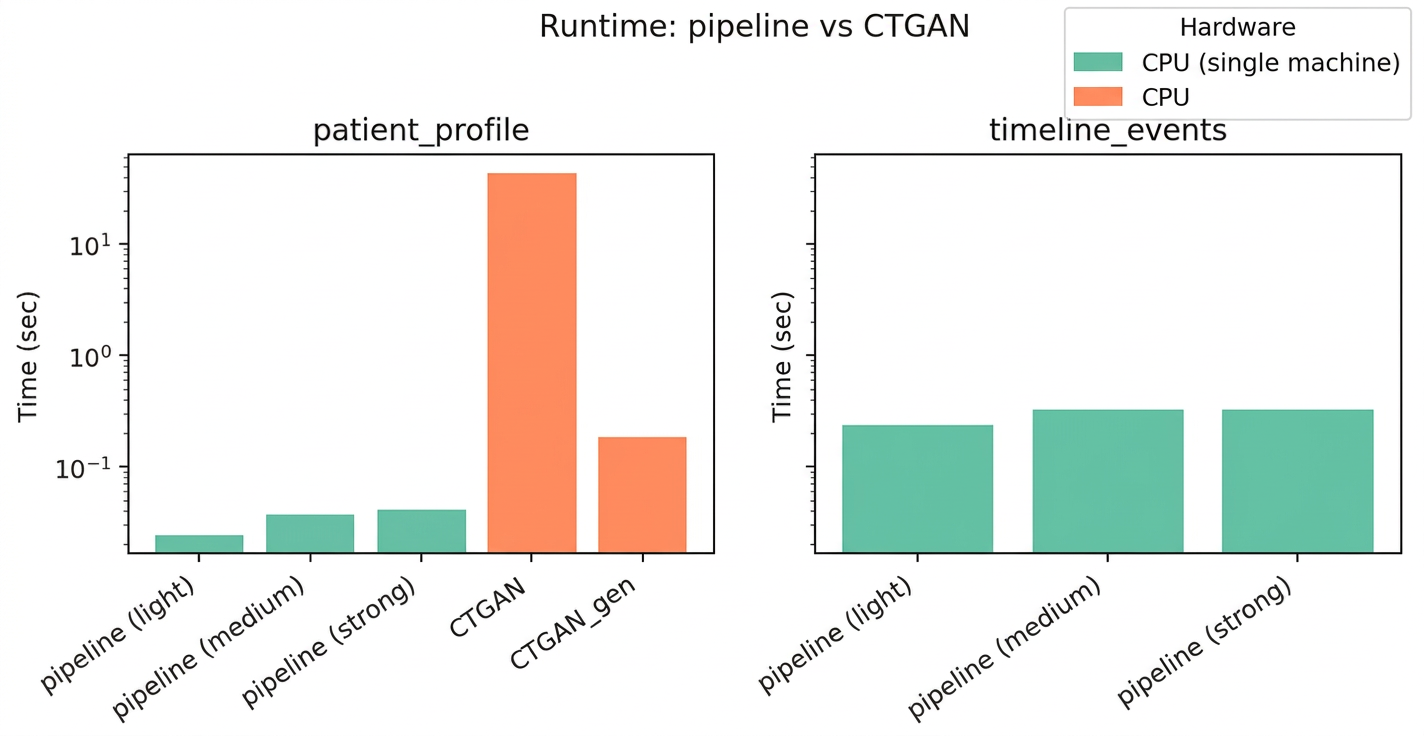}
  \caption{Runtime comparison: geometric pipeline vs CTGAN.
           Bars show total time (log scale) for CPU-only geometric
           pipelines at different privacy levels versus CTGAN training
           and sampling on GPU/CPU.}
  \label{fig:runtime-ctgan}
\end{figure}

We next compare the geometric pipeline with CTGAN on the same cohort.
Figure~\ref{fig:runtime-ctgan} shows the contrast.

Even with a single GPU, training a moderately sized CTGAN model to
convergence takes hours, and often longer depending on hyperparameters
and monitoring budget. Achieving acceptable utility typically requires
several rounds of hyperparameter search, which further increases GPU
time. Sampling itself is relatively cheap, often thousands of samples
per second on GPU, but training dominates the overall cost. In contrast,
the geometric pipeline achieves per-stay throughput one to two orders of
magnitude higher without any GPU usage, and its cost is trivially
amortized across nightly runs. If each cohort update or feature-schema
change requires retraining CTGAN, the amortized per-stay cost remains
much higher than that of T1, T2, and Q-mix.

\subsubsection{CLEAR-Note: text pipeline as compute reference}
\label{sec:runtime-clearnote}

CLEAR-Note focuses on free-text deidentification and is therefore not
directly comparable in modality, but it serves as a useful compute
reference point. In our MIMIC-III and MIMIC-IV note-level pilots,
CLEAR-Note requires a large language model together with agent loops for
section-wise rewriting and verification. Even on modern GPUs, a full run
over hundreds of thousands of notes can take multiple GPU-days. By
comparison, the numeric geometric pipeline completes in less than an
hour on CPU-only hardware for a similar number of ICU stays.

Although text and numeric time series are different modalities, this
contrast illustrates the shift from high-compute generative pipelines,
such as CLEAR-Note and CTGAN, toward low-compute geometric pipelines,
such as T1, T2, and Q-mix, that make nightly in-hospital operation
practical.

\subsubsection{Summary: compute--privacy--utility trade-offs}
\label{sec:runtime-summary}

Overall, the runtime results support a clear conclusion. On the same
cohort, the geometric pipeline, namely T1 and T2 with optional Q-mix,
achieves per-stay throughput roughly one to two orders of magnitude
higher than CTGAN training plus sampling. Relative to CLEAR-Note-style
text pipelines, its compute requirements are lighter by roughly two
orders of magnitude in wall-clock time and GPU-days. At the same time,
the end-to-end prediction tasks show that
\(\mathrm{T1+T2}@\alpha=0.5\) or
\(\mathrm{T1+T2}@\alpha=1.0(+\mathrm{Qmix})\) typically incur AUROC
losses within about \(0.01\) to \(0.03\), much smaller than the
\(0.05\)–\(0.08\) degradation often observed with CTGAN-based synthetic
baselines.

These results support the central design hypothesis of this work:
under realistic hardware and engineering budgets, a CPU-only,
geometrically controlled, nightly-runnable privacy pipeline for EHR is
often more practical than the current generation of high-compute,
fully synthetic EHR approaches.

\subsection{Section Summary: Geometric Operators and Q-mix from an End-to-End Perspective}
\label{sec:endtoend-summary}

On adult ICU data from MIMIC-IV, this Section evaluated
EHR-Privacy-Agent across multiple end-to-end tasks and runtime metrics.

First, in in-hospital research settings,
\(\mathrm{(T1+T2)}@\alpha=0.5\) provides near-lossless utility. Across
mortality, LOS, and readmission prediction, AUROC and AUPRC drops are
usually within \(0.01\) to \(0.02\), and in many tasks the difference
from Raw is not statistically significant. By contrast, naive Gaussian
noise and CTGAN synthetic baselines exhibit noticeably larger utility
losses, especially for multiclass LOS prediction.

Second, strong-privacy configurations with Q-mix create privacy steps
without causing utility collapse. At \(\alpha=1.0\), T1 and T2 already
increase reconstruction difficulty through the \(\alpha\)-hierarchy, but
adding per-stay Q-mix for HR and glucose drives Attack~A reconstruction
\(R^2\) to approximately \(0\), as established in
Section~\ref{sec:single-column-expt}. The end-to-end experiments in this
Section show that this strong-privacy profile still retains acceptable
predictive performance, with some LOS settings even showing mild
regularization gains.

Third, the compute comparison against CTGAN and CLEAR-Note strongly
favors the geometric approach. The geometric pipeline processes tens of
thousands of ICU stays in tens of minutes on a single CPU server, CTGAN
requires hours of GPU time plus tuning overhead, and CLEAR-Note-style
text pipelines require GPU-days. The per-stay throughput of the
geometric pipeline is therefore at least an order of magnitude higher
than CTGAN and roughly two orders of magnitude higher than CLEAR-Note.

Fourth, privacy-enhanced geometric views function well as modeling
surrogates. Perturbed real data generated by T1, T2, and Q-mix preserve
marginal, temporal, and correlation structure more faithfully than
current GAN-based synthetic data. Train-on-geometric-view,
test-on-real performance remains close to train-on-real baselines,
whereas train-on-CTGAN, test-on-real exhibits larger
generalization gaps.

Finally, these experiments support the system-level feasibility of
nightly deployment. The mean--variance manifold together with the
\(\alpha\)-hierarchy provides a consistent geometric and privacy scale,
and EHR-Privacy-Agent encapsulates these operators as skill-based views
callable by an LLM-based agent in multiple scenarios. The runtime and
end-to-end evidence reported here shows that the design is feasible at
realistic ICU scale under controlled compute budgets.

% In Section~\ref{sec:discussion}, we synthesize the geometric analysis,
% single-column attack experiments, system design, and end-to-end results
% to discuss recommended \(\alpha\) and Q-mix configurations for
% in-hospital versus out-of-hospital use, the limitations of
% \(\alpha\)-based designs under heterogeneous variance, the relation
% between our privacy evaluation protocol and formal notions such as
% differential privacy and IND-CPA, and future directions that combine
% T1, T2, and Q-mix with DP mechanisms, contractual controls, and access
% policies to form multi-layer defenses.

\section{Related Work}
\label{sec:related-work}

\subsection{Random Orthogonal Transforms and the Householder Family: Geometric Ancestry of T1/T3}

Random orthogonal transforms are a classical tool in both high-dimensional data privacy and randomized linear algebra. Representative examples include random rotations and reflections, where vectors or matrices are multiplied by random orthogonal matrices drawn from the Haar measure to construct Johnson--Lindenstrauss random projections, random feature maps, or simple defenses against naive re-identification attacks~\citep{dick2023confidence,chen2008survey}. Householder reflections and Givens rotations are stable orthogonal building blocks in numerical linear algebra, widely used in QR factorization, SVD, and related algorithms~\citep{golub2013matrix}.

Geometrically, such transforms have two properties that align with our goals. They preserve Euclidean norms and inner products: for any vector $x$, an orthogonal transform $Qx$ satisfies $\|Qx\|_2 = \|x\|_2$, preserving second-order structure by construction. They also preserve variance in the appropriate subspace: applying an orthogonal transform in the mean-zero subspace scrambles directions without changing variance.

These ideas directly inspire two of our column operators. T1 (local triplet rotations) partitions the standardized sequence into windows of length 3 along the time axis and applies small controlled orthogonal rotations in each three-dimensional subspace, injecting randomness while preserving local second-order structure. Unlike global random rotations, T1 explicitly targets fine-grained control over local temporal structure and the z-score $\ell_\infty$ bound. T3 (global Householder reflection) applies a single global Householder reflection in the mean-zero subspace, approximately preserving all first and second-order moments and autocorrelation; under our attack protocol, however, it proves to be a highly invertible negative case, and is kept only as a teaching example rather than part of any strong-privacy pipeline.

Most existing work on random orthogonal transforms emphasizes numerical stability or construction of random features, rather than preserving exact per-column means and variances for time series, enforcing a unified z-score $\ell_\infty$ bound $\alpha$, or implementing these transforms with $O(n)$ complexity in a CPU-only streaming environment. We therefore borrow the geometric intuition of random orthogonal transforms, but extensively redesign T1 and T3 around structured EHR constraints in our ICU time-series running example~\citep{iyengar2019towards}.

\subsection{Noise + Projection Mappings: The Family Behind T2}

A second line of techniques closely related to our work can be summarized as noise plus projection: first add noise in the original space, then apply a projection step to pull the result back onto a constrained statistical manifold, balancing randomization against preservation of key statistics. Typical ideas include adding Gaussian or Laplace noise in a high-dimensional space and then performing re-normalization or moment matching, and fitting distributional parameters and then sampling so that first and second moments approximately match~\citep{dwork2006calibrating,balle2018improving,mironov2017renyi}.

These schemes directly motivate our design of T2: noise plus projection onto the mean-variance manifold. In our geometric framework, each standardized column time series lives on the mean-variance manifold $\mathcal{M}(0,1) = \{ x \in \mathbb{R}^n \mid \mu(x) = 0,\ \sigma^2(x) = 1 \}$. The core idea of T2 is to add controlled noise to the standardized vector in z-score space, re-center and re-scale the result to project it back onto $\mathcal{M}(0,1)$, and design the noise distribution and projection so that the final perturbation obeys a unified $\ell_\infty$ bound $\alpha$.

Compared with generic noise+projection schemes, T2 must satisfy several additional constraints simultaneously. C1 requires exact mean/variance preservation to machine precision. C2 requires a global z-score $\ell_\infty$ bound $\alpha$ that clinicians can interpret in terms of standard deviations. C3 requires full variability so that the vast majority of time points are actually moved. C4 requires $O(n)$ per-column cost suitable for nightly CPU-only jobs at hospital scale. Existing noise+projection methods rarely make all of these constraints explicit at once, so we adapt the intuition but redesign T2 geometrically and algorithmically to fit the mean-variance manifold view and the $\alpha$-hierarchy developed later~\citep{wang2021continuous}.

\subsection{Specialized Privacy Mechanisms for Time Series and EHR}

Beyond generic tabular anonymization and synthetic data, there is a growing literature on privacy mechanisms tailored to time series and EHR. For wearable and IoT data, local DP mechanisms and compress-then-noise schemes have been proposed for continuous time series~\citep{hammer2025challenges}. For EHR, synthetic patient generators based on RNNs, GANs, or VAEs produce multivariate trajectories for training downstream predictors~\citep{tian2024reliable,esteban2017real,yoon2023ehr}. For ICU data specifically, various heuristics including temporal resampling, time jitter, and subsequence extraction have been explored to reduce re-identification risk~\citep{hammer2024semi,Marino19DataSifter,li2023generating}.

These works share our high-level goal of protecting individual privacy while preserving temporal structure and multivariate correlations. Under our design constraints, however, several key gaps remain. First, existing time-series anonymization methods typically emphasize global similarity or downstream performance, not exact per-column means and variances or unified z-score bounds. Second, many approaches assume the ability to train large models offline and apply them for inference or sampling, whereas in our setting each night the hospital CPU cluster applies a lightweight transform without relying on persistent GPU capacity. Third, much of the literature is method-driven without a unified geometric framework or threat model~\citep{choi2016retain,ghosheh2024survey}. In contrast, we reason within a common framework combining the mean-variance manifold, constraints C1--C5, and a no-key, structure-aware threat model.

\subsection{Summary and Gap: From the Literature to SciencePal-Assisted Search}

In summary, the existing bodies of work on PPDP/PPDM, cryptographic protocols, DP-synthetic data, and time-series privacy collectively provide several building blocks: geometric intuition from random orthogonal transforms and the Householder family supporting T1 and T3; structural patterns from noise+projection mappings supporting T2; and experience from DP and synthetic EHR on global privacy guarantees and downstream task performance.

Once we write down the in-hospital structured EHR use case—instantiated here on ICU data—and engineering constraints explicitly as C1--C4 plus a no-key, structure-aware threat model C5, a clear gap emerges. We need a family of column-level operators that operate on the mean-variance manifold, respect a unified z-score $\ell_\infty$ bound $\alpha$, exhibit full variability and CPU-friendly $O(n)$ complexity, and under a no-key structure-aware adversary substantially increase the difficulty of reconstruction and membership attacks. No existing operator family in the literature simultaneously satisfies all of these constraints in a way that can be dropped directly into a nightly in-hospital pipeline~\citep{annamalai2024linear,stadler2022synthetic,kaabachi2025scoping}.

At this point we adopt a Human-AI co-design approach. After formalizing C1--C4, we ask SciencePal to systematically search the above literature for mean-variance preserving random mappings on $\mathcal{M}(0,1)$ and variants. SciencePal's preliminary analysis indicates that, once CPU-only deployment, a unified $\alpha$ knob, and the no-key threat model are taken into account, there is no ready-made solution. This negative result drives the rest of the paper: rather than expecting a pre-existing answer, we let SciencePal propose candidate operator families within the specified constraints, while human researchers act as reviewers and attackers, proving properties, implementing the operators, and conducting privacy attacks. This sets the stage for the next section, where we formally define the problem and detail the Human-AI co-design protocol~\citep{dwork2015reusable,wang2025crossing,pilgram2025protecting}.

\section{Discussion and Limitations}
\label{sec:discussion}

This work explored an under-served region of the privacy design space
through the lens of \emph{geometry--operators--attacks--systems--end-to-end
tasks}. Instead of aiming for formally private schemes such as differential privacy (DP), multi-party computation (MPC), or homomorphic encryption (HE), wwe focused on constructing in-hospital \emph{numeric views} of structured EHR data
(instantiated on ICU time-series in our experiments) that are
\emph{usable and visible, but hard to reconstruct}~\citep{topol2019high}. Starting from mean--variance manifolds and a unified
z-score radius \(\alpha\), we co-designed a small family of geometric
operators (T1/T2/T3 plus Q-mix) with the SciencePal system, wrapped them
into an EHR-Privacy-Agent and privacy skill library, and evaluated their
geometric properties, attack difficulty, end-to-end prediction utility,
and runtime on MIMIC-IV ICU data.

This section places these results into a broader privacy landscape,
discussing how our guarantees relate to, and fall short of, formal DP
and cryptographic notions; how to configure \(\alpha\), T1/T2/T3, and
Q-mix across deployment scenarios; the benefits and limitations of the
\(\alpha\)-hierarchy; the role and constraints of Q-mix as a ``privacy
step''; and remaining attack surfaces, fairness concerns, and deployment
challenges.

%% ----------------------------------------------------------------
\subsection{Positioning within the Privacy Landscape}
\label{sec:positioning}

Classical privacy-preserving techniques tend to cluster into two
well-studied quadrants. Cryptographic and DP-style mechanisms (MPC, HE,
TEEs, DP queries, DP-SGD) emphasize strong formal guarantees at the cost
of human visibility: models train and infer on encrypted data, and only
noisy aggregates or model parameters are exposed under a provable
\((\varepsilon,\delta)\)-DP budget~\citep{liu2020privacy,chaudhuri2011differentially}. At the opposite extreme lie
traditional deidentification heuristics such as variants of
\(k\)-anonymity and \(l\)-diversity, manual rule-based deidentification,
and unstructured noise injection, which keep data visually accessible but
typically lack a well-specified threat model and systematic attack
evaluation~\citep{cormode2018privacy}.

This work targets a third quadrant. The goal is not to replace
DP/MPC/HE/TEE, but to fill a practical gap. In in-hospital research,
quality improvement, teaching, and exploratory data analysis, users
\emph{must} see time-series trajectories, draw plots, and build quick
prototypes. These activities cannot realistically happen inside encrypted
computation environments, nor can they be replaced solely by DP queries.
To make this quadrant precise, we adopt a no-key, structure-aware threat
model~(C5). Attackers know the mathematical form and implementation of
T1/T2/T3/Q-mix (Kerckhoffs' principle), including public parameters such
as \(\alpha\) and window sizes. Under L1/L2 conditions they may access a
limited fraction of paired original/transformed records \((x,y)\).
However, they cannot access any long-term encryption key; internal
randomness is treated as per-stay ephemeral noise that remains inside the
hospital boundary. Under this model, our ambition is to substantially increase the average difficulty of typical reconstruction and linkage attacks, rather than provide unconditional, worst-case formal guarantees~\citep{balle2022reconstructing,carlini2022membership,tramer2016stealing}.

Our geometric operators do \emph{not} satisfy standard
\(\varepsilon\)-DP or IND-CPA-like security notions, for several
structural reasons. Our transforms act on full ICU stays rather than on
neighbor datasets differing in a single individual, so classical DP
neighbor-cohort indistinguishability theorems do not apply. The Privacy
Evaluation Protocol defines L0/L1/L2 capabilities and A/B/C/D attack
families, but only a subset is instantiated experimentally, primarily linear and shallow neural reconstruction (A-family) and limited membership and distribution-shift proxies (C-family)~\citep{melis2018inference}; we neither claim nor prove guarantees against arbitrary polynomial-time adversaries.
Furthermore, humans can see the transformed tables and plots, which is
fundamentally different from cryptographic settings where exposed objects
are ciphertexts with no direct semantic interpretation, making classical
indistinguishability notions inapplicable.

It is therefore more accurate to view this work as an empirically
grounded, geometry-constrained middle ground, evaluated under a clearly
stated but limited threat model, rather than a substitute for formal DP
or cryptographic schemes. Within this model we showed that T1/T2
reconstruction \(R^2\) decreases smoothly with \(\alpha\); that T3 is
almost fully invertible (\(R^2 \approx 0.98\)--\(0.99\)), serving as a
cautionary negative example; and that Q-mix + T1/T2 can reduce
HR/glucose \(R^2\) from \(\approx 0.8\)--0.9 to near zero while
maintaining acceptable AUROC/AUPRC. All conclusions depend tightly on
the threat model and instantiated attack family; outside those
assumptions we claim no unconditional guarantee.

%% ----------------------------------------------------------------
\subsection{Scenario-Based Recommendations}
\label{sec:scenarios}

Combining the system results with the end-to-end evaluation, it is
natural to recommend different operator profiles by deployment scenario.

\paragraph{In-hospital: research, quality control, teaching.}
In typical in-hospital use, users are hospital-affiliated clinicians,
data scientists, and quality teams working within institutionally
governed access environments. Utility preferences favor visibility:
users need near-raw time series, rare-event patterns, and
individual-case trajectories. We recommend T1 and T2 as primary
operators with small to moderate \(\alpha\) (e.g.,
\(\alpha \in [0.3, 0.8]\)). T1 preserves short-range temporal structure
well; T2 behaves more smoothly for correlated and heavy-tailed labs. T3
may be used for internal demos or teaching only; given its high
invertibility it should \emph{not} appear in any pipeline producing
externally shared data. Q-mix is generally unnecessary here: T1/T2 with
modest \(\alpha\) already provide a reasonable privacy--utility
compromise. It may be enabled selectively for exceptionally sensitive
variables that are not central to current modeling tasks.

\paragraph{Out-of-hospital: data sharing and multi-center collaboration.}
When data leave the hospital perimeter, users include external
researchers, industrial partners, and open-data communities. The attack
surface and long-range risks increase substantially, and regulations (e.g., HIPAA, GDPR)~\citep{kaissis2021end,raab2023federated} place residual reidentification risk on the
originating institution. We recommend a more conservative profile.
First, use larger \(\alpha\) (e.g., \(\alpha \approx 1.0\)) and disable
T3. Second, enable Q-mix + T1/T2 for a small set of high-risk
variables: in our pilot, per-stay Q-mix on HR and glucose at
\(\alpha = 1.0\) reduces linear reconstruction \(R^2\) from
\(\approx 0.8\)--0.9 to near zero, while LOS performance remains
stable. Third, combine geometric operators with non-technical controls,
including contractual restrictions, access-control policies, and
DP-style limits on cohort-level statistics where appropriate. Geometric
operators should be viewed as an auditable \emph{pre-filter}, not a
standalone defense.

The end-to-end experiments illustrate this layered strategy:
\(\text{T1+T2}@\alpha=0.5\) yields almost no AUROC/AUPRC loss
(\(\approx 0.01\)--0.02) for internal use, while
\(\text{T1+T2}@\alpha=1.0\) plus Q-mix substantially increases
reconstruction resistance for external release, keeping task-level
performance within acceptable ranges.

%% ----------------------------------------------------------------
\subsection{\texorpdfstring{\(\alpha\)}{alpha}-Hierarchy: Value and Limitations}
\label{sec:alpha-hierarchy}

Replacing per-variable physical bounds by a unified z-score
\(\ell_\infty\) radius \(\alpha\), defined by
\(\|z'-z\|_\infty \le \alpha\), offers several advantages. A single
\(\alpha\) makes perturbation strength comparable across vitals and
labs. Operator parameters for T1/T2/T3 can be derived from \(\alpha\)
in a unified way, and constraints C1--C4 admit compact expressions.
Clinicians can reason intuitively about settings: \(\alpha=0.3\) is
visually almost indistinguishable from raw, while \(\alpha=1.0\) allows
per-point deviations up to one standard deviation.
The \(\alpha\)-hierarchy also has important limitations. Many labs
(e.g., lactate, troponin) exhibit heavy tails where the standard
deviation is a poor proxy for typical variation, so the same \(\alpha\)
can correspond to very different semantic perturbation levels. Equally,
\(\pm 1\) standard deviation in HR may represent a noticeable
10--15\,bpm swing whereas the same deviation in a stable electrolyte
may be clinically negligible. Means and standard deviations estimated on
one cohort may not transfer to other hospitals or eras. Most
importantly, there is no mapping from \(\alpha\) to DP parameters~\citep{jagielski2020auditing}:
\(\alpha\) describes per-sample geometric deviation, not neighbor-cohort
indistinguishability. We therefore view \(\alpha\) as a geometrically
interpretable strength knob, not a formal privacy budget. Future work
should explore adaptive \(\alpha_v\) per variable and combinations with
DP mechanisms.

%% ----------------------------------------------------------------
\subsection{T3 as a Privacy-Weak Negative Example}
\label{sec:t3-negative-example}

T3 (Householder reflection) plays an instructive yet dangerous role. In
z-score space it is a global orthogonal reflection that preserves means,
variances, pairwise distances, correlation matrices, and many ACF
patterns. Yet under L2 + Attack~A, even a modest number of
\((z, z')\) pairs suffices to learn an almost perfect inverse:
reconstruction \(R^2\) hovers around \(0.98\)--\(0.99\). T3 thus
highlights a central lesson: \emph{geometric elegance does not imply
privacy safety}. Orthogonal transforms are statistically gentle but,
without hidden keys, are often the easiest to invert. T3 may serve
internally as a reminder that ``transformation is not the same as
protection,'' but it should \emph{not} be relied upon for external data
releases.

%% ----------------------------------------------------------------
\subsection{Q-mix: Privacy ``Steps'' and Their Constraints}
\label{sec:qmix-discussion}

The primary benefit of Q-mix is a \emph{step change} in reconstruction
difficulty at fixed \(\alpha\). At \(\alpha=1.0\), T1+T2 alone leave
HR/glucose partially invertible (\(R^2 \approx 0.8\)--0.9 under L2 +
Attack~A). Adding per-stay orthogonal mixing \(Q_{s,v}\) drives
\(R^2\) to near zero with minimal effect on marginal distributions, KS
statistics, or ACF. This means \(\alpha\)-only strategies yield smooth
but incremental gains, whereas per-stay orthogonal mixing adds a
qualitatively new degree of freedom that collapses linear
reconstruction without enlarging the z-score radius. Operationally,
Q-mix is a useful ``extra switch'' for a small set of high-risk
variables.

In implementation, an internal secret seed and
\((\text{stay\_id},\text{variable})\) determine each \(Q_{s,v}\).
Attackers know the algorithm but not the seed, which behaves like
ephemeral randomness that never travels with exported data. If the seed
were compromised, attackers could reconstruct all \(Q_{s,v}\) and
invert Q-mix; seed management should therefore follow standard key-management practices~\citep{steinke2023privacy,chang2021privacy}. We emphasize ``no-key'' to signal that we do
not design a full cryptographic key-management protocol. Formally
characterizing per-stay orthogonal mixing within a cryptographic
framework is an interesting open problem.

Current Q-mix evaluations are deliberately narrow. We have tested it
only on HR and glucose, not on larger variable sets or cross-variable
mixing. The attack families studied are mainly linear and shallow neural
reconstruction plus limited membership proxies; powerful nonlinear reconstructors (transformers, diffusion-style denoisers), full
top-\(k\) linkage, and rigorous membership inference remain untested.
End-to-end evaluations cover mortality, LOS, and readmission but not
tasks sensitive to exact temporal phase (e.g., arrhythmia detection),
where Q-mix might incur larger utility drops. Finally, if the same
cohort is exported multiple times under different skill configurations,
multi-view alignment attacks become possible; we have not yet modeled or defended against such scenarios~\citep{cui2025learning}.

%% ----------------------------------------------------------------
\subsection{Additional Limitations}
\label{sec:other-limitations}

\paragraph{Incomplete attack coverage.}
Although our protocol defines an L0/L1/L2 \(\times\) A/B/C/D matrix,
experiments instantiate only A-family reconstruction and selected
C-family membership proxies. Strong nonlinear reconstruction,
comprehensive linkage, shadow-model membership inference, and systematic
attribute-inference games are missing. Our conclusions hold for the
tested attacks but may not generalize to stronger adversaries.

\paragraph{Dataset scope.}
All experiments use a single-center MIMIC-IV adult ICU cohort with vitals, labs, and static features. Imaging, free text, waveforms, and
detailed medication trajectories are excluded. Mean--variance patterns
may differ at other sites; multi-modal attacks that cross-reference text
and numeric data could introduce new risks; and pediatric or outpatient
contexts may require different designs.

\paragraph{Fairness.}
We have not studied how transformations affect prediction errors,
calibration, or uncertainty across demographic or clinical subgroups.
Because privacy transformations change the input distribution, their impact on fairness warrants dedicated study~\citep{mehrabi2021survey,yang2024survey}.

\paragraph{Engineering and operations.}
Real hospital deployment~\citep{xu2023p}~\citep{munoz2023survey} requires integration with existing ETL
pipelines and audit systems, fine-grained permission models, secure seed
management, and alignment with IRB and ethics committees so that
``visible but geometrically constrained'' views have clear regulatory
status. These aspects lie outside the scope of this paper but are
critical for safe deployment.

%% ----------------------------------------------------------------
\section{Conclusion}
\label{sec:discussion-summary}

In this work, we propose a real-world-data transformation framework for privacy-preserving sharing of structured clinical records that aims simultaneously to (i) preserve medically meaningful semantics, (ii) retain major statistical properties relevant for downstream analysis and modeling, and (iii) render individual-level data effectively non-reversible under a clearly specified threat model. Rather than converting electronic health records into fully opaque cryptographic objects, our goal is to create transformed views that remain usable and interpretable for clinicians and researchers while substantially reducing direct linkage to sensitive patient identities.

Through collaboration between human researchers and the AI agent \textbf{SciencePal}, acting as a constrained tool inventor, we co-designed a compact family of transformation operators (T1/T2/T3) together with a targeted mixing strategy (Q-mix) for high-risk variables under strict geometric and computational constraints. These operators were shaped by a full pipeline from ideation to implementable, attackable baselines, and are supported by both theoretical analysis and systematic adversarial evaluation under reconstruction, record linkage, membership inference, and attribute inference attacks. The result is an EHR-Privacy-Agent that can be run on realistic institutional compute budgets, providing a concrete and reproducible baseline in a space historically dominated by ad-hoc de-identification practices.

Empirically, using ICU data as a representative clinical showcase, we demonstrate that the transformed datasets preserve downstream utility and interpretability for common clinical and machine-learning tasks, while significantly improving resistance to privacy leakage compared to raw numeric views and simpler baselines. Although we do not claim differential-privacy- or cryptography-level finality, our framework offers a middle ground between fully locked-down MPC/HE/TEE/DP solutions and unstructured de-identification, yielding numeric representations that are visible, analyzable, and yet substantially harder to reconstruct at the record level within our threat model.

More broadly, the proposed framework is intended for structured EHR and real-world clinical datasets in general, and can in principle support multi-center real-world-evidence (RWE) studies, cross-hospital training of large-scale and foundation models, and emerging clinical data-sharing and data-transaction scenarios, thereby lowering barriers for privacy-preserving clinical and life-science research. By providing a practical, attack-evaluated, and semantically faithful transformation pipeline, we hope to lower the operational barriers to privacy-preserving data reuse across institutions, and to offer privacy researchers a concrete platform on which to build stronger attacks and sharper formalizations. Finally, this work also serves as a case study in human--AI co-design with SciencePal, illustrating a collaborative paradigm in which the AI system proposes candidate operators, attack sketches, and draft text while human researchers formalize, prove, attack, adjudicate, and bear full responsibility for the final scientific claims, ultimately yielding systems that can plausibly be deployed in complex real-world clinical environments~\citep{mitchell2019model}.

% References
% \newpage
\bibliographystyle{config/antgroup}
\bibliography{my}

% ============================================================
%  Updated Author Block
% ============================================================
\author{
  Beining Bao\textsuperscript{1,\dag},
  Maolin Wang\textsuperscript{1,\dag},
  SciencePal\textsuperscript{1,\ddag},
  Gan Yuan\textsuperscript{2},
  Hongyu Chen\textsuperscript{1},
  Bingkun Zhao\textsuperscript{1},
  Baoshuo Kan\textsuperscript{1},
  Yao Wang\textsuperscript{1},
  Qi Shi\textsuperscript{3},
  Yinggong Zhao\textsuperscript{3},
  Wei-Ying Ma\textsuperscript{1},
  Jiming Xu\textsuperscript{4},
  Jun Yan\textsuperscript{1,*}
}

{\renewcommand\thefootnote{}
\footnotetext{%
  \textsuperscript{\dag}Equal contribution.\quad
  \textsuperscript{\ddag}AI co-author (SciencePal Agent); see
  Appendix~\ref{sec:appendix-human-ai} for a detailed role statement.\quad
  \textsuperscript{*}Corresponding author.}
}

% Affiliations (adjust as needed)
% 1 = Your main institution
% 2 = Gan Yuan's institution
% 3 = SciencePal team affiliation
% 4 = YIDU TECH

% ============================================================
%  Appendix A — Contribution Statement
% ============================================================
\section*{Appendix A: Author Contribution Statement}
\label{sec:appendix-human-ai}

A distinguishing feature of this work is that we explicitly include
SciencePal, an AI agent system, as a co-author and document every
contributor's role with full transparency.
We believe that openly crediting AI contributions, rather than
concealing them behind vague acknowledgments, is both more honest and
more reproducible.
To the best of our knowledge, this is among the first efforts to grant
formal co-authorship to an AI system while simultaneously providing a
granular, per-author contribution breakdown that places the AI on equal
footing with human contributors in terms of disclosure.
We hope that this practice may serve as a reference point for future
human\textendash AI collaborative research across the broader scientific
community.
All scientific judgments, ethical responsibilities, and final decisions
rest solely and entirely with the human authors listed in this paper.

\subsection*{Per-author contributions}
\label{sec:appendix-per-author}

\paragraph{Beining Bao.}
Led the execution of all experiments reported in this paper, spanning
geometric validation, statistical divergence analysis,
adversarial-attack evaluation, and end-to-end downstream prediction
tasks on the MIMIC-IV dataset.
Took primary responsibility for hyperparameter tuning across all
operator configurations, conducted systematic sanity checks to verify
numerical correctness, and performed detailed error analyses to trace
unexpected behaviors back to their root causes.
Managed runtime benchmarking across CPU and GPU settings, generated all
figures and tables presented in the experimental sections, and ensured
that every reported number was reproducible from the committed codebase.
Co-drafted substantial portions of the initial manuscript text, with a
particular focus on the method details, experimental setup, results
description, and result interpretation sections.
Also contributed to iterative rounds of revision by incorporating
feedback from other co-authors and refining the narrative flow between
sections.

\paragraph{Maolin Wang.}
Managed the overall project timeline from initial conception through
final manuscript preparation, and coordinated task allocation and
communication among all human authors and between the human team and
the SciencePal AI system throughout the project lifecycle.
Served as the central hub for cross-team collaboration, organizing
regular progress meetings, consolidating feedback from clinical
advisors, engineering contributors, and the SciencePal platform team,
and ensuring that decisions and action items were clearly communicated
and tracked across all parties involved.
On the engineering side, contributed substantially to the codebase by
implementing and maintaining core data processing pipelines, including
the ETL modules that extracted and standardized MIMIC-IV tables into
the feature matrices consumed by downstream operator and attack
experiments.
Wrote and reviewed significant portions of the experimental scripting
infrastructure, such as batch job orchestration, result aggregation,
and automated logging, which enabled the team to run large-scale
experiments efficiently and reproducibly.
Designed and maintained the human\textendash AI collaboration workflow,
which included crafting structured prompts that translated high-level
research goals into actionable AI queries, systematically reviewing and
filtering AI-generated code snippets and text fragments for relevance
and correctness.

\paragraph{SciencePal (AI co-author).}
Contributed to mathematical operator design and formal analysis under
human-specified constraints, including conditions C1 through C5,
CPU-friendliness, and the requirement of no long-term key storage.
Given high-level privacy and utility requirements articulated by human
authors, SciencePal explored a broad space of candidate column-level
transforms, including local rotations, orthogonal projections,
piecewise-linear maps, Householder reflections, permutation-based
shuffles, and various orthogonal mixing schemes, and proposed initial
operator formulations together with analytical sketches for properties
such as mean and variance preservation, invertibility behavior, and
reconstruction difficulty under different adversarial knowledge levels.
A carefully curated subset of these proposals, specifically the T1, T2,
and T3 operators and the per-stay Q-mix mechanism, was subsequently
selected, simplified, and rigorously formalized by the human authors,
who also constructed counterexamples and tightened the arguments where
needed.
Beyond operator design, SciencePal contributed to attack and evaluation
protocol exploration by proposing variants of reconstruction, record
linkage, membership inference, and attribute inference attacks, as well
as suggesting experimental flows for each attack family; human authors
then formalized, simplified, or substantially modified these suggestions
into the final attack matrix and protocols used in the evaluation
sections.
SciencePal also generated prototype code for operations such as z-score
transformation pipelines, simple linear and ridge-regression
reconstruction baselines, statistical divergence computation, and basic
plotting utilities; all of these prototypes were re-implemented,
extended, hardened, and optimized by human authors in a local
environment, and were subjected to security and scalability review
before any use on real clinical data.
In addition, SciencePal produced draft text fragments for selected
method and discussion sections in both English and Chinese, and
suggested alternative structural organizations for the manuscript, such
as splitting the Discussion Section into positioning, clinical
scenarios, operator analysis, Q-mix analysis, limitations, and future
work subsections.
All such textual output was treated as raw material; human authors
thoroughly reviewed, rewrote, trimmed, reorganized, and integrated these
fragments into the final manuscript, ensuring consistency of style,
technical accuracy, and alignment with the overall narrative.

\paragraph{Gan Yuan and Yao Wang.}
Validated the AI-generated intermediate results, including operator
outputs, attack success metrics, and downstream task performance
numbers, against domain knowledge and clinical plausibility throughout
the project.
Contributed to scenario-level validation by assessing whether the
proposed privacy-preserving views remain meaningful and actionable in
realistic hospital analytics workflows, for instance verifying that
transformed vital-sign distributions retain clinically interpretable
ranges and that length-of-stay predictions built on protected views do
not introduce systematic bias in patient subgroups.
Provided feedback on operator parameter choices from the perspective of
practitioners who routinely work with clinical data, and helped identify
edge cases, such as extreme outlier patients or rare diagnosis codes,
where additional robustness testing was warranted.
Also participated in manuscript review by checking that the clinical
motivation sections accurately reflected current hospital data
governance practices and regulatory expectations.

\paragraph{Hongyu Chen, Bingkun Zhao, and Baoshuo Kan.}
Assisted in the day-to-day execution and support of the project across
multiple stages.
Helped prepare and preprocess the MIMIC-IV data tables used in the
experiments, including cleaning raw extracts, aligning stay-level
identifiers across modules, and producing the standardized feature
matrices consumed by the operator and attack pipelines.
Assisted in running batch experiments, collecting and organizing
intermediate outputs, and maintaining version-controlled records of
experimental configurations so that results could be reliably
reproduced.
Contributed to literature survey efforts by collecting, categorizing,
and summarizing related work on differential privacy, synthetic data
generation, and clinical NLP de-identification, which informed the
positioning and related-work sections of the manuscript.
Also assisted in proofreading draft sections, cross-checking notation
consistency across sections, verifying reference formatting, and
preparing supplementary materials such as additional tables and plots
that supported the main experimental narrative.

\paragraph{Wei-Ying Ma, Qi Shi, and Yinggong Zhao.}
Responsible for the design, development, deployment, and ongoing
maintenance of the SciencePal agent platform that served as the AI
backbone of this collaboration.
Provided the underlying large-language-model infrastructure, API
services, prompt orchestration layer, and tool-use capabilities that
enabled the human\textendash AI co-design workflow used throughout this
project.
Ensured system reliability and response quality during the intensive
experimentation and writing phases, iterated on the agent's mathematical
reasoning and code generation modules in response to feedback from the
research team, and supported the integration of domain-specific
knowledge sources that improved the relevance of SciencePal's operator
and attack suggestions.
Their engineering efforts made it possible for the research team to
interact with SciencePal as a responsive, context-aware collaborator
rather than a text completion tool.

\paragraph{Jiming Xu.}
Provided deep domain expertise in clinical data systems, electronic
health record architectures, and health-care information technology
infrastructure.
Verified the clinical value and practical relevance of the proposed
privacy framework by evaluating whether the operator-based view
generation approach is compatible with the data pipelines, access
control models, and audit requirements found in production hospital
environments.
Served as a consulting advisor on real-world deployment considerations,
including computational budget constraints in resource-limited hospital
IT departments, regulatory alignment with data protection standards, and
the practical trade-offs between privacy strength and data utility that
clinical stakeholders care about most.
Also contributed insights on how the proposed framework could
interoperate with existing commercial de-identification solutions and
hospital data warehouse systems, and reviewed the manuscript sections on
system architecture and clinical scenarios for factual accuracy.

\paragraph{Jun Yan.}
Defined the core research problem, specifically the notion of
``in-hospital usable and visible, but hard to reconstruct'' numeric
views, and formulated the overall modeling framework that structures the
entire paper.
This framework includes the mean\textendash variance manifold
perspective that provides the geometric foundation, the z-score
\(\alpha\)-hierarchy that organizes operators by privacy strength, and
the threat model with adversarial capability levels L0, L1, and L2 and
attack families A, B, C, and D.
Independently validated key theoretical results, including the
conditions under which each operator satisfies or violates properties
C1 through C5, and cross-checked critical experimental conclusions
against back-of-the-envelope calculations to ensure consistency.
Contributed to the design and partial implementation of several
SciencePal agent capabilities that were used in this collaboration,
bridging the gap between general-purpose language model reasoning and
the specialized mathematical and privacy-engineering demands of the
project.
Supervised the overall scientific direction, resolved disagreements on
modeling choices, made final decisions on which results to include or
exclude, and performed the last round of manuscript review to ensure
that all claims are adequately supported by evidence.

\subsection*{Boundaries and responsibility}
\label{sec:appendix-boundaries}

\paragraph{Independent verification.}
All scientific conclusions and quantitative results reported in this
paper were independently verified by at least one human author before
inclusion in the manuscript.
This verification covers, but is not limited to, attack success rates,
\(R^{2}\) reconstruction curves, AUROC and AUPRC scores for downstream
prediction tasks, Kolmogorov\textendash Smirnov statistics for
distributional fidelity, runtime benchmarks across CPU and GPU
configurations, and the correctness of all figures and tables.
Any preliminary interpretation or explanation provided by SciencePal
during the research process was treated strictly as a hypothesis to be
checked and, where necessary, re-derived or corrected by human authors;
no AI-generated interpretation was accepted at face value.

\paragraph{Responsibility.}
Although SciencePal is credited as a co-author to transparently and
accurately reflect its substantive contributions to operator design,
code prototyping, and text drafting, all theoretical claims,
methodological choices, threat model specifications, experimental
designs, parameter settings, interpretive conclusions, and policy
recommendations presented in this paper are the sole responsibility of
the human authors.
The human authors collectively accept full accountability for the
correctness, reproducibility, and ethical compliance of the work, and
any errors or omissions should be attributed to them rather than to the
AI system.

\paragraph{Referencing this collaboration.}
We encourage other researchers who wish to reference the
human\textendash AI collaboration model adopted in this work to do so in
a way that clearly distinguishes the roles of human authors and the AI
system.
For example, one may write:
\emph{``This work credits SciencePal (AI agent) as a co-author for its
contributions to operator design, analytical sketching, and drafting
support; all scientific responsibility rests with the human authors.''}
We also welcome future studies that build upon, refine, or critique the
co-authorship and disclosure framework presented here, as we believe
that establishing community norms for transparent AI attribution is an
evolving process that benefits from open discussion and diverse
perspectives.

\end{document}